\documentclass[a4paper,12pt,oneside]{book}
\usepackage{multicol}
\usepackage{epsfig}
\usepackage{amsmath}
\usepackage{tikz}
\usepackage{amsfonts}
\usepackage{amssymb}
\usepackage{dsfont}
\usepackage{graphicx}
\usepackage[all]{xy}
\usepackage{colortbl}
\usepackage{color}
\usepackage{array}
\usepackage{float}
\usepackage{caption}
\usepackage[latin1]{inputenc}
\usepackage{wasysym}
\usepackage{fancyhdr}
\begin{document}
\pagestyle{empty}
\begin{figure}[h]
\begin{center}
\includegraphics[scale=1]{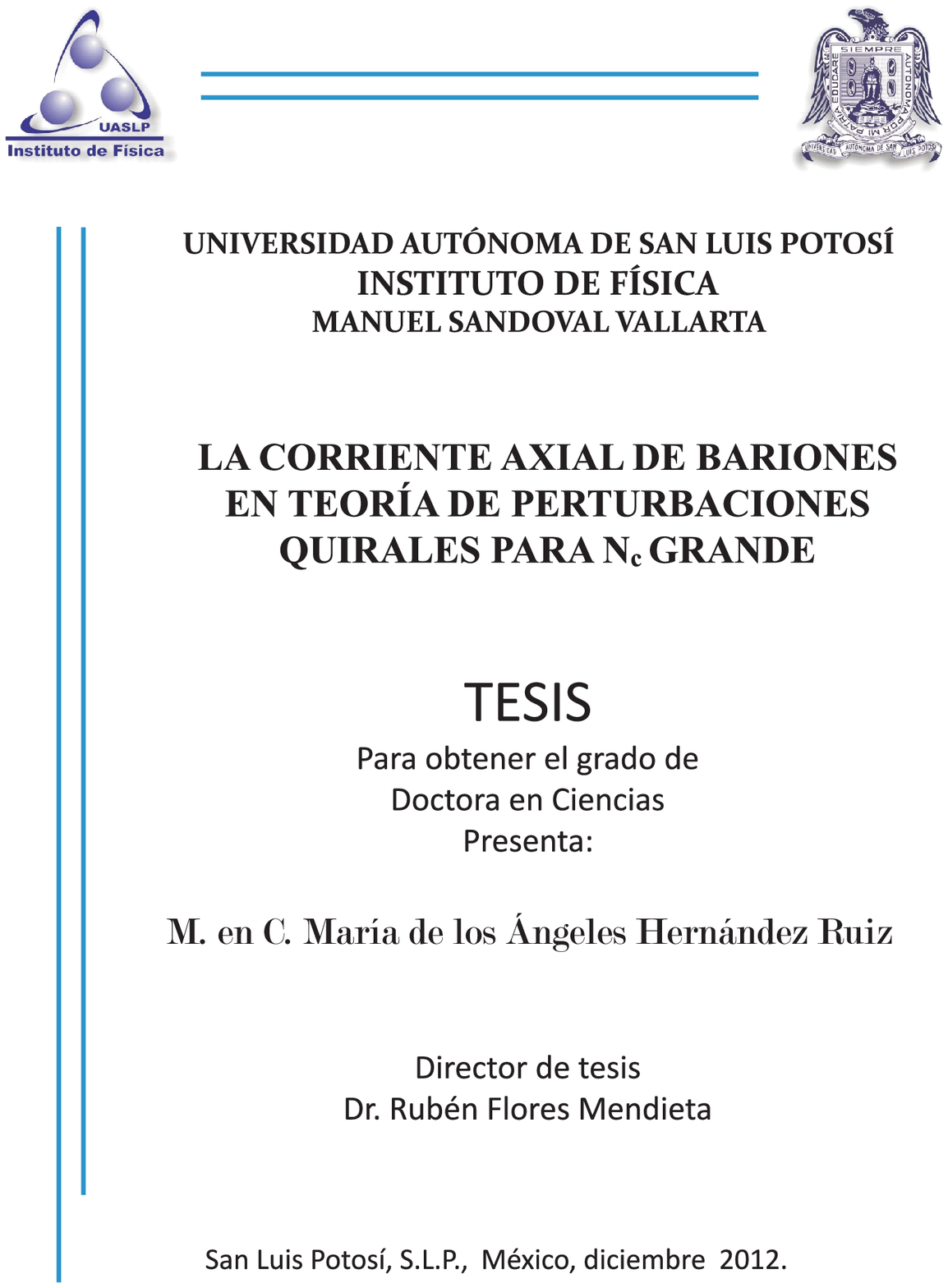}
\end{center}
\end{figure}

\chapter*{}
\thispagestyle{plain}
\begin{center}
{\Large \bf La corriente axial de bariones en teor{\'\i}a de perturbaciones quirales para ${\mathbf \it N}_c$ grande}
\end{center}
\vspace{1.5cm}
\centerline{\Large \bf Resumen}
\vspace{1cm}
En este trabajo de tesis calculamos el operador de corriente axial para bariones en un formalismo combinado entre la teor{\'\i}a de perturbaciones quirales y la expansi\'on $1/N_c$, donde $N_c$ es el n\'umero de colores. En este c\'alculo se consideran diagramas de Feynman a orden de un loop, con estados intermediarios octete y decuplete. Obtenemos correccio\-nes al operador de corrien\-te axial a nivel de un loop y con rotura de simetr{\'\i}a de sabor perturbativa. Las primeras correcciones vienen de los diagramas de Feynman, entonces hablamos de una rotura de simetr{\'\i}a impl{\'\i}cita en el l{\'\i}mite quiral $m_q \rightarrow 0$, donde $m_q$ es la masa del quark y las segundas correcciones se obtienen al ignorar la rotura de isoesp{\'\i}n y en ese caso se incluye la rotura de simetr{\'\i}a SU(3) perturbativa a primer orden, conduciendo a una rotura de simetr{\'\i}a expl{\'\i}cita. Los elementos de matriz  de las componentes espaciales del opera\-dor axial entre los estados de la simetr{\'\i}a esp{\'\i}n sabor, dan los valores usuales de los acoplamientos axial vector. Para el octete de bariones, los acoplamientos axial vector son $g_A$, tal como est\'an definidos en los experimentos en decaimientos semilept\'onicos de bario\-nes, donde $g_A \approx 1.27$ para el decaimiento beta del neutr\'on. Para los decaimientos fuertes de bariones los acoplamientos axial vector son $g$, los cuales son extra{\'\i}dos de las anchuras de los decaimientos fuertes del decuplete de bariones al octete de bariones y piones. El c\'alculo de este trabajo nos permite realizar diferentes ajustes por m{\'\i}nimos cuadrados, es decir, ajustando nuestras expresiones anal{\'\i}ticas con los datos experimentales podemos hacer la comparaci\'on entre la teor{\'\i}a y el experimento. Encontramos que la comparaci\'on de nuestros resultados te\'oricos con el experimento tiene una total consistencia.

\pagestyle{fancy}
\tableofcontents
\chapter*{Introducci\'on}
En la actualidad el estudio de las propiedades est\'aticas de los bariones en teor{\'\i}a de perturbaciones quirales y en el l{\'\i}mite de $N_c$ grande,
ha sido impulsado por el desarrollo de las teor{\'\i}as de campo efectivas. Con esta motivaci\'on, en este trabajo de tesis se estudia la renormalizaci\'on de la corriente axial vector de bariones a orden de un loop, en el contexto de una expansi\'on combinada: Este poderoso m\'etodo consiste, en considerar la combinaci\'on entre la teor{\'\i}a de perturbaciones quirales con la expansi\'on $1/N_c$ \cite{JenkinsPRD53, RFM2006}. As{\'\i}, podemos describir las interacciones, entre el octete de bariones de esp{\'\i}n $\frac{1}{2}$ y el decuplete de bariones de esp{\'\i}n $\frac{3}{2}$ con el octete de bosones pseudoescalares de Goldstone, m\'as la contribuci\'on del bos\'on $\eta'$.

El objetivo principal de este trabajo, es determinar correcciones a nivel de un loop y con rotura de simetr{\'\i}a SU(3) perturbativa, sobre los acopla\-mien\-tos axial vector $g_A$ y $g$, para los decaimientos semilept\'onicos del octete de bariones y para los decaimientos fuertes del decuplete de bariones a piones respectivamente, cuando $N_c = 3$.
La contribuci\'on m\'as importante en este trabajo es el c\'alculo de la corriente axial vector para el caso no degenerado, es decir, $\Delta/m_\Pi \neq 0$, donde $\Delta$ es la diferencia de masas octete-decuplete y $m_\Pi$ son las masas de los piones, kaones y $\eta$ (los bosones pseudoescalares de Goldstone). Las expresiones que se obtienen a este orden, contienen inserciones de masa de bariones a trav\'es del operador $\cal{M}$ y de esta manera los t\'erminos conducen al caso no degenerado. Para el desarrollo de este trabajo, seguimos el for\-ma\-lis\-mo como se ha presentado en la Ref. \cite{RFM2006} y utilizamos el Lagrangiano quiral $1/N_c$ para bariones al orden m\'as bajo, el cual est\'a dado en Ref. \cite{JenkinsPRD53}. Esta formulaci\'on Lagrangiana da origen a importantes desarrollos en el estudio de la estructura esp{\'\i}n sabor de bariones en la expansi\'on $1/N_c$.

La generalizaci\'on de la Cromodin\'amica Cu\'antica (QCD), desde $N_c = 3$ hasta $N_c\gg 3$ co\-lo\-res es lo que conocemos como el l{\'\i}mite de $N_c$ grande.
El sector de bariones para el l{\'\i}mite de $N_c$ grande presenta una simetr{\'\i}a contra{\'\i}da esp{\'\i}n sabor SU($2N_f$), donde $N_f$ es el n\'umero de sabores de los tres quarks ligeros, $u$, $d$ y $s$. La simetr{\'\i}a contra{\'\i}da esp{\'\i}n sabor contenida en SU(6) nos permite clasificar estados de bariones y elementos de matriz de operadores bari\'onicos \cite{JenkinsPRD53, {JenkinsNucl}}, de esta manera, es posible calcular las propiedades est\'aticas de bariones en la expansi\'on $1/N_c$ \cite{DashenPRD49,DashenPRD51}. Las cantidades f{\'\i}sicas que se han calculado en este formalismo son, las masas de bariones \cite{JenkinsPRD53,Bedaque,Y.Oh}, los momentos magn\'eticos de bariones \cite{LutyD51, RMFPRD80(2009), tesisGiovanna, Giovanna, Jen2012} y la corriente axial vector de bariones \cite{RFM2006,tesisra,tesisJR,RFM61(2000),RFM2000}. En el l{\'\i}mite de $N_c$ grande los estados bari\'onicos octete y decuplete pueden ser degenerados, la diferencia de masas $\Delta$ entre las masas del octete de bariones $M_B$ y decuplete de bariones $M_T$ est\'an dadas por $\Delta \equiv M_T - M_B \propto 1/N_c$.

Como sabemos, la QCD se comporta de forma muy distinta a altas y  a bajas energ{\'\i}as; es decir, mientras los quarks disfrutan de libertad asint\'otica a cortas distancias y la teor{\'\i}a de perturbaciones es aplicable; a largas distancias quedan confinados dentro de hadrones y su evoluci\'on se vuelve altamente no perturbativa. Esto hace muy dif{\'\i}cil realizar un an\'alisis de la din\'amica de QCD en t\'erminos de quarks y gluones. Para dar soluci\'on a este problema,
G. \'{}t Hooft \cite{Hooft}, se di\'o cuenta que QCD tiene un par\'ametro oculto $N_c$, el n\'umero de color y que la teor{\'\i}a se simplifica, en el l{\'\i}mite $N_c \rightarrow \infty$. En el l{\'\i}mite de $N_c$ grande, un bari\'on es un estado confinado de $N_c$ quarks y llaga a ser un estado con un n\'umero infinito de quarks.  Por otra parte, las predicciones de QCD en el l{\'\i}mite $N_c \rightarrow \infty$ sa\-tis\-fa\-cen relaciones de simetr{\'\i}a esp{\'\i}n sabor contenidas en SU(6) para los quarks ligeros. Estas relaciones de simetr{\'\i}a son las mismas que aquellas obtenidas en el modelo de quarks no relativista, el cual da un grupo id\'entico y as{\'\i} podemos ob\-te\-ner resultados te\'oricos en el l{\'\i}mite de $N_c$ grande.

Por otra parte, la teor{\'\i}a de perturbaciones quirales explota la simetr{\'\i}a del Lagrangiano de QCD bajo transformaciones de SU(3)$_L$ $\times$ SU(3)$_R$ $\times$ U(1) de los tres sabores de quarks ligeros en el l{\'\i}mite $m_q \rightarrow 0$, donde $m_q$ es la masa del quark. El vac{\'\i}o de QCD se alinea en una cierta direcci\'on en el espacio de simetr{\'\i}a interno que representan las transformaciones quirales, entonces hablamos de una rotura espont\'anea del grupo de simetr{\'\i}a global
SU(3)$_L$ $\times$ SU(3)$_R$ $\times$ U(1) a un subgrupo diagonal SU(3) $\times$  U(1). La rotura espont\'anea tiene una consecuencia muy importante, la aparici\'on del octete pseudoescalar de bosones de Goldstone. Puesto que los quarks $u$, $d$ y $s$ tienen una peque\~na masa, la simetr{\'\i}a del Lagrangiano no es exacta. Los observables f{\'\i}sicos pueden ser expandidos orden por orden en potencias de $p^2/\Lambda_\chi^2$ y $m_\Pi^2/\Lambda_\chi^2$, o equivalentemente, $m_q/\Lambda_\chi$ donde $p$ es el momento del mes\'on y $\Lambda_\chi$ es la escala de rotura de la simetr{\'\i}a quiral.

Adicionalmente, la teor{\'\i}a de perturbaciones quirales puede extenderse para incluir grados de li\-ber\-tad fermi\'onicos, tales
como los bariones, cuyas propiedades bajo la transformaci\'on quiral fijan sus acoplamientos a los mesones y es conveniente
considerar al campo de bariones est\'aticos dependiente de la velocidad, este formalismo se conoce como, la teor{\'\i}a de perturbaciones quirales para bariones pesados \cite{Jenkins255(91), {Jenkins259(91)}}.

A partir de la informaci\'on experimental disponible en los decaimientos semilept\'onicos de
bariones y de los decaimientos fuertes de bariones a piones \cite{PDG} podemos determinar las constantes de acoplamiento mes\'on-bari\'on,
el an\'alisis num\'erico del c\'alculo consiste en hacer ajustes, mediante t\'ecnicas de minimizaci\'on, con los datos experimentales y los par\'ametros de la expansi\'on $1/N_c$.

Este trabajo est\'a estructurado de la siguiente manera. En el Cap{\'\i}tulo 1, presentamos los fundamentos de la teor{\'\i}a de las interacciones fuertes.
En el Cap{\'\i}tulo 2, incorporamos la teor{\'\i}a de campo efectiva con el Lagrangiano quiral efectivo para bariones $1/N_c$ al orden m\'as bajo. En el Cap{\'\i}tulo 3, presentamos el formalismo combinado entre la teor{\'\i}a de perturbaciones quirales y la expansi\'on $1/N_c$. Calculamos el operador de corriente axial vector y lo presentamos en t\'erminos de las contribuciones de sabor singulete, octete, ${\bf 10} + \overline{\bf 10}$ y {\bf 27}, a un loop y con rotura de simetr{\'\i}a SU(3) perturbativa a primer orden. En el Cap{\'\i}tulo 4, hacemos un an\'alisis num\'erico detallado, realizando ajustes por m{\'\i}nimos cuadrados de las expresiones te\'oricas con la informaci\'on experimental disponible. Finalmente, en el Cap{\'\i}tulo 5 presentamos las conclusiones correspondientes de este trabajo. Los detalles t\'ecnicos se consideran en cinco ap\'endices. El Ap\'endice A, contiene las integrales a un loop para los diagramas de Feynman respectivos. El Ap\'endice B, contiene el c\'alculo de uno de los t\'erminos de las estructuras conmutador/anticonmutador de operadores de bariones. El Ap\'endice C, la reducci\'on de los operadores de bariones, en el Ap\'endice D, escribimos las ecuaciones completas de las contribuciones de sabor {\bf 8} y {\bf 27} para el acoplamiento axial $g_A$ y finalmente el Ap\'endice E, contiene los elementos de matriz de los operadores bari\'onicos.

\chapter{La Cromodin\'amica Cu\'antica QCD}

La parte del Modelo Est\'andar que describe las interacciones fuertes entre quarks a trav\'es del intercambio de gluones es la Cromodin\'amica Cu\'antica (QCD). \'Esta es una teor{\'\i}a de norma no abeliana basada en el grupo SU(3). La QCD postula que la fuerza es debida al color de los quarks y los gluones son los bosones de norma, los cuales son part{\'\i}culas sin masa y de esp{\'\i}n $1$; debido a la naturaleza no abeliana del grupo de norma, los gluones tambi\'en tienen color como los quarks. Un glu\'on acopla a dos quarks con color, pero conserva el sabor (la interacci\'on responsable del cambio de sabor es la interacci\'on d\'ebil de corriente cargada). Adem\'as, tambi\'en existen acoplamientos entre gluones, lo que explica el corto alcance de la interacci\'on. Todas estas propiedades quedan perfectamente descritas por la QCD. El contenido de este cap{\'\i}tulo se basa en las Refs. \cite{Gasiorowicz,ChengLi}.

\section{El desarrollo de la QCD}
Actualmente el espectro hadr\'onico est\'a compuesto por cientos de part{\'\i}culas que pueden clasificarse en dos grandes grupos dependiendo de su esp{\'\i}n: los {\bf mesones} con esp{\'\i}n entero (bosones) y los {\bf bariones} con esp{\'\i}n semientero (fermiones).

El estudio de la interacci\'on fuerte, responsable de la existencia de los n\'ucleos at\'omicos, ha ido cambiando nuestra concepci\'on de la materia al nivel m\'as b\'asico. Las mediciones experimentales se han mejorado y comprueban las interacciones entre los quarks; de esta manera permitieron las primeras pruebas cuantitativas de la QCD, es decir, lo que empez\'o siendo f{\'\i}sica nuclear, se desarroll\'o r\'apidamente en una nueva disciplina, al descubrirse experimentalmente la existencia de una numerosa familia de part{\'i}culas que interaccionan fuertemente, los llamados hadrones de los que el prot\'on y el neutr\'on solo son los miembros m\'as representativos. Despu\'es de una serie de experimentos en colisiones muy inel\'asticas se estableci\'o la realidad f{\'\i}sica de los quarks.

Los quarks se unen en {\bf estados ligados} para formar hadrones. \'Estos permiten entender de forma sencilla la totalidad del espectro hadr\'onico a partir de dos reglas b\'asicas:
\begin{enumerate}
  \item Los {\bf mesones} son estados compuestos por un quark y un antiquark $q\bar{q}$. Son bosones con esp{\'\i}n entero $J = 0,1,2,...$
  \item Los {\bf bariones} est\'an formados por tres quarks $qqq$. Son fermiones con esp{\'\i}n simientero $J = \frac{1}{2}, \frac{3}{2}, ...$
\end{enumerate}
Si bien los quarks son los constituyentes fundamentales de la materia, nunca se han detectado en estado libre. S\'olo se pueden observar combinaciones de quarks y de acuerdo con la simetr{\'\i}a SU(3), son estados neutros de color. Conocemos seis tipos o sabores distintos de quarks, m\'as sus correspondientes antiquarks con cargas opuestas. En la Tabla {\ref{quarks}} listamos los seis sabores de quarks con su correspondiente carga el\'ectrica $Q$, la cual, es una fracci\'on de la carga el\'ectrica del electr\'on.

\begin{table}[h]
\begin{center}
\begin{tabular}{cccc}
\hline
\hline
\vspace*{0.25cm}
 $Q= +\frac{2}{3}e$ &  u (up) & c (charm) & t (top) \\
  \hline
  $Q= -\frac{1}{3}e$ & d (down) & s (strange) & \vspace{0.25cm} b (beauty) \\
\hline
\hline
\end{tabular}
\caption{Tipos o sabores de quarks. \label{quarks}}
\end{center}
\end{table}

El Modelo de Quarks establece la relaci\'on entre los estados hadr\'onicos propuestos a partir de la combinaci\'on de quarks (o antiquarks). Por otra parte, la funci\'on de onda orbital de los quarks en los bariones debe ser totalmente sim\'etrica. Esto es porque corresponden a los sistemas m\'as ligados de tres quarks, al menos los que aparecen en multipletes a energ{\'\i}as m\'as bajas, como el octete con esp{\'\i}n $\frac{1}{2}$ y el decuplete con esp{\'\i}n $\frac{3}{2}$. Si la funci\'on no fuera sim\'etrica, existir{\'\i}an nodos en la funci\'on de onda cuando las coordenadas de dos de los quarks coincidieran. Estos nodos aumentar{\'\i}an la energ{\'\i}a cin\'etica y reducir{\'\i}an el efecto de atracci\'on de los quarks. Por otro lado, el momento angular orbital de los tres quarks, en su centro de masas, debe ser $L=0$. Si no es as{\'\i}, la funci\'on de onda se anular{\'\i}a en ciertas direcciones, de forma an\'aloga a los arm\'onicos esf\'ericos con $L\neq0$.

Gell-Mann (1961, 1964) y Neeman (1961) clasificaron a los hadrones conocidos de acuerdo a representaciones del grupo SU(3) de sabor.
En la Fig. \ref{Fig:8_9_quarks} tenemos el octete de bariones de esp{\'\i}n $\frac{1}{2}$ y paridad positiva. Los ocho bariones m\'as ligeros pueden organizarse en un hex\'agono con dos part{\'\i}culas en su centro. En la Fig. \ref{Fig:8_9_quarks} el octete de mesones son de esp{\'\i}n cero. Los ocho mesones m\'as ligeros pueden distribuirse en un hex\'agono con dos part{\'\i}culas en el centro. Por otra parte, en la Fig. \ref{Fig:decuplete} se presenta el decuplete de bariones de esp{\'\i}n $\frac{3}{2}$. Diez bariones pesados se organizan en un tri\'angulo.

\begin{figure}[h]
\begin{center}
\setlength{\unitlength}{5mm}

\begin{picture}(11.5,10.0)(1.85,-2.5981)

\put(-7.5,-5.5){\framebox(30.5,13)}

\put(0,6.0){\makebox(0,0){{\sf \large Octete de bariones}}}
\put(15,6.0){\makebox(0,0){{\sf \large Octete de mesones}}}

\put(-7.5,4.8){\line(90,0){30.5}}
\put(7.5,-5.5){\line(0,90){13}}

\put(-2,3.5981){\makebox(0,0){$n$}}
\put(2,3.5981){\makebox(0,0){$p$}}
\put(-4,1.166){\makebox(0,0){$\Sigma^{-}$}}
\put(0,1.166){\makebox(0,0){$\Lambda, \Sigma^{0}$}}
\put(4,1.166){\makebox(0,0){$\Sigma^{+}$}}
\put(-2,-1.366){\makebox(0,0){$\Xi^{-}$}}
\put(2,-1.366){\makebox(0,0){$\Xi^{0}$}}

\put(13,3.5981){\makebox(0,0){$K^0$}}
\put(17,3.5981){\makebox(0,0){$K^+$}}
\put(11,1.166){\makebox(0,0){$\pi^-$}}
\put(15,1.166){\makebox(0,0){$\pi^0, \eta$}}
\put(19,1.166){\makebox(0,0){$\pi^+$}}
\put(13,-1.366){\makebox(0,0){$K^-$}}
\put(17,-1.366){\makebox(0,0){$\overline{K}^0$}}

\end{picture}
\vspace*{1.2truecm}
\caption{\label{Fig:8_9_quarks}Bariones de esp{\'\i}n $\frac{1}{2}$ y mesones de
esp{\'\i}n cero.}
\end{center}
\end{figure}

\begin{figure}[h]
\begin{center}
\setlength{\unitlength}{5mm}

\begin{picture}(11.5,10.0)(1.85,-2.5981)

\put(-7.5,-5.5){\framebox(30.5,13)}

\put(0,6.0){\makebox(0,0){{\sf \large Decuplete de bariones}}}
\put(15,6.0){\makebox(0,0){{\sf \large Contenido de quarks}}}

\put(-7.5,4.8){\line(90,0){30.5}}
\put(7.5,-5.5){\line(0,90){13}}

\put(-6,3.5981){\makebox(0,0){$\Delta^-$}}
\put(-2,3.5981){\makebox(0,0){$\Delta^0$}}
\put(2,3.5981){\makebox(0,0){$\Delta^+$}}
\put(6,3.5981){\makebox(0,0){$\Delta^{++}$}}
\put(-4,1.166){\makebox(0,0){$\Sigma^{*-}$}}
\put(0,1.166){\makebox(0,0){$\Sigma^{*0}$}}
\put(4,1.166){\makebox(0,0){$\Sigma^{*+}$}}
\put(-2,-1.366){\makebox(0,0){$\Xi^{*-}$}}
\put(2,-1.366){\makebox(0,0){$\Xi^{*0}$}}
\put(0,-3.5981){\makebox(0,0){$\Omega^{-}$}}

\put(9,3.5981){\makebox(0,0){$ddd$}}
\put(13,3.5981){\makebox(0,0){$ddu$}}
\put(17,3.5981){\makebox(0,0){$duu$}}
\put(21,3.5981){\makebox(0,0){$uuu$}}
\put(11,1.166){\makebox(0,0){$dds$}}
\put(15,1.166){\makebox(0,0){$dus$}}
\put(19,1.166){\makebox(0,0){$uus$}}
\put(13,-1.366){\makebox(0,0){$dss$}}
\put(17,-1.366){\makebox(0,0){$uss$}}
\put(15,-3.5981){\makebox(0,0){$sss$}}

\end{picture}
\vspace*{1.2truecm}
\caption{\label{Fig:decuplete} Representaci\'on del decuplete de bariones con esp{\'\i}n $\frac{3}{2}$ de acuerdo a la clasificaci\'on del grupo SU(3) de sabor.}
\end{center}
\end{figure}
\vspace{0.5cm}

El prot\'on, por ejemplo, corresponde a la combinaci\'on {\bf uud}, mientras que el neutr\'on
es un estado {\bf udd}. Sin embargo, el bari\'on $ \Delta^{++}$ est\'a formado por tres quarks
{\bf u} con momento angular orbital relativo $L=0$; tiene esp{\'\i}n $J =\frac{3}{2}$ y, por tanto, el
estado con polarizaci\'on $J_z = +\frac{3}{2}$ corresponde a {\bf u$^\uparrow$ u$^\uparrow$ u$^\uparrow$}
con los espines de los tres quarks alineados en la misma direcci\'on. Esta funci\'on de onda es sim\'etrica,
dando lugar a un estado $\Delta^{++}$ con la estad{\'i}stica equivocada (bos\'on en lugar de fermi\'on).

El problema de la estad{\'\i}stica es general y afecta a todos los bariones. Su resoluci\'on exige
introducir un nuevo n\'umero cu\'antico, el ``color''. Cada sabor de quark tiene $N_c = 3$ posibles colores:
q$^\alpha$, $\alpha = $ 1, 2, 3 (rojo, verde, azul). Esto permite antisimetrizar las funciones de onda
bari\'onicas en el espacio de color, restituyendo as{\'\i} la estad{\'\i}stica correcta. Los bariones y mesones
corresponden a las combinaciones neutras de color

\begin{equation}
B = \frac{1}{\sqrt{6}} \epsilon_{\alpha\beta\gamma} \mid q^\alpha q^\beta q^\gamma >,
\end{equation}

\begin{equation}
M=\frac{1}{\sqrt{3}} \delta_{\alpha\beta} \mid q^\alpha q^\beta >.
\end{equation}

La raz\'on por la que los quarks son part{\'\i}culas que no se pueden observar libres, sino que est\'an siempre
confinados al interior de bariones o mesones, hay que buscarla en las caracter{\'\i}sticas peculiares de la interacci\'on fuerte, es decir, la interacci\'on fuerte est\'a caracterizada por dos importantes propiedades que a primera vista parecen contradictorias:

\begin{enumerate}
  \item {\bf Confinamiento.} Lo que observamos de forma directa son hadrones y no quarks. Por tanto, la interacci\'on entre los quarks tiene que ser sumamente fuerte para mantenerlos siempre confinados en el interior de los hadrones. Este comportamiento implica que es necesaria una enorme cantidad de ener\-g{\'\i}a para separar dos quarks. Por ejemplo, el par quark-antiquark que forma un mes\'on. Se crea una cuerda de fuerza fuerte entre ellos hasta el punto de que llegado un cierto momento es ener\-g\'eticamente favorable la creaci\'on de un nuevo par quark-antiquark, por lo que el estado final es de dos mesones, en lugar de conseguir quarks libres. De hecho creemos que es literalmente imposible tener quarks como estados asint\'oticos.
      El color es un n\'umero cu\'antico escondido ya que todos los hadrones conocidos son neutros de color.

  \item {\bf Libertad asint\'otica.} Los experimentos realizados a altas energ{\'\i}as nos muestran que los quarks se comportan como part{\'\i}culas casi libres. Es decir, cuando la distancia que separa dos quarks se hace muy pequeña, la intensidad de la interacci\'on, en lugar de hacerse mayor, disminuye. Por ello, cuando est\'an muy pr\'oximos, los quarks se comportan como si estuvieran libres. Este peculiar comportamiento se denomina libertad asint\'otica y fue el descubrimiento de David Gross, Frank Wilezek y David Politzer por el que fueron galardonados con el premio Nobel de f{\'\i}sica en 2004. No existe contradicci\'on entre esta propiedad y la anterior porque el confinamiento se activa s\'olo a largas distancias, del orden del radio de un hadr\'on t{\'\i}pico. A muy cortas distancias los quarks son casi libres.
\end{enumerate}

\section{El Lagrangiano de la QCD}

La evidencia a favor de una teor{\'\i}a de campo de las interacciones fuertes tiene argumentos muy s\'olidos. Entre ellos se encuentran los siguientes:
\begin{itemize}
    \item Los hadrones est\'an compuestos de quarks y tienen carga fraccionaria.
    \item Los quarks son fermiones con esp{\'\i}n $\frac{1}{2}$ y de distintos colores.
    \item Sabemos que el color exhibe una simetr{\'\i}a SU(3).
    \item Los quarks sienten la interacci\'on fuerte.
    \item Adem\'as de los quarks hay partones adicionales en el
    n\'ucleo.
    \item Los partones no sienten la interacci\'on
    electromagn\'etica ni la d\'ebil.
\end{itemize}

El Lagrangiano de la QCD est\'a dado por la ecuaci\'on siguiente
\begin{eqnarray}\label{ec:L-QCD}
\mathcal{L}_\mathrm{QCD}= -\frac{1}{4}G_{\mu\nu}^{a}
G^{a\mu\nu}+\bar{\psi}(x)(i\gamma^{\mu} D_{\mu}-\hat{m})\psi(x),
\label{eq:qcdlag}
\end{eqnarray}
donde
\begin{equation}
\psi(x)= \left( \begin{array}{c} u(x)\\ d(x)\\ s(x)\\ c(x)\\ t(x)\\ b(x) \end{array} \right),
\end{equation}
representa el campo de quarks con seis sabores y con tres colores
impl{\'\i}citos. Las masas de los quarks est\'an agrupadas en la
matriz de masa $\hat{m}$ definida por
\begin{equation}
\hat{m} = \mathrm{diag}(m_{u}, m_{d}, m_{s}, m_{c}, m_{t}, m_{b})
\end{equation}
en el espacio de sabor. La derivada covariante
\begin{equation}\label{Ta}
D_\mu=\partial_\mu-ig_s T^a  A_\mu^{a},
\end{equation}
est\'a relacionada con los campos glu\'onicos $A_\mu^{a}$ (con el {\'\i}ndice de color $a = 1,\ldots,8$) y
\begin{equation}\label{f(abc)}
G_{\mu\nu}^{a}=\partial_\mu A_\nu^a-\partial_\nu A_\mu^a + g_s f^{abc} A_\mu^{b}A_\nu^c,
\end{equation}
es el tensor de campo glu\'onico y $g_s$ la constante de acoplamiento de la interacci\'on fuerte. Podemos darnos cuenta que la masa de un quark es independiente de su color.

En la Ec.~(\ref{Ta}), $T^a$ representa a los 8 generadores del grupo SU(3) contenidos en el Lagrangiano de QCD, es decir, $T^a=\lambda^{a}/2$, con $\lambda^{a}$ las matrices de Gell-Mann, las cuales son,

\begin{eqnarray}
\lambda^1 = \left(\begin{array}{ccc} 0 & 1 & 0 \\ 1 & 0 & 0
\\ 0 & 0 & 0 \end{array} \right),
\hspace{1cm} \lambda^2 = \left(
\begin{array}{ccc} 0 & -i & 0 \\ i & 0 & 0\\
0 & 0 & 0 \end{array} \right), \hspace{1cm}
\lambda^3 = \left( \begin{array}{ccc} 1 & 0 & 0 \\ 0 & -1 & 0 \\
0 & 0 & 0
\end{array} \right), \nonumber
\end{eqnarray}

\begin{eqnarray}
\lambda^4 = \left( \begin{array}{ccc} 0 &0 &1 \\0 & 0 & 0\\
1& 0 & 0\end{array} \right),
\hspace{1cm} \lambda^5 = \left(\begin{array}{ccc} 0 & 0 & -i\\ 0 & 0 & 0\\
i & 0 & 0 \end{array} \right), \hspace{1cm} \lambda^6 =
\left(\begin{array}{ccc} 0 & 0 & 0 \\ 0 & 0 & 1 \\ 0 & 1 & 0
\end{array} \right), \nonumber
\end{eqnarray}

\begin{eqnarray}
\lambda^7 = \left(\begin{array}{ccc} 0 & 0 & 0 \\ 0 & 0 & -i \\
0 & i & 0\end{array} \right), \hspace{1cm} \lambda^8 =
\sqrt{\frac13} \left(\begin{array}{ccc} 1 &0 & 0 \\ 0 & 1 &0\\ 0 &
0 & -2\end{array} \right). \nonumber
\end{eqnarray}

El grupo SU(3) es de rango dos, los generadores diagonales son
$\lambda^3$ y $\lambda^8$ con vectores
propios simult\'aneos,

\begin{eqnarray}
{\rm R} = \left( \begin{array}{l} 1 \\ 0 \\ 0 \end{array} \right)
\, , \qquad {\rm G} = \left( \begin{array}{l} 0 \\ 1 \\ 0
\end{array} \right) \, , \qquad {\rm B} = \left( \begin{array}{l}
0 \\ 0 \\ 1 \end{array} \right). \nonumber
\end{eqnarray}

Las constantes de estructura del grupo SU(3) est\'an dadas por $f^{abc}$ y est\'an contenidas en la Ec.~(\ref{f(abc)}); \'estas se
definen a trav\'es de las relaciones (\ref{fabcT})
\begin{equation}\label{fabcT}
    \left[ T^a , T^b \right] = i f^{abc}T^c,
\end{equation}
y son totalmente antisim\'etricas bajo el intercambio de dos de sus
{\'\i}ndices \cite{Gasiorowicz}. Sus valores son listados en la Tabla \ref{t:f(abc)}.
Podemos derivar tambi\'en relaciones de anticonmutaci\'on,

\begin{equation}\label{}
    \left \{ T^a, T^b \right \} = \frac{1}{3} \delta^{ab}\,
    {\mathds{I}} + \frac{1}{2} d^{abc} \lambda^c,
\end{equation}
donde $\mathds{I}$ denota la matriz identidad de $3 \times 3$ y las constantes $d^{abc}$ son sim\'etricas en los tres {\'\i}ndices. Los valores
de las constantes de estructura se listan en la Tabla \ref{t:f(abc)}.

\begin{table}[h]
\begin{center}
\begin{tabular}{cccc}
\hline\hline
$(abc)$ & $f^{abc}$ & $(abc)$ & $d^{abc}$ \\
\hline
123 & 1        &    118 & $1/\sqrt{3}$ \\
147 & $1/2$ & 146 & $1/2$\\
156 & $-1/2$ & 157 & $1/2$\\
246 & $1/2$ & 228 & $1/\sqrt{3}$\\
257 & $1/2$ & 247 & $-1/2$\\
345 & $1/2$ & 256 & $1/2$\\
367 & $-1/2$ & 338 & $1/\sqrt{3}$\\
458 & $\sqrt{3}/{2}$ & 344 & $1/2$\\
678 & $\sqrt{3}/{2}$ & 355 & $1/2$\\
&&366& $-1/2$\\
&&377& $-1/2$\\
&&448& $-1/2\sqrt{3}$\\
&&558& $-1/2\sqrt{3}$\\
&&668& $-1/2\sqrt{3}$\\
&&778& $-1/2\sqrt{3}$\\
&&888& $-1/\sqrt{3}$ \\
\hline\hline
\end{tabular}
\caption{Los valores no nulos de las constantes $f^{abc}$ y
$d^{abc}$. \label{t:f(abc)}}
\end{center}
\end{table}
El Lagrangiano de la QCD dado en la Ec. (\ref{ec:L-QCD}) es invariante ante transformaciones de norma SU(3) en el espacio de color. Por el car\'acter no abeliano del grupo de norma, la QCD tiene ciertas caracter{\'\i}sticas que la diferencian de las teor{\'\i}as de norma abelianas como la Electrodin\'amica Cu\'antica (QED). Algunas particularidades son:

\begin{enumerate}
\item El Lagrangiano de la QCD contiene acoplamientos glu\'onicos (v\'ertices de tres y cuatro gluones). Los gluones transportan carga de color.
\item La QCD es asint\'oticamente libre, es decir, el acoplamiento se debilita a cortas distancias o equivalentemente a momentos grandes.
\item Inversamente, el acoplamiento se vuelve intenso a bajos momentos. Por tal motivo, no es
posible aplicar teor{\'\i}a de perturbaciones en la QCD para
describir hadrones de masas menores que 2 GeV.
\end{enumerate}

Los acoplamientos entre bosones de norma son caracter{\'\i}sticos de una teor{\'\i}a de norma basada en un grupo no abeliano donde los bosones de norma llevan la carga de interacci\'on. $-$El color en el caso de QCD$-$ y de este modo es posible acoplarse directamente ellos mismos. La parte fermi\'onica del lagrangiano es una suma sobre todos los sabores de quarks, un t\'ermino de campo libre y un t\'ermino para el acoplamiento quark-glu\'on.

La QCD es una teor{\'\i}a simple y en principio, enormemente predictiva, ya que tiene un \'unico par\'ametro libre: la constante fundamental de la interacci\'on fuerte
\begin{equation}
\alpha_s \equiv g_s^2 / (4\pi).
\end{equation}
A pesar de su simplicidad e indiscutible elegancia matem\'atica, parece a primera vista una teor{\'\i}a extraordinariamente alejada del mundo real. ¿C\'omo podemos pretender describir las interacciones entre hadrones y la estructura de los n\'ucleos, donde los quarks y gluones son totalmente invisibles? ¿Por qu\'e una teor{\'\i}a conceptualmente tan simple exhibe un espectro tan extraordinariamente complejo a largas distancias? De hecho, si despreciamos las masas de los quarks, la QCD no contiene ning\'un par\'ametro con dimensiones.

\subsection{Efectos cu\'anticos}

Cu\'anticamente, el fot\'on intermediario puede dar lugar a la creaci\'on y posterior ani\-qui\-la\-ci\'on de pares virtuales electr\'on-positr\'on, que act\'uan como pequeños dipolos modificando la in\-te\-ra\-cci\'on. Las correcciones cu\'anticas m\'as importantes est\'an asociadas con esta auto-energ{\'\i}a del fot\'on

\begin{eqnarray}\label{ec:TQ2}
  T(Q^2) & \sim & \frac{\alpha}{Q^2} \,\, \left\{ 1 - \Pi(Q^2) + \Pi(Q^2)^2 + \cdots \right\} = \frac{\alpha}{Q^2} \frac{1}{1 + \Pi(Q^2)} \nonumber \\
         & \sim & \frac{\alpha(Q^2)}{Q^2}.
\end{eqnarray}
donde $Q^2$ es el momento transferido en la amplitud de colisi\'on $T(Q^2) \sim \alpha/Q^2$.
\noindent Esto define un acoplamiento efectivo (``running'') que depende logar{\'\i}tmicamente de $Q^2$

\begin{equation}\label{ec:alphaQ2}
  \alpha(Q^2) = \frac{\alpha(Q_0^2)}{1 - \frac{\beta _1}{2\pi} \alpha(Q_0^2){\rm ln}(Q^2/Q_0^2)}.
\end{equation}

La constante usual de estructura fina, medida a bajas energ{\'i}as, corresponde a $\alpha = \alpha(m_e^2).$ En Electrodin\'amica Cu\'antica (QED) $\beta_1 = 2/3 > 0$. Por lo tanto, la intensidad de la interacci\'on electromagn\'etica aumenta (disminuye) con la energ{\'\i}a (distancia). In\-tui\-tivamente, el vac{\'\i}o cu\'antico de QED se comporta como un medio diel\'ectrico polarizado, donde los dipolos formados por los pares virtuales $e^- e^+$ apantallan la carga el\'ectrica. Aunque $\alpha$ es pequeña, la enorme diferencia entre la masa del electr\'on ($m_e = 0.51$ MeV) y la del bos\'on electrod\'ebil Z ($M_Z = 91$ GeV) hace que este efecto cu\'antico sea muy relevante para los experimentos de precisi\'on realizados en el acelerador europeo LEP (CERN, Ginebra),

\begin{equation}\label{ec:alphame}
\alpha^{-1}(m_e^2) = 137.036 > \alpha^{-1} (M_Z^2) = 128.95 \pm 0.05
\end{equation}

En QCD ocurre algo muy similar. El glu\'on que media la interacci\'on fuerte entre dos quarks genera pares quark-antiquark que apantallan la intensidad de la fuerza de color. Sin embargo, debido a las auto-interacciones del campo glu\'onico, tambi\'en se producen pares virtuales glu\'on-glu\'on con el efecto contrario. El acoplamiento efectivo resultante para la interacci\'on fuerte tiene formalmente la misma dependencia en $Q^2$ dada por la Ec. (\ref{ec:TQ2}), pero con un coeficiente $\beta_1$ negativo

\begin{equation}\label{beta1}
  \beta_1 = \frac{2N_f - 11N_c}{6} < 0.
\end{equation}

La contribuci\'on positiva proporcional al n\'umero de sabores de quarks $N_f$ est\'a ge\-ne\-ra\-da por los pares quark-antiquark, mientras que las auto-energ{\'\i}as glu\'onicas introducen el t\'ermino negativo adicional proporcional al n\'umero de colores $N_c$. Como $\beta_1 < 0$, la intensidad
de la interacci\'on disminuye (aumenta) al aumentar la energ{\'\i}a (distancia). Esto demuestra que la libertad asint\'otica de la interacci\'on fuerte a altas energ{\'\i}as es una consecuencia de la din\'amica de QCD. Este hecho crucial, caracter{\'\i}stico de las teor{\'\i}as de Yang-Mills, fue descubierto por Gross, Wilczek y Politzer en 1973. El fuerte aumento de la interacci\'on a bajas energ{\'\i}as hace tambi\'en muy plausible el confinamiento de los quarks. No obstante, como los c\'alculos perturbativos dejan de tener validez cuando $\alpha_s$ se hace muy grande, la demostraci\'on matem\'atica de esta \'ultima propiedad sigue plan\-tean\-do enormes dificultades t\'ecnicas, aunque ha sido establecida mediante simulaciones num\'ericas.

Dado que $\alpha_s$ es una cantidad adimensional, su dependencia en $Q^2$ debe venir nor\-ma\-li\-zada por otro par\'ametro dimensional. \'Este se denota convencionalmente como $\Lambda_{QCD}$ y representa f{\'\i}sicamente aquella escala energ\'etica donde el acoplamiento ``runing'' a que antes nos hemos referido se hace de orden unidad. Por lo tanto $\Lambda_{QCD}$ establece un l{\'\i}mite inferior absoluto  a la validez de la teor{\'\i}a de perturbaciones, que aproximadamente es del orden de una escala hadr\'onica t{\'\i}pica. \'Este es el par\'ametro dimensional que la mec\'anica cu\'antica introduce en una teor{\'\i}a que cl\'asicamente es independiente de cualquier otra escala y que nos permite hablar de cortas y largas distancias.


\section{Simetr{\'\i}as y leyes de conservaci\'on}
Partiendo de las simetr{\'\i}as que posee el Lagrangiano que
describe el sistema es posible construir la teor{\'\i}a y
describir las interacciones que hay entre las part{\'\i}culas del
espectro. Es decir, las simetr{\'\i}as juegan un papel fundamental
en f{\'\i}sica te\'orica dado que est\'an detr\'as de las
propiedades que caracterizan un sistema f{\'\i}sico. Por ello, en
esta secci\'on se presentan algunos aspectos importantes sobre
simetr{\'\i}as y leyes de conservaci\'on.

El lenguaje matem\'atico natural de las simetr{\'\i}as es la teor{\'\i}a
de grupos. De la isotrop{\'\i}a y homogeneidad del espacio-tiempo,
se supone como grupo de simetr{\'\i}a de los sistemas f{\'\i}sicos
fundamentales el grupo de Poincar\'e, es decir, invarianza bajo
traslaciones y transformaciones de Lorentz. Las funciones de
estado de las part{\'\i}culas deben llevar entonces
representaciones de este grupo  y estar\'an clasificadas por sus
operadores de Casimir \cite{Wigner}.

De hecho las simetr{\'\i}as son una herramienta muy valiosa para construir una Teor{\'\i}a de Campo Efectiva como se presenta en el siguiente cap{\'\i}tulo.

\subsubsection{Teorema de Noether}
La simetr{\'\i}a tiene una relaci\'on directa con las
leyes de conservaci\'on. Fue en 1918 cuando Emmy Noether
(1982-1935) pudo probar que {\it simetr{\'\i}a implica leyes de
conservaci\'on}. El teorema de Noether se basa en las propiedades
de invariancia del Lagrangiano de un sistema bajo la acci\'on de
ciertas transformaciones de simetr{\'\i}a. A las leyes de
conservaci\'on que obedece dicho sistema, se les llama tambi\'en
``principios''.

Para un sistema descrito por el Lagrangiano $L$,
\begin{equation}
L = \int  d^{3}x{\cal L}(\phi_{i}(x),\partial_{\mu}\phi_{i}(x)),
\end{equation}
donde ${\cal L}$ es la densidad Lagrangiana, con la ecuaci\'on de
movimiento
\begin{eqnarray}
\partial_{\mu}\frac{\delta {\cal L}}
{\delta(\partial_{\mu}\phi_{i})} - \frac{\delta {\cal
L}}{\delta\phi_{i}}=0.
\end{eqnarray}
\noindent La cantidad m\'as importante para un sistema mec\'anico
es la acci\'on,
\begin{equation}
S=\int L dt.
\end{equation}
Cualquier transformaci\'on de simetr\'{\i}a continua para la cual
la acci\'on  es invariante implica la existencia de una corriente
conservada
\begin{eqnarray}
\partial^{\mu}J_{\mu}(x)=0,
\end{eqnarray}
con la carga definida por
\begin{eqnarray}
Q(t)= \int d^{3}x J_{0}(x)
\end{eqnarray}
y que es una constante de movimiento
\begin{eqnarray}
\frac{dQ}{dt}=0,
\end{eqnarray}
porque el t\'ermino de la superficie en el infinito es despreciablemente peque\~no.

Por otra parte, los campos $\phi_{i}(x)$ se transforman de la
misma manera para todos los puntos $x$ del espacio-tiempo. La
densidad Lagrangiana ${\cal L}$ es invariante bajo algunos grupos
de simetr\'{\i}a, es decir, bajo las transformaciones
infinitesimales
\begin{eqnarray}
\phi_{i}(x)\rightarrow\phi'_{i}(x) = \phi_{i}(x)+\delta\phi_{i}(x),
\end{eqnarray}
con
\begin{eqnarray}
\delta\phi_{i}(x) = i{\cal E}^{a}T^{a}_{ij}\phi_{j}(x),
\end{eqnarray}
\noindent donde ${\cal E}^{a}$ (independientes de $x$) son
par\'ametros peque\~{n}os y $T^{a}$ son un conjunto de matrices
que satisfacen el \'algebra de Lie del grupo de norma,
\begin{eqnarray}
\lbrack T^{a}, T^{b} \rbrack = i f^{abc}T^{c}
\end{eqnarray}
\noindent y  $f^{abc}$ son las constantes de estructura del grupo de norma. El cambio en la densidad Lagrangiana est\'a dado por
\begin{eqnarray}
\delta{\cal L} = \frac{\delta{\cal
L}}{\delta\phi_{i}}\delta\phi_{i} + \frac{\delta{\cal
L}}{\delta(\partial_{\mu}\phi_{i})}\delta(\partial_{\mu}\phi_{i}).
\label{eq:trans}
\end{eqnarray}
Usando la ecuaci\'on de movimiento y el hecho de que
$\partial_\mu$ y $\delta$ conmutan
\begin{eqnarray}
\delta(\partial_{\mu}\phi_{i}) \equiv
\partial_{\mu}\phi'_{i}-\partial_{\mu}\phi_{i}=\partial_{\mu}(\delta\phi_{i}),
\end{eqnarray}
podemos escribir $\delta{\cal L}$ como
\begin{eqnarray}
\delta{\cal L} &=& \partial_{\mu}\frac{\delta {\cal
                   L}}{\delta(\partial_{\mu}\phi_{i})}
                   \delta\phi_{i} +
                   \frac{\delta{\cal
                   L}}{\delta(\partial_{\mu}\phi_{i})}
                   \partial_{\mu}(\delta\phi_{i})\nonumber\\
               &=& \partial_{\mu}\Biggl[\frac{\delta {\cal
                   L}}{\delta(\partial_{\mu}\phi_{i})}
                   \delta\phi_{i}\Biggr]\nonumber\\
               &=& {\cal E}^{a}\partial_{\mu}\Biggl[\frac{\delta
                   {\cal L}}{\delta(\partial_{\mu}\phi_{i})}iT^{a}_{ij}\phi_{j}\Biggr].
\end{eqnarray}
Claramente si el Lagrangiano es invariante bajo la
transformaci\'on
\[
\phi_{i}(x) \rightarrow \phi'_{i}(x) = \phi_{i}(x)  +
            \delta\phi_{i}(x),
\]
donde
\[
\delta\phi_{i}(x) = i {\cal E}^{a}T^{a}_{ij}\phi_{j}(x),
\]
entonces $\delta{\cal L} = 0$ implica una corriente conservada dada por
\begin{equation}
\partial^{\mu}J^{a}_{\mu}=0
\end{equation}
con
\begin{equation}
J^{a}_{\mu}=-i\frac{\delta {\cal L}}{\delta(\partial^{\mu}
\phi_{i})}T^{a}_{ij}\phi_{j}.
\end{equation}
Con este procedimiento se ha encontrado una corriente conservada, $J^a_\mu$, cuya
existencia es consecuencia de la invariancia de la acci\'on bajo
las transformaciones de $\phi_i(x)$. Lo anterior implica la
conservaci\'on de una carga $Q(t)$.
\[
\frac{dQ(t)}{dt}=0,\hspace{1cm} \mbox{donde} \hspace{1cm}
Q(t)=\int d^3 x J_0(x).
\]
Este resultado se conoce con el nombre de Teorema de
Noether, el cual juega un papel muy importante en teor{\'\i}a de
campo y f{\'\i}sica de part{\'\i}culas. Este teorema da cuenta de
la conservaci\'on de la energ{\'\i}a, momento lineal, momento
angular y de n\'umeros cu\'anticos que poseen las
part{\'\i}culas tales como carga, isosp{\'\i}n, color, etc.

\subsection{La simetr{\'\i}a quiral}

En la simetr{\'\i}a SU(3) de sabor los mesones
pseudoescalares, los mesones vectoriales y los bariones son
ordenados en representaciones de singuletes y octetes, es decir,
las ra{\'\i}ces de la simetr{\'\i}a quiral hay que buscarlas en
{\it el \'algebra de corrientes} y en la clasificaci\'on de hadrones
de Gell-Mann.

No se sabe cuales son las fuerzas fundamentales que rompen las
simetr{\'\i}as de los ba\-rio\-nes. Lo que generalmente se hace, es ver
como se transforma bajo un determinado grupo G (simetr{\'\i}a de norma) la
parte no sim\'etrica del Hamiltoniano y de esa manera
obtener informaci\'on sobre la din\'amica del sistema. Debido a
esto es conveniente descomponer el operador Hamiltoniano en dos partes:
\begin{equation}\label{Hamiltoniano}
    {\cal H} = {\cal H}_0 + \lambda {\cal H}',
\end{equation}
\noindent donde ${\cal H}_0$ es la parte sim\'etrica y $\cal H'$
la parte que rompe esta simetr{\'\i}a. De acuerdo con el Teorema
de Noether podemos relacionar la conservaci\'on de la corriente
$J^\mu_a(x)$ con la invariancia del Hamiltoniano ${\cal H}_0$,
\begin{equation}\label{}
    Q_a (x_0) = \int J^0_a(x)d^3 x.
\end{equation}
En el caso de que $\lambda = 0$, las cargas $Q_a(x_0)$ son
independientes del tiempo y por tanto se conservan.

La teor{\'\i}a m\'as importante sobre las simetr{\'\i}as aproximadas fue postulado por Gell-Mann \cite{Gell-Mann1}.
De acuerdo a esta teor{\'\i}a los operadores forman una representaci\'on del grupo de simetr{\'\i}a considerando
\begin{equation}\label{}
    \left [ Q_a(x_0), Q_b(x_0) \right ] = if_{abc}Q_c(x_0).
\end{equation}
Existen dos maneras de que una simetr{\'\i}a del Hamiltoniano
${\cal H}_0$ sea rea\-li\-za\-da
\begin{itemize}
    \item La realizaci\'on de Wigner-Weyl,
    \begin{equation}\label{}
    Q_a(x_0)|0\rangle = 0.
    \end{equation}
    \item La realizaci\'on de Nambu-Goldstone
    \begin{equation}\label{}
    Q_a(x_0)|0\rangle \neq 0.
    \end{equation}
\end{itemize}

En la realizaci\'on de Wigner-Weyl la simetr{\'\i}a se manifiesta
directamente en el espectro del Hamiltoniano ${\cal H}_0$ como una
degeneraci\'on de los multipletes, si $\lambda = 0$.

De la realizaci\'on de Nambu-Goldstone se sigue la existencia de
bosones con masa cero si $\lambda = 0$ y no mostrar\'a ninguna
degeneraci\'on en la estructura de los multipletes. A este tipo de
realizaci\'on se le denomina rotura espont\'anea de la simetr{\'\i}a.
En este caso particular los operadores de carga $Q_a(x_{\mu})$ no
son definidos o mejor dicho no son normalizables.

Gell-Mann hizo un postulado adicional el cual establece que los
generadores $Q_a(x_0)$ de las simetr{\'\i}as hadr{\'o}nicas
coinciden con las cargas, las cuales est\'an relacionadas con las
corrientes en las interacciones d\'ebiles y electromagn\'eticas.
Adem\'as postul\'o que las corrientes d\'ebiles tambi\'en
cumplen relaciones similares de conmutaci\'on.
\begin{equation}\label{corriente}
    \delta (x_0 - y_0)\left[ J^0_a(x), J^0_b(y) \right] = i \delta^4(x-y) f_{abc}J^0_c(x),
\end{equation}
donde  $J^0_a(x)$ son las corrientes d\'ebiles. Esto
significa que las interacciones d\'ebiles nos pueden servir para
informarnos sobre las simetr{\'\i}as y sobre la rotura de las
simetr{\'\i}as de las interacciones fuertes. De estas ideas se
origin\'o el {\it \'algebra de corrientes} en f{\'\i}sica de
part{\'\i}culas.

Del decaimiento $\beta$ fueron postuladas la existencia de las
corrientes vectoriales y co\-rrien\-tes axiales (Teor{\'\i}a V-A) por
Feynman, Gell-Mann, Marshak y Sudarshan.

Gell-Mann Postul\'o tambi\'en para las corrientes vectoriales y
axiales un \'algebra similar a la ecuaci\'on (\ref{corriente})

\begin{eqnarray}
  \delta (x_0 - y_0)\left [ V^0_a(x), V^0_b(y) \right ] &=& i f_{abc}V^0_c(x)\delta^4(x-y), \nonumber \\
  \delta (x_0 - y_0)\left [ V^0_a(x), A^0_b(y) \right ] &=& i f_{abc}A^0_c(x)\delta^4(x-y), \\
  \delta (x_0 - y_0)\left [ A^0_a(x), A^0_b(y) \right ] &=& i
  f_{abc}V^0_c(x)\delta^4(x-y). \nonumber
\end{eqnarray}
Ahora podemos definir nuevos operadores de carga,
\begin{eqnarray}
  F_a (x) &=& \int d^3 x V^0_a(x), \nonumber\\
  F^5_a(x_0) &=& \int d^3 x A^0_a(x).
\end{eqnarray}
Estas cargas generan el \'algebra de Lie de SU(3)$_L$ $\times$ SU(3)$_R$.
\begin{eqnarray}\label{ec. 5}
  \left [F_a(x_0), F_b(x_0) \right ] &=& i f_{abc}F(x_0), \nonumber \\
  \left [F_a(x_0), F^5_b(x_0) \right ] &=& i f_{abc}F^5_c(x_0), \\
  \left [F^5_a(x_0), F^5_b(x_0) \right ] &=& i f_{abc}F_c(x_0).\nonumber
\end{eqnarray}

Para observar mejor la estructura de esta \'algebra de
Lie SU(3)$_L$ $\times$ SU(3)$_R$ podemos definir unos nuevos
operadores $F^\pm (x_0)$ donde

\begin{equation}
    F^\pm_a(x_0) = \frac{1}{2} \left [ F_a(x_a) \pm F^5_a (x_0)\right ];
\end{equation}
reemplazando en las relaciones de conmutaci\'on Ec. (\ref{ec. 5}) obtendremos las siguientes relaciones
\begin{eqnarray}
  \left [F^\pm_a(x_0), F^\pm_b(x_0)\right ] &=& if_{abc}F^\pm_c(x_0), \nonumber \\
  \left [F^+_a(x_0), F^-_b(x_0)\right ] &=& 0.
\end{eqnarray}
Esto demuestra que los generadores $F_a(x_0)$ y $F^5_a(x_0)$ generan el producto directo SU(3)$_L$ $\times$ SU(3)$_R$ y en este
caso se habla de {\bf las simetr{\'\i}as quirales.}

La simetr{\'\i}a SU(3) exige que todas las part{\'\i}culas que pertenecen al mismo multiplete posean la misma masa. Las simetr{\'\i}as quirales SU(3)$_L$ $\times$ SU(3)$_R$ no exigen multipletes de part{\'\i}culas pero en cambio poseen ocho mesones pseudoescalares (Teorema de Golds\-tone), es decir, tres piones ($\pi^0$, $\pi^+$ y $\pi^-$), cuatro kaones ($K^0$, $\overline{K}^0$, $K^+$ y $K^-$) y $\eta$, como se muestra en la Fig. 1.1(b).

Supongamos que el Hamiltoniano ${\cal H}$ posea la simetr{\'\i}a quiral. Esto significa

\begin{equation}\label{}
    \left[F^\pm_a, {\cal H} \right]=0.
\end{equation}
Supongamos adicionalmente que el operador de carga
$F_a(x_0)$ deja invariante el vac{\'i}o, esto es

\begin{equation}\label{}
    F_a(x)|0\rangle = 0.
\end{equation}

Pero nosotros no podemos hacer la misma suposici\'on para las cargas axiales ya que en el espectro hadr\'onico no se encuentran en el mismo multiplete part{\'\i}culas con diferente paridad. En lugar de hacer esta suposici\'on, supondremos que el vac{\'\i}o no es invariante bajo la simetr{\'\i}a quiral SU(3)$_L$ $\times$ SU(3)$_R$. Esto significa que:

\begin{equation}\label{}
    F^5_a(x)|0\rangle \neq 0.
\end{equation}

De aqu{\'\i} se obtiene el teorema de Goldstone por el cual en la
simetr{\'\i}a quiral SU(3)$_L$ $\times$ SU(3)$_R$ los mesones
pseudoescalares deben de tener masa igual a cero. Naturalmente
nosotros sabemos que en el mundo f{\'\i}sico todos los mesones
poseen masa. Esto significa que la simetr{\'\i}a quiral
SU(3)$_L$ $\times$ SU(3)$_R$ no puede ser exacta.

\subsubsection{La simetr{\'\i}a quiral en el vac{\'\i}o de QCD}

Para ampliar el tema de que el vac{\'\i}o de la QCD no es invariante bajo la simetr{\'\i}a quiral, fijemos nuestra atenci\'on en el papel fundamental que juegan las simetr{\'\i}as globales de QCD, en particular la simetr{\'\i}a quiral asociada con el n\'umero cu\'antico de sabor. Si despreciamos las masas de los quarks, lo cual es una aproximaci\'on excelente para los quarks $u$ y $d$ y bastante buena para el $s$, el Lagrangiano de QCD asociado a estos tres quarks ligeros
\begin{equation}\label{LQCDq}
  {\cal L} = \bar{q} i \gamma^\mu D_\mu q = \bar{q}_L i \gamma^\mu D_\mu q _L + \bar{q}_R i \gamma^\mu D_\mu q_R, \hspace{1cm}  q\equiv(u, d, s),
\end{equation}
es invariante bajo transformaciones unitarias en el espacio de los tres sabores de los quarks, ya que sus interacciones son id\'enticas.
Los hadrones que pertenecen a un mismo multiplete tienen aproximadamente la misma masa e id\'enticos espines, paridades, n\'umero bari\'onico, etc.

Sin embargo, el Lagrangiano Ec. (\ref{LQCDq}) posee demasiadas simetr{\'\i}as. Al separar los quarks sus quiralidades izquierda y derecha (la quiralidad es un n\'umero cu\'antico que en el caso de fermiones sin masa coincide exactamente con la helicidad, o proyecci\'on del momento angular en la direcci\'on del movimiento), observamos que hay dos simetr{\'\i}as SU(3) independientes para las dos quiralidades. Esto implicar{\'\i}a una duplicidad de multipletes  hadr\'onicos con paridades opuestas que no existe en el espectro observado.

La discrepancia entre las simetr{\'\i}as del Lagrangiano (la interacci\'on) y del espectro se debe a que el estado fundamental $-$el vac{\'\i}o$-$ de QCD no respeta la simetr{\'\i}a quiral. El vac{\'\i}o de QCD se alinea en una cierta direcci\'on en el espacio de simetr{\'\i}a interno que representan las transformaciones quirales. En t\'erminos matem\'aticos hablamos de una rotura espont\'anea del grupo de simetr{\'\i}a global SU(3)$_L$ $\times$ SU(3)$_R$ a su subgrupo diagonal SU(3)$_V$.

En t\'erminos f{\'i}sicos, la rotura espont\'anea tiene una consecuencia muy importante, la aparici\'on de ocho mesones pseudoescalares (paridad negativa) sin masa. Puesto que los quarks $u$, $d$ y $s$ tienen una pequeña masa, la simetr{\'\i}a del Lagrangiano no es exacta y estas ocho part{\'\i}culas adquieren una masa si bien muy pequeña en comparaci\'on con los restantes hadrones. La rotura espont\'anea de simetr{\'\i}a implica propiedades muy peculiares para estas part{\'\i}culas, cuya din\'amica a bajas energ{\'\i}as puede predecirse de forma rigurosa mediante argumentos basados \'unicamente en la simetr{\'\i}a subyacente. La confirmaci\'on experimental de todas estas predicciones ha demostrado inequ{\'\i}vocamente el fen\'omeno de rotura espont\'anea de la simetr{\'\i}a quiral en el vac{\'\i}o de QCD.

\section{Corrientes de simetr{\'\i}a como corrientes f{\'\i}sicas}
El \'algebra de corriente que representa simetr{\'\i}as de la interacci\'on fuerte puede ser probada directamente en
electromagnetismo o en procesos de interacci\'on d\'ebil que involucren hadrones. Las corrientes de simetr{\'\i}a son
justamente las corrientes f{\'\i}sicas que aparecen en electromagnetismo e interacciones d\'ebiles, es decir, las mismas
co\-rrien\-tes de Noether o combinaciones lineales de ellas aparecen en el Lagrangiano de interacci\'on.

\subsubsection{Corriente electromagn\'etica}
La corriente electromagn\'etica $J_{\lambda}^\mathrm{em}(x)$
est\'a acoplada al campo del fot\'on $A_{\lambda}(x)$ en el
Lagrangiano de interacci\'on dado por
\begin{eqnarray}
\mathcal{L}^\mathrm{em} = eJ^\mathrm{em}_{\lambda}A^{\lambda},
\end{eqnarray}
donde $e$ es la constante de acoplamiento (carga el\'ectrica).
Separando la corriente en dos partes, una lept\'onica y otra
hadr\'onica, tenemos
\begin{eqnarray}
J^\mathrm{em}_{\lambda} = J^\mathrm{em}_{l\lambda}
+J^\mathrm{em}_{h\lambda}.
\end{eqnarray}
La corriente lept\'onica se puede escribir directamente en
t\'erminos de los campos de los leptones cargados mientras que la
parte hadr\'onica se puede expresar en t\'erminos de campos de
quarks.

La corriente electromagn\'etica es una corriente de simetr{\'\i}a
y  se conserva en todas las interacciones conocidas. El operador
de carga hadr\'onico
\begin{eqnarray}
Q^\mathrm{em}_{h} = \int J^\mathrm{em}_{h0}d^{3}x
\end{eqnarray}
obedece las relaciones de Gell-Mann-Nishijima
\begin{eqnarray}
Q^{\rm{em}}_{h} =  {\rm I}_z + \frac{\rm Y}{2},
\end{eqnarray}
donde ${\rm I}_{z}$ es la tercera componente de isoesp{\'\i}n y {\rm Y} es la hipercarga.
Esto implica una relaci\'on similar para las corrientes correspondientes
\begin{eqnarray}
J^{\rm{em}}_{h\lambda} = J^{3}_{\lambda} +
\frac{1}{2}J^{Y}_{\lambda}.
\end{eqnarray}

\subsubsection{Corriente d\'ebil}
En interacciones d\'ebiles la corriente juega un papel similar a
$J^\mathrm{em}_{\lambda}$ en la inte\-racci\'on
electromagn\'etica. Estas dos interacciones est\'an unificadas en
teor{\'\i}as de norma modernas y son miembros de un mismo
multiplete. La corriente d\'ebil cargada $J_\lambda$ se acopla al
campo $W_\lambda$ del bos\'on vectorial intermediario cargado como
\begin{eqnarray}
\mathcal{L}^{W} = g_s J_{\lambda}W^{\lambda} + \mbox{h.c.},
\end{eqnarray}
donde $g_s$ es la constante de acoplamiento, es decir, $g_s$
sen$\theta_W = e$. De aqu{\'\i} tenemos el Lagrangiano efectivo a
bajas energ{\'\i}as para una interacci\'on corriente-corriente,
\begin{eqnarray}
\mathcal{L}_\mathrm{eff} =
-\frac{G_{F}}{\sqrt{2}}J^{\dagger}_{\lambda}J^{\lambda} +
\mbox{h.c.},
\end{eqnarray}
donde $G_{ F} = \frac{\sqrt{2}g_s^2}{8M_W^2}\simeq 1.17 \times
10^{-5}$ GeV$^{-2}$, es la constante de Fermi. La corriente d\'ebil
$J_{\lambda}$ puede separarse tambi\'en en partes lept\'onica y
hadr\'onica
\begin{eqnarray}
J^{\lambda} = J^{\lambda}_{l} + J^{\lambda}_{h},
\end{eqnarray}
con
\begin{eqnarray}
J^\lambda_l = \bar{\nu}_e \gamma^{\lambda}
(1-\gamma_{5})\mathrm{e} +
\bar{\nu}_\mu\gamma^{\lambda}(1-\gamma_{5})\mu + \ldots
\label{eq:corrl}
\end{eqnarray}
La Ec.~(\ref{eq:corrl}) posee expl{\'\i}citamente  la estructura
$V-A$ tal y como fue postulada por Feynman y Gell-Mann e
independientemente por Sudarshan y Marshak. En dicha ecuaci\'on,
$\nu_e(x)$ y $\nu_\mu(x)$ son los operadores de campo para los
neutrinos mientras que e y $\mu$ representan los correspondientes
operadores de campo para los leptones cargados electr\'on y
mu\'on, respectivamente. La corriente hadr\'onica $J^{\lambda}_{h}$, escrita tambi\'en en
la forma $V-A$, est\'a dada por
\begin{equation}
J^{\lambda}_{h} = \lbrack(V^{\lambda}_{1} + i V^{\lambda}_{2})-(A^{\lambda}_{1} + iA^{\lambda}_{2})\rbrack \cos\theta_{c}+
\lbrack(V^{\lambda}_{4} + iV^{\lambda}_{5})-(A^{\lambda}_{4} + iA^{\lambda}_{5}) \rbrack \sin \theta_{c},
\end{equation}
donde $\theta_c$ es el \'angulo de Cabibbo. Los
sub{\'\i}ndices en los operadores de corrientes vector y
axial vector son {\'\i}ndices del octete de SU(3).
Existen varias reglas de selecci\'on y relaciones de simetr{\'\i}a
implicadas por las propiedades de transformaci\'on bajo SU(2) y
SU(3) de las corrientes hadr\'onicas (Modelo de
Cabibbo) que han sido probadas en procesos semilept\'onicos
d\'ebiles. Las corrientes de isosp{\'\i}n son aproximadamente conservadas.
Este hecho se conoce como \textit{hip\'otesis de la corriente
vectorial conservada} (CVC).

Conociendo los elementos de matriz de la corriente electromagn\'etica, los cuales est\'an relacionados directamente con los factores de forma, podemos predecir los correspondientes factores de forma del sector $\Delta {S}=0$ en procesos semilept\'onicos de bariones a trav\'es de rotaciones de isosp{\'\i}n. An\'alogamente, los factores de forma d\'ebiles para el sector $|\Delta {S}|=1$ se pueden fijar por rotaciones de SU(3), dado que esas corrientes vectoriales son miembros del mismo multiplete. Las corrientes vector y axial vector satisfacen el \'algebra de Lie SU(3)$_L$ $\times$ SU(3)$_R$ Ec. (\ref{ec. 5}).


\chapter{Teor{\'\i}as de Campo Efectivas}

Las teor{\'\i}as de campo efectivas proporcionan una herramienta muy poderosa para analizar las consecuencias de la simetr{\'\i}a quiral.
La idea, que en realidad es aplicable a cualquier descripci\'on efectiva de QCD o de cualquier otra teor{\'\i}a de campos, es la siguiente:
Si estamos interesados en describir la f{\'\i}sica hasta una escala de energ{\'\i}a y momento $\Lambda_\chi$, solamente necesitamos retener como grados de libertad expl{\'\i}citos aquellos que pueden producirse cuando en un proceso los 4-momentos externos son menores que $\Lambda_\chi$; es decir, aquellas part{\'\i}culas que son ligeras comparadas con $\Lambda_\chi$. La informaci\'on sobre los grados de libertad m\'as pesados queda incluida en las constantes de acoplamiento del Lagrangiano efectivo local que describe la f{\'\i}sica de bajas energ{\'\i}as. Si las simetr{\'\i}as son suficientemente restrictivas, s\'olo podr\'an existir en el Lagrangiano efectivo un n\'umero muy pequeño de t\'erminos con la dimensi\'on adecuada a una densidad Lagrangiana.

En teor{\'\i}as no perturbativas (como la QCD a bajas energ{\'\i}as) es posible construir una teor{\'\i}a de campo efectiva predictiva para
fen\'omenos a bajas energ{\'\i}as combinando una serie de potencias de operadores con las restricciones de simetr{\'\i}a bajo la teor{\'\i}a dada (como el Lagrangiano quiral para piones f{\'\i}sicos). Las teor{\'\i}as de campo efectivas nos sirven para estudiar el comportamiento de un sistema a una escala determinada de forma m\'as simple que la teor{\'\i}a fundamental que est\'a por encima. Se necesita ir a\~na\-dien\-do m\'as y m\'as t\'erminos a la teor{\'\i}a efectiva, generando as{\'\i} una expansi\'on controlada por un par\'ametro $\Lambda_\chi$, caracter{\'\i}stico de una escala m\'as alta. A medida que aumentamos la potencia de $\Lambda_\chi$ en el denominador, operadores de dimensiones m\'as altas son introducidos en la teor{\'\i}a efectiva. Consideremos ahora que nuestro problema f{\'\i}sico involucra part{\'\i}culas ligeras que interaccionan
entre s{\'\i} a trav\'es del intercambio de otra part{\'\i}cula de masa $M$ mucho mayor. Si la energ{\'\i}a caracter{\'\i}stica es mucho menor que $M$, podemos reemplazar el intercambio no-local de la part{\'\i}cula pesada (es decir el propagador) por una torre de interacciones locales entre las part{\'\i}culas ligeras. El conjunto de operadores as{\'\i} generado es en principio infinito en n\'umero y de dimensi\'on creciente, de manara que se requieren infinitos contrat\'erminos para absorber las divergencias. A un determinado orden en $1/\Lambda_\chi$ s\'olo un n\'umero finito de operadores de la expansi\'on del Lagrangiano efectivo contribuyen al proceso f{\'\i}sico, con lo que un n\'umero finito de contrat\'erminos es suficiente \cite{Pich}. Entonces, podemos recuperar el poder predictivo de la teor{\'\i}a obteniendo
resultados con una precisi\'on limitada.


\section{Teor{\'\i}a de perturbaciones quirales para bariones pesados}
La teor{\'\i}a de perturbaciones quirales es la teor{\'\i}a de campo efectiva para los mesones ligeros de la QCD. Dado que estos estados son los seudobosones de Goldstone de la simetr{\'\i}a quiral SU(3)$_L$ $\times$ SU(3)$_R$ espont\'aneamente rota, es decir, la teor{\'\i}a de perturbaciones quirales explota la simetr{\'\i}a del Lagrangiano de QCD bajo transformaciones de SU(3)$_L$ $\times$ SU(3)$_R$ $\times$ U(1) de los tres sabores de quarks ligeros en el l{\'\i}mite $m_q \rightarrow 0$, donde $m_q$ es la masa del quark. El vac{\'\i}o de QCD se alinea en una cierta direcci\'on en el espacio de simetr{\'\i}a interno que representan las transformaciones quirales, entonces hablamos de una rotura espont\'anea del grupo de simetr{\'\i}a global SU(3)$_L$ $\times$ SU(3)$_R$ $\times$ U(1) a un subgrupo diagonal SU(3) $\times$ U(1). La rotura espont\'anea tiene una consecuencia muy importante, la aparici\'on del octete pseudoescalar de bosones de Goldstone. La simetr{\'\i}a del Lagrangiano no es exacta,
puesto que los quarks $u$, $d$ y $s$ tienen una peque\~na masa.

Adicionalmente, la teor{\'\i}a de perturbaciones quirales puede extenderse para incluir grados de li\-ber\-tad fermi\'onicos, tales
como los bariones, cuyas propiedades bajo la transformaci\'on quiral fijan sus acoplamientos a los mesones y es conveniente
considerar al campo de bariones est\'aticos dependiente de la velocidad, este formalismo introducido por Jenkins y Manohar se conoce como {\it Teor{\'\i}a de Perturbaciones Quirales para Bariones Pesados} \cite{Jenkins255(91),Jenkins259(91)}.

El Lagrangiano efectivo al m\'as bajo orden \cite{Jenkins255(91)} est\'a dado por la siguiente ecuaci\'on,
\begin{eqnarray} \label{eq:lchiral}
{\cal L}_{\rm bari\acute{o}n} & = & i \, \mathrm{Tr} \,
                                    \bar{B}_v\bigl(v \cdot {\cal D}\bigr) B_v  - i \, \bar{T}_v^\mu
                                    \bigl(v \cdot {\cal D}\bigr) T_{v\mu}
                              + \Delta \, \bar{T}_v^\mu T_{v\mu} + 2 D \, \mathrm{Tr} \, \bar{B}_v S_v^\mu \{{\cal A}_\mu, B_v\} \nonumber \\
                            && + \,2 F \, \mathrm{Tr} \, \bar{B}_v S_v^\mu \bigl[ {\cal A}_\mu, B_v \bigr]
                            + {\cal C} \, \bigl( \bar{T}_v^\mu {\cal A}_\mu
                            B_v + \bar{B}_v {\cal A}_\mu T_v^\mu \bigr) + 2 \, {\cal H} \, \bar{T}_v^\mu
                            S_v^\nu {\cal A}_\nu T_{v\mu},\nonumber\\
\end{eqnarray}
donde $D$, $F$, ${\cal C}$ y ${\cal H}$ son los coeficientes quirales, es decir, los acoplamientos pi\'on-bari\'on y $\Delta$ es la diferencia de masas entre las masas invariantes de SU(3) del octete de bariones, $M_B$ y decuplete de bariones $M_T$ dadas por $\Delta \equiv M_T - M_B$, adem\'as $B_v$ y $T_v^\mu$ son campos nuevos con 4-velocidad definida $v^\mu$; as{\'\i}, tenemos que los campos de bariones dependen de la velocidad \cite{Jenkins255(91)}.

Para construir el Lagrangiano efectivo Ec. (\ref{eq:lchiral}) que describe las interacciones de mesones y bariones se consideran los campos siguientes:
\begin{enumerate}
\item El campo de pseudobosones de Goldstone contenido en la matriz
\begin{equation}
\xi=\exp\, \left(\frac{i\Pi}{f}\right),
\end{equation}
donde $f \approx 93$ MeV es la constante de decaimiento del
pi\'on; expl{\'\i}citamente
\begin{eqnarray}
\Pi = \frac{1}{\sqrt 2} \left(\begin{array}{ccc} \displaystyle
\frac{1}{\sqrt{2}} \pi^0 + \frac{1}{\sqrt 6} \eta & \pi^+ & K^+ \\
\pi^- & \displaystyle
-\frac{1}{\sqrt 2} \pi^0 + \frac{1}{\sqrt 6} \eta & K^0 \\
\displaystyle K^- & \overline{K}^0 & \displaystyle -\frac{2}{\sqrt
6} \eta
\end{array} \right). \label{eq:mp}
\end{eqnarray}
\item El octete de bariones, el cual est\'a contenido en la matriz
\begin{eqnarray}
B = \left(\begin{array}{ccc} \displaystyle
{\Sigma}^0 + \frac{1}{\sqrt 6} {\Lambda} & {\Sigma}^+ & p \\
\Sigma^- & \displaystyle
-\frac{1}{\sqrt 2} \Sigma^0 + \frac{1}{\sqrt 6} \Lambda & n \\
\displaystyle -\Xi^- & \Xi^0 & \displaystyle -\frac{2}{\sqrt 6}
\Lambda
\end{array} \right). \label{eq:mb}
\end{eqnarray}
\item La matriz $\bar{B}$ de antibariones
\begin{eqnarray}
\bar{B} = \left(\begin{array}{ccc} \displaystyle
\bar{\Sigma}^0 + \frac{1}{\sqrt 6} \bar{\Lambda} & \bar{\Sigma}^- & -\bar{\Xi}^- \\
\bar{\Sigma}^+ & \displaystyle
-\frac{1}{\sqrt{2}} \bar{\Sigma}^0 + \frac{1}{\sqrt 6} \bar{\Lambda} & \bar{\Xi}^0 \\
\displaystyle \bar{p} & \bar{n} & \displaystyle -\frac{2}{\sqrt 6}
\bar{\Lambda}
\end{array} \right). \label{eq:antib}
\end{eqnarray}

\item El decuplete de bariones de esp{\'\i}n $\frac{3}{2}$,
descrito por un campo de Rarita-Schwinger $T^\mu$, el cual
satisface la restricci\'on $\gamma_\mu T^\mu = 0$ y es
completamente sim\'etrico en sus {\'\i}ndices de SU(3). Sus componentes
son
\begin{eqnarray}
&& T_{uuu}=\Delta^{++}, \hspace{1cm}  T_{uud} = \frac{1}{\sqrt 3} \Delta^+,\hspace{1cm} T_{udd} = \frac{1}{\sqrt 3} \Delta^0,\nonumber\\
&& T_{ddd} = \Delta^- \hspace{1cm} T_{uus} = \frac{1}{\sqrt 3} {\Sigma^{*+}}, \hspace{1cm} T_{uds}=\frac{1}{\sqrt 6} {\Sigma^*}^0,\nonumber\\
&& T_{dds} = \frac{1}{\sqrt 3} {\Sigma^*}^- \hspace{1cm} T_{uss} = \frac{1}{\sqrt 3} {\Xi^*}^0,\hspace{1cm} T_{dss} = \frac{1}{\sqrt 3} {\Xi^*}^-,\nonumber\\
&& T_{sss} = \Omega^-. \label{eq:decu}
\end{eqnarray}
\end{enumerate}

Los nuevos campos dependientes de la velocidad, se relacionan con los de la teor{\'i}a original a trav\'es de \cite{Jenkins255(91)}
\begin{equation}\label{eq:camposNew}
B_v(x)= \exp \left(i\, M_N {\not \! v} v_\mu x^\mu \right) B(x), \hspace{1cm} T_v^\mu(x) = \exp \left(i\, M_\Delta {\not \! v}
v_\mu x^\mu \right) T^\mu (x),
\end{equation}
y el operador de esp{\'\i}n tiene la siguiente expresi\'on,
\begin{equation}\label{eq:espin}
S^\mu_s = i \gamma_5 \sigma^{\mu\nu}v_\nu/2.
\end{equation}
De la ecuaci\'on (\ref{eq:lchiral}) se obtiene el propagador del bari\'on \cite{Jenkins255(91)}
\begin{equation}\label{eq:propagadorbarion}
\frac{i}{k\cdot v -\Delta + i\varepsilon},
\end{equation}
donde $\Delta = 0$ corresponde al propagador del bari\'on del
octete y $\Delta \neq 0$ corresponde al propagador del bari\'on
del decuplete. Aqu{\'\i} $k \sim {\cal O}(p)$ es el momento fuera
de la capa de masa portado por los componentes ligeros de $B_v$
a este orden y est\'a relacionado con el momento $p^\mu$ del
bari\'on f{\'\i}sico a trav\'es de
\begin{equation}\label{momentoB}
p^\mu = M_0 v^\mu + k^\mu,
\end{equation}
donde $M_0$ es la masa del bari\'on en el l{\'\i}mite quiral y
$p^2 = M_0^2$ sobre la capa de masa.

La teor{\'\i}a de perturbaciones quirales para bariones pesados es
la teor{\'\i}a de campo efectiva de los mesones ligeros de la
QCD. El octete de mesones pseudoescalares puede identificarse
con el multiplete de bosones Goldstone (aproximadamente sin masa)
asocia\-dos con la rotura espont\'anea de la simetr{\'\i}a quiral.
Los bosones de Goldstone obedecen teoremas de baja energ{\'\i}a,
los cuales resultan en las predicciones conocidas del \'algebra de
corriente.

Adicionalmente, en teor{\'\i}a de perturbaciones quirales para
bariones pesados se tiene una doble expansi\'on: la expansi\'on
quiral en potencias de $p/\Lambda_\chi$ y la expansi\'on de
bari\'on pesado $p/M_B$ (donde $p$ es el momento del mes\'on, $M_B$
es la masa del bari\'on y $\Lambda_\chi$ es la escala de
rotura de la simetr{\'\i}a quiral). Dado que $M_B \sim
\Lambda_\chi$ las dos expansiones pueden ser incorporadas
simult\'aneamente.

La aplicaci\'on m\'as inmediata de la teor{\'\i}a de
perturbaciones quirales para bariones pesados fue para la
corriente axial vector de bariones \cite{Jenkins255(91),
Jenkins259(91)}. Dos resultados importantes fueron obtenidos de
este an\'alisis. En el primero se encontr\'o que las razones del
acoplamiento axial vector es cercano a sus valores en SU(6) con
una raz\'on $F/D$ cercana a $2/3$, el valor predicho por el modelo
de quark no relativista. Segundo, se encontraron grandes cancelaciones en
las correcciones a un loop para la corriente axial vector de
bariones entre las gr\'aficas de loop con estados intermediarios
del octete de bariones de esp{\'\i}n $\frac{1}{2}$ y el decuplete de
bariones de esp{\'\i}n $\frac{3}{2}$. Esto fue probado m\'as tarde usando
la expansi\'on $1/N_c$: las razones de los acoplamientos axial
vector de bariones tendr{\'\i}an valores de SU(6) con $F/D =
2/3$, haciendo correcciones a orden $1/N^2_c$ para piones
\cite{DashenPRD49, DashenB315(1993)b}.

\subsection{La corriente axial para bariones en teor{\'\i}a de perturbaciones quirales}
\begin{figure}[h]
\begin{center}
\includegraphics[scale=0.78]{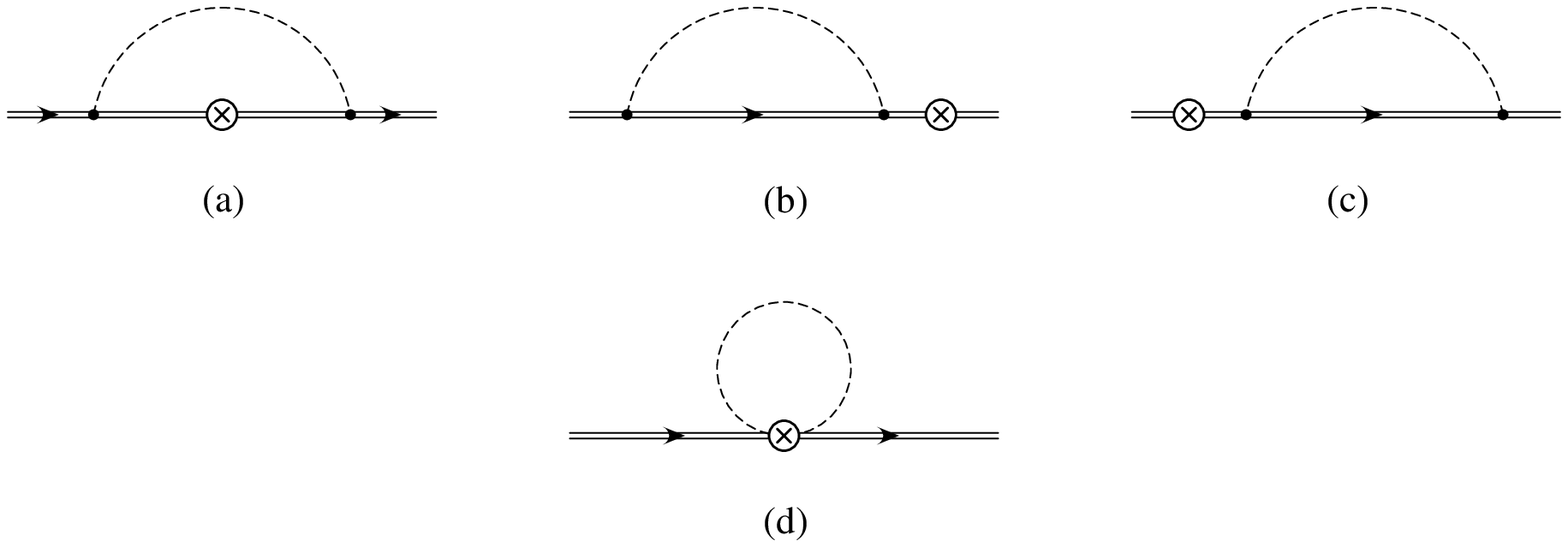}
\caption{Diagramas de Feynman a un loop que contribuyen a la
renormalizaci\'on de la corriente axial. \label{Fig:diagramas}}
\end{center}
\end{figure}

\begin{figure}[h]
\begin{center}
\includegraphics[scale=0.25]{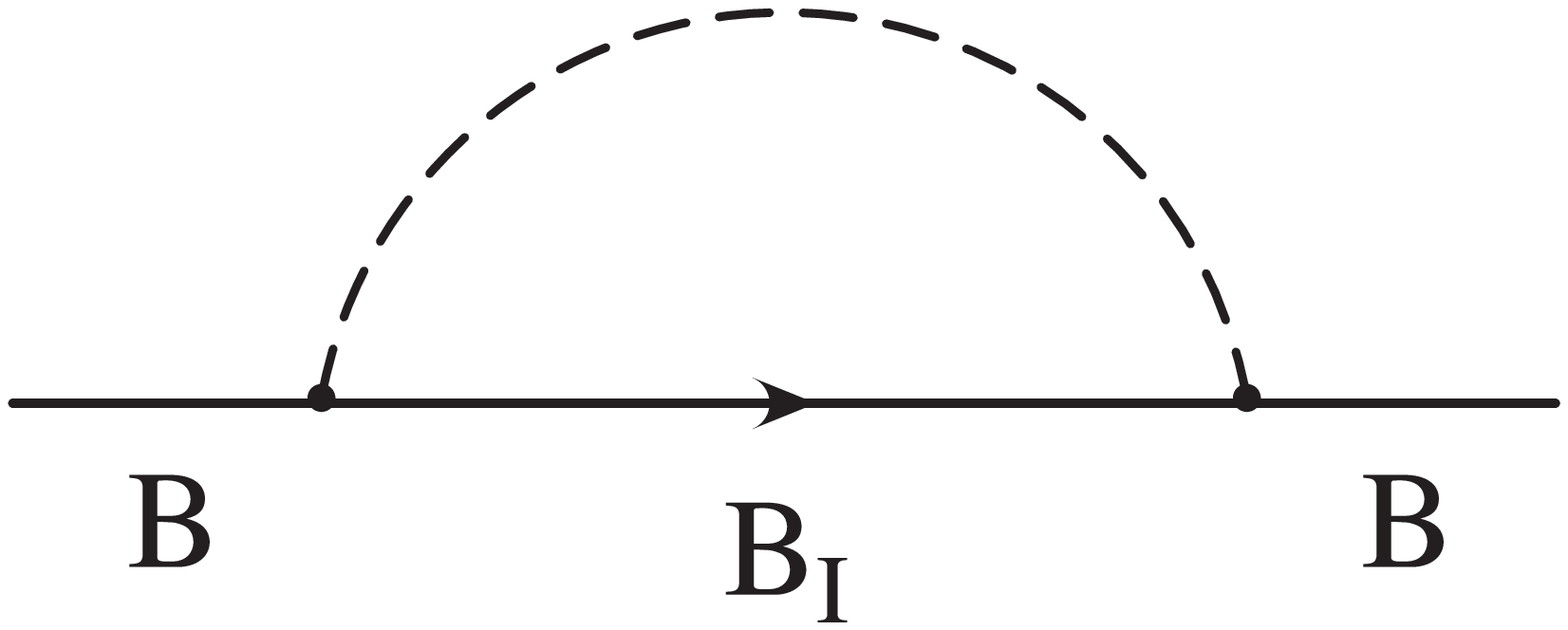}
\caption{Diagrama a un loop de la funci\'on de onda. \label{fun-onda}}
\end{center}
\end{figure}

Las correcciones a un loop para la corriente axial se originan de
los diagramas de Feynman que se presentan en la Fig.
\ref{Fig:diagramas}. En esta figura, las l{\'\i}neas continuas
dobles representan a un bari\'on, mientras que la l{\'\i}nea
segmentada representa a un mes\'on. El s{\'\i}mbolo $\otimes$ indica la
inserci\'on de la co\-rrien\-te axial. El s{\'\i}mbolo $\bullet$
representa un v\'ertice de interacci\'on fuerte proveniente del
Lagrangiano.
Primero consideremos la renormalizaci\'on de la funci\'on de onda a un loop Fig. \ref{fun-onda}. Este diagrama de Feynman es parte de los diagramas 2.1(b) y 2.1(c). En esta secci\'on nuestro c\'alculo consiste en las correcciones del octete de mesones. El diagrama de Feynman de la Fig. \ref{fun-onda} depende de la funci\'on $F(m_\Pi,\Delta,\mu)$ la cual esta definida por la integral a un loop
\begin{equation}\label{delta}
\delta^{ij} F(m_\Pi, \Delta, \mu) =
\frac{i}{f^{2}}\int \frac{d^4k}{(2\pi)^4} \frac{d^{4}k ({\bf k}^{i})(-{\bf k}^{j})}{(k^{2}-m^{2}_{\Pi}) (k\cdot v -\Delta + i\epsilon)}.
\end{equation}
Esta integral fue resuelta usando regularizaci\'on dimensional \cite{RFM2000,JenkinsManohar}; as{\'\i} $\mu$ en la Ec. (\ref{delta}) denota el par\'ametro de escala. La
expresi\'on que se obtiene al evaluar esta integral esta dada con m\'as detalle en el Ap\'endice A de esta tesis.

En la Fig. \ref{fun-onda} se presenta el diagrama de Feynman que contribuye a la renor\-ma\-li\-za\-ci\'on de la funci\'on de onda para el bari\'on $B$, donde uno suma sobre todos los posibles bariones intermediarios $B_I$.
Las combinaciones lineales de sabor singulete, octete y {\bf 27} de las integrales sobre loop son
\begin{eqnarray}
F_{\bf 1}^{(n)} & = & \frac18 \left[3F^{(n)}(m_\pi,0,\mu) + 4F^{(n)}(m_K,0,\mu) + F^{(n)}(m_\eta,0,\mu) \right],\label{eq:Fn1}\\
F_{\bf 8}^{(n)} & = & \frac{2\sqrt 3}{5} \left[\frac32 F^{(n)}(m_\pi,0,\mu) - F^{(n)}(m_K,0,\mu) - \frac12 F^{(n)}(m_\eta,0,\mu) \right],\label{eq:Fn8}\\
F_{\bf 27}^{(n)} & = & \frac13 F^{(n)}(m_\pi,0,\mu) - \frac43
F^{(n)}(m_K,0,\mu) + F^{(n)}(m_\eta,0,\mu). \label{eq:Fn27}
\end{eqnarray}
Las Ecs. (\ref{eq:Fn1}), (\ref{eq:Fn8}) y (\ref{eq:Fn27}) son
combinaciones lineales de $F^{(n)}(m_\pi,0,\mu)$,
$F^{(n)}(m_K,0,\mu)$ y $F^{(n)}(m_\eta,0,\mu)$, las cuales son el
resultado de efectuar las integrales a un loop. Aqu{\'\i}, la
funci\'on $F^{(n)}(m_\Pi, 0, \mu)$ representa el l{\'\i}mite de
degenaraci\'on $\Delta/m_\Pi \rightarrow 0$ de la funci\'on
general $F^{(n)}(m_\Pi,\Delta,\mu)$, definida como
\begin{equation}
F^{(n)}(m_\Pi, \Delta, \mu)\equiv
\frac{\partial^{n}F(m_\Pi,\Delta,\mu)}{\partial \delta^{n}},
\end{equation}
donde $\mu$ es el par\'ametro de escala de la regularizaci\'on
dimensional. La funci\'on $F(m_\Pi, \Delta, \mu)$ junto con sus
derivadas son dadas expl{\'\i}citamente en el Ap\'endice A.

An\'alogamente, las combinaciones lineales de las integrales a un loop $I(m_\pi, \mu)$, $I(m_K, \mu)$ y $I(m_\eta, \mu)$ para la contribuci\'on de sabor singulete $I_{\bf 1}$, sabor octete $I_{\bf 8}$ y sabor {\bf 27} $I_{27}$ son
\begin{equation}\label{Ec:I(d)}
I(m_\Pi, \mu) = \frac{i}{f^2} \int \frac{d^4 k}{(2\pi)^4} \frac{1}{k^2 - m_\Pi^2} = \frac{m_\Pi^2}{16 \pi^2 f^2} \left[{\rm ln}\frac{m_\Pi^2}{\mu^2} - 1 \right].
\end{equation}
Estos forman las combinaciones lineales Ecs. (\ref{eq:I1}) - (\ref{eq:I27})
\begin{eqnarray}
I_{\bf 1} &=& \frac{1}{8} \left[ 3 I(m_\pi, \mu) + 4I(m_K, \mu) + I(m_\eta, \mu) \right],\label{eq:I1}\\
I_{\bf 8} &=& \frac{2\sqrt{3}}{5} \left[ \frac{3}{2} I(m_\pi, \mu) - I(m_K, \mu) - \frac{1}{2}I(m_\eta, \mu) \right],\label{eq:I8}\\
I_{\bf 27} &=& \frac{1}{3} I(m_\pi, \mu) - \frac{4}{3}I(m_K, \mu) + I(m_\eta, \mu).\label{eq:I27}
\end{eqnarray}

Las correcciones quirales logar{\'\i}tmicas para la corriente axial en decaimientos se\-mi\-lep\-t\'o\-ni\-cos de bariones se pueden calcular considerando estados intermediarios octete y decuplete, se ha comprobado que existen grandes cancelaciones en las co\-rrec\-cio\-nes a un loop para la co\-rrien\-te axial entre las contribuciones de los estados bari\'onicos, decuplete de esp{\'\i}n $\frac{3}{2}$ y octete de esp{\'\i}n $\frac{1}{2}$ Refs. \cite{RFM2006, RFM2000, Jenkins259(91)}.

La corriente axial renormalizada puede escribirse de la forma
siguiente \cite{Jenkins255(91), Jenkins259(91)}
\begin{eqnarray}
\langle B_j | J_\mu^A | B_i \rangle & = & \left[\alpha_{B_jB_i} - \sum_\Pi \left(\bar{\beta}_{B_jB_i}^\Pi - \bar{\lambda}_{B_jB_i}^\Pi \alpha_{B_jB_i}\right)
F(m_\Pi,\mu) + \sum_\Pi \gamma_{B_jB_i}^\Pi I(m_\Pi,\mu) \right]\nonumber \\
&  & \times \bar{u}_{B_j} \gamma_\mu \gamma_5u_{B_i},\label{eq:axreno}
\end{eqnarray}
donde $\alpha_{B_jB_i}$ es el resultado a nivel \'arbol. Es importante puntualizar que los elementos de matriz  de las componentes espaciales del operador axial vector entre los estados de simetr{\'\i}a SU(6) dan los valores usuales de los acoplamientos axial vector. Para el octete de bariones, los acoplamientos axial vector son $g_A$, tal como est\'an definidos en los experimentos en decaimientos semilept\'onicos de bario\-nes, donde $g_A \approx 1.27$ para el decaimiento beta del neutr\'on. Para los decaimientos fuertes de bariones los acoplamientos axial vector son $g$, los cuales son extra{\'\i}dos de las anchuras de los decaimientos fuertes del decuplete de bariones al octete de bariones y piones.

Al orden m\'as bajo, tenemos
\begin{equation}\label{garbol}
(g_A)_{\small{\acute{A}rbol}}^{B_i B_j} = \langle B_j | J_\mu^A | B_i \rangle = F + D.
\end{equation}
donde $F$ y $D$ so los coeficientes quirales.

Continuando con las contribuciones a nivel de un loop. La contribuci\'on del diagrama de
Feynman de la Fig. 2.1(a) tiene la siguiente expresi\'on
\begin{equation}
\bar{\beta}_{B_jB_i}^\Pi = \beta_{B_jB_i}^\Pi + {\beta^\prime}_{B_jB_i}^\Pi \label{eq:prima1}
\end{equation}
Los diagramas (b) y (c) de la Fig. 2.1 corresponden a la correcci\'on debida a la renor\-ma\-lizaci\'on de la funci\'on de
onda, dada por la siguiente ecuaci\'on
\begin{equation}
\bar{\lambda}_{B_jB_i}^\Pi = \lambda_{B_jB_i}^\Pi + {\lambda^\prime}_{B_jB_i}^\Pi \label{eq:prima2}
\end{equation}
Finalmente, $\gamma_{B_j B_i}^\Pi$ es la correcci\'on a un loop que se origina en el diagrama Fig. 2.1 (d),
\begin{eqnarray}
\sqrt{Z_{B_j}Z_{B_i}} = 1 - \sum_\Pi \bar{\lambda}_{B_jB_i}^\Pi
F(m_\Pi,\mu), \qquad \bar{\lambda}_{B_jB_i}^\Pi = \frac12
(\bar{\lambda}_{B_i}^\Pi + \bar{\lambda}_{B_j}^\Pi),
\end{eqnarray}
$\Pi$ representa a los mesones $\pi$, $K$ y $\eta$.

Las cantidades no primadas y primadas en Ecs. (\ref{eq:prima1}) y (\ref{eq:prima2}) son contribuciones con estados
intermediarios de octete y decuplete, respectivamente. Las
f\'ormulas expl{\'\i}citas para los coeficientes quirales
$\alpha_{B_jB_i}$, $\bar{\beta}_{B_jB_i}^\Pi$,
$\bar{\lambda}_{B_jB_i}^\Pi$ y $\gamma_{B_jB_i}^\Pi$ se
encuentran publicadas en la Ref. \cite{RFM2006}. Notemos que al restringirnos al
caso de correcciones no anal{\'\i}ticas en el l{\'\i}mite
$m_u=m_d=0$ y al usar la f\'ormula de masa de Gell-Mann-Okubo para
reescribir $m_\eta^2$ como $(4/3)m_K^2$, la Ec.~(\ref{eq:axreno})
se reduce a los resultados ya obtenidos \cite{Jenkins255(91), Jenkins259(91)}.

Ahora podemos reescribir la corriente axial renormalizada Ec.~(\ref{eq:axreno}) en una forma m\'as conveniente
\begin{eqnarray}\label{eq:corriente}
\langle B_j | J_\mu^A | B_i \rangle & = & \left[\alpha_{B_jB_i} + b_\mathbf{1}^{B_jB_i} F_\mathbf{1} + b_\mathbf{8}^{B_jB_i}F_\mathbf{8} + b_\mathbf{27}^{B_jB_i}F_\mathbf{27} + c_\mathbf{1}^{B_jB_i}I_\mathbf{1} + c_\mathbf{8}^{B_jB_i}I_\mathbf{8} + c_\mathbf{27}^{B_jB_i}I_\mathbf{27} \right] \nonumber \\
&  & \mbox{} \times \bar{u}_{B_j} \gamma_\mu \gamma_5 u_{B_i},
\end{eqnarray}
escrita en t\'erminos de las combinaciones lineales Ecs. (\ref{eq:Fn1}) - (\ref{eq:Fn27}) y Ecs. (\ref{eq:I1}) - (\ref{eq:I27})

donde los nuevos coeficientes son
\begin{eqnarray}
b_\mathbf{1}^{B_jB_i} & = & -(a_{B_jB_i}^\pi + a_{B_jB_i}^K + a_{B_jB_i}^\eta), \label{eq:b01} \\
b_\mathbf{8}^{B_jB_i} & = & -\frac{1}{\sqrt 3} \left(a_{B_jB_i}^\pi - \frac12 a_{B_jB_i}^K - a_{B_jB_i}^\eta\right), \label{eq:b08} \\
b_\mathbf{27}^{B_jB_i} & = & -\frac{3}{40} \left(a_{B_jB_i}^\pi - 3a_{B_jB_i}^K + 9 a_{B_jB_i}^\eta\right), \label{eq:b27} \\
c_\mathbf{1}^{B_jB_i} & = & \gamma_{B_jB_i}^\pi + \gamma_{B_jB_i}^K + \gamma_{B_jB_i}^\eta, \label{eq:c01} \\
c_\mathbf{8}^{B_jB_i} & = & \frac{1}{\sqrt 3} \left(\gamma_{B_jB_i}^\pi - \frac12 \gamma_{B_jB_i}^K - \gamma_{B_jB_i}^\eta\right), \label{eq:c08} \\
c_\mathbf{27}^{B_jB_i} & = & \frac{3}{40}
\left(\gamma_{B_jB_i}^\Pi - 3\gamma_{B_jB_i}^{K} + 9
\gamma_{B_jB_i}^{\eta}\right), \label{eq:c27}
\end{eqnarray}
y los diferentes coeficientes $a_{B_jB_i}^\Pi$ son expresados en
t\'erminos de los coeficientes quirales como
\begin{eqnarray}
a_{B_jB_i}^\Pi = \bar{\beta}_{B_jB_i}^\Pi -
\bar{\lambda}_{B_jB_i}^\Pi \alpha_{B_jB_i}.
\end{eqnarray}
Num\'ericamente, se determina que la correcci\'on de sabor
${\bf 27}$ est\'a suprimida con respecto a la de sabor octete
${\bf 8}$ y \'esta a su vez esta suprimida con respecto a la de
sabor singulete ${\bf 1}$.

\section{QCD en el l{\'\i}mite $N_c \rightarrow \infty$}
La expansi\'on $1/N_c$ es un esquema sistem\'atico para acercarse al estudio de aspectos no pertubativos de la Cromodin\'amica Cu\'antica desde una teor{\'\i}a que se asemeja a QCD en un determinado l{\'\i}mite.

La generalizaci\'on de QCD desde $N_c = 3$ hasta $N_c \gg 3$ colores se conoce como {\it el l\'imite de $N_c$ grande}. En este l{\'\i}mite surge en el sector de bariones una simetr{\'\i}a contra{\'\i}da esp{\'\i}n sabor SU(2N$_f$) \cite{DashenPRD49, DashenPRD51, JenkinsNucl}, donde $N_f$ es el n\'umero de sabores de los 3 quarks ligeros $u$, $d$ y $s$. En el l{\'\i}mite $N_c \rightarrow \infty $ la f{\'\i}sica se simplifica notablemente, dado que las cantidades f{\'\i}sicas en consideraci\'on adquieren correcciones de orden re\-la\-ti\-vo $1/N_c$, $1/N^{2}_c$, $1/N^3_c$, etc., lo que da origen a la expansi\'on $1/N_c$ \cite{Hooft}. El l{\'\i}mite $N_c \rightarrow \infty$ para bariones organiza estados bari\'onicos a orden m\'as bajo en la representaci\'on SU(2$N_f$) completamente sim\'etrica. Bajo la descomposici\'on [SU($2N_f$) $\rightarrow$ SU(2) $\times$ SU($N_f$)], esta representaci\'on esp{\'\i}n-sabor se descompone en una torre de estados de bariones con espines $\frac{1}{2}$, $\frac{3}{2}$, $\cdots$, $\frac{1}{N_c}$. Para $N_c = 3$ estos multipletes de sabor se reducen al octete y decuplete de bariones. Sin embargo, para $N_c > 3$, los multipletes contienen estados bari\'onicos adicionales que no existen para $N_c = 3$. Dada la complejidad de las representaciones de sabor para $N_f > 2$, es m\'as sencillo enfocarse en los operadores que en los estados.

La simetr{\'\i}a contraida viene de las condiciones de consistencia en las amplitudes de dispersi\'on mes\'on-bari\'on, las cuales se deben satisfacer para que la teor{\'i}a sea unitaria \cite{JenkinsNucl}. El estudio del \'algebra esp{\'\i}n-sabor SU(6) para bariones en el l{\'\i}mite de $N_c$ grande es conveniente efectuarlo en la representaci\'on de quarks. Sin embargo, esto no significa que los quarks dentro del bari\'on se traten como no relativistas. El \'algebra del modelo de quarks no relativista proporciona una forma conveniente de expresar los resultados de c\'alculos en $1/N_c$, los cuales son v\'alidos incluso para bariones con quarks no masivos. En la
re\-pre\-sen\-ta\-ci\'on de quarks se define un conjunto de operadores de creaci\'on y aniquilaci\'on $q_\alpha^{\dag}$ y $q^{\alpha}$, donde $\alpha = 1, \ldots, N_f$ representa los $N_f$ sabores de quarks con esp{\'\i}n hacia arriba y $\alpha = N_f + 1,\ldots, 2N_f$, los $N_f$ sabores de quarks con esp{\'\i}n hacia abajo.

Cualquier operador de QCD que se transforme de acuerdo a una cierta representaci\'on SU(2) $\times$ SU(3) tiene una expansi\'on en t\'erminos de operadores de $n$-cuerpos Refs. \cite{{JenkinsPRD53}, DashenPRD51} dados de la siguiente forma
\begin{equation}
{\cal O}_{\rm QCD} = \sum_n c_{(n)} \frac{1}{N_c^{n-1}}
{\cal O}_{n}, \label{eq:OQCD}
\end{equation}
donde la base de operadores ${\cal O}_n$ consiste de polinomios en los generadores: esp{\'\i}n $J^i$, sabor $T^a$ y esp{\'\i}n sabor $G^{ia}$. Los coeficientes de los operadores $c_n (1/N_c)$ tienen expansiones en series de potencias en $1/N_c$ comenzando con orden 1. El problema de encontrar un conjunto completo e independiente de operadores de cualquier representaci\'on esp{\'\i}n sabor fue resuelto por Dashen, {\it et. al.}, en la Ref. \cite{DashenPRD51}.

En el l{\'\i}mite de $N_c$ grande los bariones tienen masas del orden $N_c$ y llegan a ser muy pesados con respecto a los mesones.
Para 3 sabores de quarks ($u$, $d$ y $s$), co\-rres\-pon\-dien\-tes al sector ligero, la simetr{\'\i}a esp{\'\i}n sabor es SU(6) y los
bariones se agrupan en la representaci\'on {\bf 56}, cuya descomposici\'on bajo SU(2) $\times$ SU(3) es \cite{JenkinsNucl}
\begin{eqnarray}
   {\bf 56} &=& \left( S=\frac12, {\bf 8} \right) +
\left( S=\frac32, {\bf 10} \right).\nonumber
\end{eqnarray}
Los generadores de la simetr{\'\i}a esp{\'\i}n sabor SU(6), expresados en t\'erminos de ope\-ra\-do\-res bos\'onicos
de quarks de creaci\'on $q_\alpha^\dagger$ y de aniquilaci\'on $q^\alpha$, se expresan como
\begin{equation}\label{ec:op}
  J^k = q^\dagger \frac{\sigma^k}{2} q, \hspace{0.5cm}  T^c = q^\dag \frac{\lambda^c}{2} q, \hspace{0.5cm}  G^{kc} = q^\dag \frac{\sigma^k}{2}\frac{\lambda^c}{2} q,
\end{equation}
donde $\sigma^k$ y $\lambda^c$ son las matrices de esp{\'\i}n de Pauli y de sabor de Gell-Mann, respectivamente. Los generadores de la simetr{\'\i}a esp{\'\i}n sabor SU(6) satisfacen las relaciones de conmutaci\'on listadas en la Tabla \ref{su(2nf)}.
\begin{table}[h]
\begin{center}
\begin{tabular}{ccc}
\hline \hline
& $\left[J^i,T^a \right]=0$& \\
\hline
  $\left[J^i,J^j\right]=i\epsilon^{ijk} J^k$ &&$\left[T^a,T^b\right]=i f^{abc} T^c$ \\
  $\left[J^i,G^{ja}\right]=i\epsilon^{ijk} G^{ka}$ && $\left[T^a,G^{ib}\right]= i f^{abc} G^{ic}$ \\
\end{tabular}
\end{center}
\begin{center}
\begin{tabular}{c}
  $\left[G^{ia},G^{jb}\right] = \frac{i}{4}\delta^{ij} f^{abc} T^c + \frac{i}{2N_F} \delta^{ab}
  \epsilon^{ijk} J^k + \frac{i}{2} \epsilon^{ijk} d^{abc} G^{kc}$\\
  \hline
  \hline
\end{tabular}
\caption{Relaciones de conmutaci\'on $SU(2N_f).$\label{su(2nf)}}
\end{center}
\end{table}
Las relaciones de conmutaci\'on del \'algebra de Lie
para la descomposici\'on esp{\'\i}n sabor SU(2$N_f$) $\rightarrow$ SU(2) $\times$ SU($N_f$) entre operadores de un cuerpo, se encuentran listadas en la Tabla \ref{su(2nf)}.
Las constantes de estructura $f^{abc}$ y $d^{abc}$ para SU(2$N_f$) pueden escribirse en
t\'erminos de las trazas de los generadores $\lambda^a$ del grupo
SU(2$N_f$) \cite{DashenPRD51} en la representaci\'on fundamental,
\begin{eqnarray}
f^{abc} &=& -2i Tr \lambda^a[\lambda^b, \lambda^c],\nonumber\\
d^{abc} &=& 2 Tr \lambda^a \{\lambda^b, \lambda^c\}.\label{fyd}
\end{eqnarray}
Las identidades para contraer los {\'\i}ndices de las constantes de estructura $f^{abc}$ y $d^{abc}$ de la Tabla \ref{identi} son utilizadas para derivar las
identidades de los operadores de quarks \cite{DashenPRD51}. Estas pueden demostrarse usando las Ecs. (\ref{fyd}).
\begin{table}[h]
\begin{center}
\[
\begin{array}{|l|l|}
\hline
d^{aab}=0 & d^{abc}d^{abd}=\left(N_f-\frac{4}{N_f}\right)\delta^{cd} \\
f^{abc}f^{abd}=N_f\delta^{cd} & f^{acd}d^{bcd}=0 \\
f^{abc}f^{ade}d^{bdf}=\frac{N_f}{2}d^{cef} &
d^{abc}d^{ade}d^{bdf}=\left(\frac{N_f}{2}-\frac{6}{N_f}\right)d^{cef}\\
d^{abc}d^{ade}f^{bdf}=\left(\frac{N_f}{2}-\frac{2}{N_f}\right)f^{cef}
& \delta^{aa}=N_f^{2}-1\\
\delta^{ab}\delta^{ab}=N_f^{2}-1 & \delta^{ii}=3\\
\delta^{il}\epsilon^{ijk}\epsilon^{klm}=\epsilon^{ljk}\epsilon^{klm}=\epsilon^{klj}\epsilon^{klm}=2\delta^{jm}
&\epsilon^{ijk}\epsilon^{klm}=\epsilon^{ijk}\epsilon^{lmk}=\delta^{il}\delta^{jm}-\delta^{jl}\delta^{im}\\
\hline
\end{array}
\]
\caption{Identidades.  \label{identi}}
\end{center}
\end{table}
Todos los productos de operadores para los cuales dos {\'\i}ndices de sabor se contraen usando $f^{abc}$ y/o $d^{abc}$ o dos indices de esp{\'\i}n usando $\delta^{ij}$ y $\epsilon^{ijk}$ pueden ser eliminados \cite{Manohar}.

\section{Propiedades est\'aticas de bariones}
En esta secci\'on se estudian algunas propiedades est\'aticas de los bariones, es decir, los espectros de masas, acoplamientos en decaimientos semilept\'onicos, momentos magn\'eticos.

\subsection{Masa de los bariones}
Un hecho muy importante en QCD es que las masas de los bariones
son par\'ametros esenciales en la descripci\'on de estados ligados
de quarks. Pero las masas de los quarks $u$, $d$ y $s$ son muy
peque\~nas, as{\'\i} que pueden considerarse cero. Por otra parte,
las masas de los bariones son de orden ${\cal O}(N_c)$.

Los bariones en QCD son estados singuletes de color de $N_c$
quarks. Esto se logra al contraer los {\'\i}ndices de color de los
$N_c$ quarks con el s{\'\i}mbolo $\epsilon$ de SU($N_c$), es
decir,
\begin{equation}\label{ec:Nc}
    \epsilon_{i_1 i_2 i_3 \ldots i_{N_c}} q^{i_1} q^{i_2} q^{i_3}\ldots
    q^{i_{N_c}},
\end{equation}
as{\'\i} que en el l{\'\i}mite de $N_c$ grande un bari\'on es
totalmente antisim\'etrico en los {\'\i}ndices de color de sus
$N_c$ quarks de valencia, y debe ser sim\'etrico en los otros
n\'umeros cu\'anticos tales como esp{\'\i}n y sabor.

Un argumento importante que debemos considerar es que la masa del
bari\'on crece con $N_c$ \cite{Manohar},

\begin{equation}\label{masa-barion}
    {\cal M}_{\rm bari\acute{o}n} \sim {\cal O}(N_c),
\end{equation}
y los bariones son infinitamente pesados en el l{\'\i}mite de $N_c$ grande. Para quarks no masivos, el \'unico par\'ametro dimensional
de QCD para $N_c$ grande es $\Lambda_{\rm QCD}$, ($\Lambda_{\rm QCD}$ establece un l{\'\i}mite inferior absoluto para la validez de la teor{\'\i}a de perturbaciones, que aproximadamente es del orden de una escala hadr\'onica t{\'\i}pica),

\begin{equation}
{\cal M}_{\rm bari\acute{o}n} \sim N_c \Lambda_{\rm QCD}.
\end{equation}
Notemos tambi\'en que aunque en este l{\'\i}mite el n\'umero de quarks crece dentro del bari\'on a medida que $N_c$ crece, su tamaño se rige por $\Lambda_{\rm QCD}$ y permanece fijo.

\subsubsection{Operador de masa}
El operador de masa de bariones en la expansi\'on $1/N_c$ se obtiene usando la regla de reducci\'on de operadores
\cite{DashenPRD51}. Esta regla permite encontrar criterios para
eliminar los operadores redundantes. La expansi\'on $1/N_c$
involucra operadores de cero cuerpos $\mathds{1}$ y de un cuerpo
$J^i$, $T^a$ y $G^{ia}$. Para la reducci\'on de operadores se
utilizan tensores invariantes en el espacio de sabor SU($N_f$),
tales como $\delta^{ab}$, $f^{abc}$ y $d^{abc}$. El operador de
masa est\'a dado por \cite{DashenPRD51}
\begin{equation}\label{ec:masa-barion}
    {\cal M}_B = N_c {\cal P} \left( \frac{J^2}{N_c^2}\right),
\end{equation}
donde ${\cal P}$ es un polinomio en $J^2 / N_c^2$.
Expl{\'\i}citamente \cite{JenkinsPRD53},
\begin{equation}\label{ec:2masa-barion}
{\cal M} = m_{0} N_c \mathds{1} + \sum_{n=2,4}^{N_c-1}m_n \frac{1}{N_c^{n-1}}J^{n},
\end{equation}
donde los coeficientes $m_n$ son par\'ametros
adimensionales de ${\cal O}(\Lambda_{\rm QCD})$. El primer t\'ermino en la
expansi\'on (\ref{ec:2masa-barion}), es el t\'ermino global de
masa independiente de esp{\'\i}n del multiplete de bariones. \'Este
es eliminado desde el Lagrangiano quiral por la redefinici\'on del
campo de bariones pesados \cite{Jenkins255(91), {Jenkins259(91)}}.
Los t\'erminos dependientes de esp{\'\i}n en la Ec.
(\ref{ec:2masa-barion}) definen ${\cal M}_{\rm hiperfina}$, los
cuales aparecen expl{\'\i}citamente en el Lagrangiano. La expansi\'on
de masa hiperfina se reduce a un simple operador
\cite{JenkinsB315c, JenkinsB315d}
\begin{equation}\label{ec:3masa-barion}
    {\cal M}_{\rm hiperfina} = m_2 \frac{1}{3}J^2,
\end{equation}
para $N_c=3$.

El operador ${\cal M}_{\rm hiperfina}$ es el responsable de
particiones finas entre las masas de multi\-ple\-tes diferentes. La
expansi\'on (\ref{ec:2masa-barion}) es v\'alida en el l{\'\i}mite
de simetr{\'\i}a de sabor SU(3) exacta. Su interpretaci\'on
f{\'\i}sica es la siguiente: establece que todos los bariones de
un mismo multiplete son degenerados en masa; por ejemplo, para las part{\'\i}culas $N, \Lambda, \Sigma, \Xi$
del octete de bariones, tenemos la siguiente expresi\'on

\begin{eqnarray}
&&\langle N| {\cal M}_B | N \rangle =\langle \Lambda| {\cal M}_B | \Lambda \rangle =\langle \Sigma| {\cal M}_B | \Sigma \rangle = \langle \Xi| {\cal M}_B |
\Xi\rangle\nonumber\\
&&\hspace{2.3cm} = N_c \left[m_0 + \frac{3}{4}\frac{1}{N_c^2}m_2 + \frac{9}{16}\frac{1}{N_c^4}m_4 + \ldots \right].\label{ec:e-matriz}
\end{eqnarray}
An\'alogamente para las part{\'\i}culas $\Delta$, $\Sigma^*$, $\Xi^*$, $\Omega$ del decuplete de bariones, obtenemos la expresi\'on siguiente
\begin{eqnarray}\label{ec:e-matrizdecuplete}
&&\langle \Delta| {\cal M}_B | \Delta \rangle =\langle \Sigma^*| {\cal M}_B | \Sigma^* \rangle =\langle \Xi^* | {\cal M}_B | \Xi^* \rangle =\langle \Omega| {\cal M}_B | \Omega \rangle \nonumber\\
&&\hspace{2.3cm} = N_c \left[ m_0 + \frac{15}{4}\frac{1}{N_c^2}m_2 + \frac{225}{16} \frac{1}{N_c^4}m_4 + \ldots \right].\nonumber\\
\end{eqnarray}
Las masas de bariones se pueden calcular en la expansi\'on $1/N_c$ y con rotura de simetr{\'\i}a de sabor SU(3) debida a $m_s$. En este caso, la forma general del operador de masa a todos los \'ordenes en la expansi\'on $1/N_c$ tiene la forma

\begin{equation}\label{ec:4masa-barion}
    {\cal M}_B = N_c {\cal P} \left(\frac{N_s}{N_c},\, \frac{J^2}{N_c^2},\, \frac{I^2}{N_c^2}
    \right),
\end{equation}
donde $N_s$ es el operador de n\'umero de quarks $s$ en el bari\'on \cite{DashenPRD51}. La forma general del operador tiene contribuciones de las representaciones de sabor SU(3) dadas por

\begin{equation}\label{ec:5masa-barion}
    {\cal M}_B = {\cal M}^1 + {\cal M}^8 + {\cal M}^{27} +
    {\cal M}^{64},
\end{equation}
donde el singulete {\bf 1}, octete {\bf 8}, ${\bf 27}$ y ${\bf
64}$ corresponden a orden cero, primer orden, segundo orden y tercer orden en la
rotura de simetr{\'\i}a de sabor, respectivamente. Cada una
de estas re\-pre\-sen\-ta\-cio\-nes esp{\'\i}n sabor tienen una
expansi\'on en t\'erminos de operadores suprimidos por $1/N_c$
\cite{JenkinsPRD52}.

\subsection{El acoplamiento axial}

El acoplamiento axial de bariones ha sido estudiado extensamente.
Dashen y Manohar \cite{DashenB315(1993)a,DashenB315(1993)b}
analizaron el acoplamiento axial para $N_f = 2$ de quarks ligeros, donde
$N_f$ es el n\'umero de sabor. La extensi\'on al n\'umero de sabor
$N_f = 3$ de quarks ligeros con rotura de simetr{\'\i}a de
sabor SU(3) fue estudiado por Dashen \textit{et. al.}
\cite{DashenPRD49}, as{\'\i} como el an\'alisis con simetr{\'\i}a
de sabor exacta SU(3) impuesta a orden principal en $1/N_c$. Luty
\cite{Luty} analiz\'o la simetr{\'\i}a de sabor SU(3) a orden
principal en $1/N_c$ usando operadores de quarks. Dashen \textit{et.
al.} \cite{DashenPRD51} deriv\'o la expansi\'on completa $1/N_c$
con simetr{\'\i}a de sabor exacta SU(3) y rotura de simetr{\'\i}a
SU(3) perturbativa. Dai \textit{et al} \cite{Dai} realiz\'o en
forma detallada mediciones y ajustes de datos para los decaimientos de
hiperones a los acoplamientos decuplete $\rightarrow$ octete
$+$ pi\'on.

Por otra parte, para tres sabores de quarks ligeros se han obtenido varios
resultados. Primero en el l{\'\i}mite de simetr{\'\i}a de sabor SU(3),
los acoplamientos del pi\'on con el octete de bariones de esp{\'\i}n
$\frac{1}{2}$ y con el decuplete de bariones de esp{\'\i}n $\frac{3}{2}$ est\'a descrito por
la expansi\'on $1/N_c$ Ec. (\ref{Akc3}) como \cite{DashenPRD49, DashenPRD51, Luty}

Los cuatro coeficientes $a_1$, $b_2$, $b_3$ y $c_3$ parametrizan a
los cuatro acoplamientos de piones $D$ y $F$ con el octete de
bariones, $\cal C$ con las transiciones octete-decuplete, y $\cal
H$  con el decuplete de bariones \cite{Jenkins255(91),
Jenkins259(91)}. A orden principal en $1/N_c$, la expansi\'on puede
truncarse despu\'es de dos operadores \cite{DashenPRD49, DashenPRD51, Luty}, con los resultados siguientes,

\begin{equation}\label{}
    {\cal{C}} = -2D, \hspace{1cm} {\cal{H}} = 3D - 9F,
\end{equation}
el cual es v\'alido hasta correcciones a orden relativo $1/N_c^2$.
La primera relaci\'on es una relaci\'on de SU(6), la cual
explica por qu\'e un acoplamiento de SU(6) se encuentra en los
c\'alculos de teor{\'\i}a de perturbaciones quirales
\cite{Jenkins255(91), Jenkins259(91)}.

La raz\'on $F/D$ puede ser extra{\'\i}da analizando los
acoplamientos pi\'on-bari\'on para $N_c$ grande arbitrario
\cite{DashenPRD49, DashenPRD51}.

Posteriormente, Flores-Mendieta, {\it et. al.} en el 2006 estudiaron las correcciones a orden de un loop para
la corriente axial de bariones en un formalismo combinado entre la teor{\'\i}a de perturbaciones
quirales y el l{\'\i}mite de $N_c$ grande para el caso degenerado.
Ahora, en este trabajo de tesis doctoral se realiza un estudio para la corriente axial de bariones a orden de un loop en el contexto combinado entre la
la teor{\'\i}a de perturbaciones quirales para bariones y la expansi\'on $1/N_c$, pero ahora, consideramos el caso m\'as realista, el caso no degenerado con lo cual, obtenemos resultados mucho m\'as interesantes y los cuales presentamos en los siguientes cap{\'\i}tulos de esta tesis y en la Ref. \cite{PRD2012}.

\subsection{Momento magn\'etico de bariones}

Existen en la literatura varios trabajos sobre el momento magn\'etico de bariones en diferentes contextos, por ejemplo, el modelo de quarks \cite{Ha, Faessler},
el m\'etodo de reglas de suma de QCD \cite{AlievPRD66},
la expansi\'on $1/N_c$, \cite{Dai,JenkinsB335,JenkinsPRLett89,LebedPRD70}
y la teor{\'\i}a de perturbaciones quirales \cite{GengPRD80,Caldi-Pagels,Gasser,Krause,JenkinsPRLett302,Banerjee,DurandPRD58,Puglia,GengPRLett101}.

En la expansi\'on $1/N_c$, los momentos
magn\'eticos de bariones tienen las mismas propiedades
cinem\'aticas que los acoplamientos axiales y en consecuencia
pueden describirse por los mismos operadores \cite{Dai}. El
operador de momento magn\'etico, al igual que el operador de
corriente axial $A^{kc}$, es un objeto de esp{\'\i}n 1 y es un
octete bajo SU(3). Por lo tanto, el operador que conduce a los
momentos magn\'eticos de bariones \cite{RMFPRD80(2009)} es

\begin{equation}\label{ec:1m-magnetico}
    M^{kc} = m_1 G^{kc} + m_2 \frac{1}{N_c} {\cal D}_2^{kc}+ m_3
    \frac{1}{N_c^2} D_3^{kc} + m_4\frac{1}{N_c^2} {\cal O}_3^{kc},
\end{equation}
donde la serie se ha truncado para el valor f{\'\i}sico $N_c = 3$.
Si consideramos simetr{\'\i}a SU(3), los coeficientes
desconocidos $m_i$ son independientes de $k$ y no
est\'an relacionados con los de la expansi\'on dada por la Ec. (\ref{Akc3}) 
en este l{\'\i}mite. Los momentos magn\'eticos son proporcionales
a la matriz de carga de los quarks ${\cal Q} = diag(2/3, -1/3,
-1/3)$, as{\'\i} que pueden ser separados en componentes
isovectoriales e isoescalares, $M^{k3}$ y $M^{k8}$,
respectivamente. Entonces, el operador de momento magn\'etico de
bariones es

\begin{equation}\label{ec:magneticos}
    M^k = M^{kQ} \equiv M^{k3} + \frac{1}{\sqrt{3}} M^{k8},
\end{equation}
donde, para fines pr\'acticos y en lo sucesivo, al
{\'\i}ndice de esp{\'\i}n $k$ le asignamos el valor $3$
(componente $z$) y el {\'\i}ndice de sabor $Q$ representar\'a al
valor $Q = 3 + (1/\sqrt{3})8$, as{\'\i} que cualquier operador de
la forma $X^Q$ deber\'a entenderse como $X^3 + (1/\sqrt{3})X^8$.

En la evaluaci\'on de elementos de matriz de los operadores
contenidos en la Ec. (\ref{ec:1m-magnetico}) con frecuencia aparecen los
operadores a un cuerpo $T^c$ y $G^{ic}$, $c = 3, 8$; estos
operadores pueden expresarse en t\'erminos de los operadores de
n\'umero de quarks $N_q$ y de esp{\'\i}n de quarks $J_q$ de la siguiente forma \cite{DashenPRD51}.
\begin{eqnarray}
  T^3 &=& \frac{1}{2} (N_u - N_d), \\
  T^8 &=& \frac{1}{2\sqrt{3}}(N_c - 3N_s), \\
  G^{i3}&=&\frac{1}{2}(J_u^i - J_d^i),\\
  G^{i8}&=&\frac{1}{2\sqrt{3}}(J^i - 3J_s^i),
\end{eqnarray}
donde
\begin{eqnarray}\label{}
N_c &=& N_u + N_d + N_s, \,\,\,  {\rm y} \nonumber\\
J^i &=& J_u^i+J_d^i+J_s^i.\nonumber
\end{eqnarray}

El operador $M^k$ de la Ec. (\ref{ec:magneticos}), es un operador que
puede utilizarse para calcular los momentos magn\'eticos de bariones del octete,
decuplete y de transiciones decuplete-octete. En total es posible obtener $27$
momentos magn\'eticos. En este procedimiento s\'olo se deben
obtener los elementos de matriz de los operadores presentes en
la Ec. (\ref{ec:magneticos}) entre estados de simetr{\'\i}a SU(6). Los
elementos de matriz en (\ref{ec:magneticos}) han sido calculados y
publicados en las Refs. \cite{RMFPRD80(2009), Giovanna}.

\section{El Lagrangiano quiral para bariones en el l{\'\i}mite $N_c \rightarrow \infty$}
En esta secci\'on presentamos el Lagrangiano quiral para bariones en el l{\'\i}mite de $N_c$ grande, finito, e impar presentado por primera vez en la Ref. \cite{JenkinsPRD53}.

El Lagrangiano quiral $1/N_c$ para bariones est\'a formulado considerando a los bario\-nes como campos est\'aticos pesados con una velocidad fija $v^\mu$.
Para $N_c$ arbitrario este Lagrangiano toma la siguiente forma
\begin{equation}\label{ec:Lagrangiano1Nc}
    {\cal L}_{\rm bari\acute{o}n} = i {\cal D}^0 - {\cal M}_{\rm{hiperfina}} + {\rm Tr}\left({\cal A}^k \lambda^c\right)A^{kc}
                                   + \frac{1}{N_c} Tr\left({\cal A}^k\frac{2I}{\sqrt{6}}\right)A^k +\ldots,
\end{equation}
donde
\begin{equation}\label{deri}
    D^0 = \partial^0 {\mathds{1}} + Tr\left( {\cal V}^0 \lambda^c \right) T^c.
\end{equation}
Observemos que las Ecs. (\ref{ec:Lagrangiano1Nc}) y (\ref{deri}) son muy compactas y cada t\'ermino implica un operador bari\'onico. El operador de masa hiperfina bari\'onico describe el desdoblamiento de esp{\'\i}n de la torre de bariones. Los campos de piones aparecen en el Lagrangiano quiral a trav\'es de las combinaciones vector y axial vector son
\begin{equation}\label{campos-vectorialyaxial}
{\cal V}^0= \frac{1}{2}\left( \xi\partial^0\xi^\dag + \xi^\dag\partial^0\xi \right), \hspace{1.3cm} {\cal A}^k =
\frac{i}{2}\left( \xi\nabla^k\xi^\dag - \xi^\dag \nabla^k\xi
\right),
\end{equation}
las cuales dependen no linealmente del campo $\xi = {\rm exp}[i\Pi(x)/f]$ donde $\Pi (x)$ representa el nonete de piones $\pi$, $K$, $\eta$ y $\eta'$ \cite{JenkinsPRD53}.

Las combinaciones vectoriales de piones se acoplan a las cargas vectoriales de bariones: la combinaci\'on del octete de piones se acopla a la carga de sabor del octete de bariones\footnote{Esta carga se identifica con el operador cuyos elementos de matriz conducen al factor de forma vectorial $g_V$ discutido en la Ref. \cite{RFM(1998)}.}
\begin{equation}\label{acopla-vectorialOctete}
    V^{0a}=\left\langle {\cal B}^\prime \bigg| \left(\bar{q}\gamma^0 \frac{\lambda^a}{2}
    q\right)_{\rm QCD} \bigg| {\cal B}\right\rangle,
\end{equation}
mientras, que la combinaci\'on de piones, para la
contribuci\'on de sabor singulete
\begin{equation}\label{singulete}
    {\rm Tr} \left( {\cal V}^0 \frac{2I}{\sqrt{6}}\right),
\end{equation}
se acopla a la carga de la contribuci\'on de sabor
singulete
\begin{equation}\label{acopla-vectorialSingulete}
    V^{0}=\left\langle {\cal B}^\prime \bigg| \left(\bar{q}\gamma^0 \frac{I}{\sqrt{6}}
    q\right)_{\rm QCD} \bigg| {\cal B} \right\rangle.
\end{equation}
En el l{\'\i}mite de simetr{\'\i}a de sabor SU(3) la carga vectorial bari\'onica es
\begin{eqnarray}
    V^{0a} &=& v^0T^a\, = \,T^a,\nonumber\\
    V^0 &=& v^0\frac{1}{\sqrt{6}}N_c{\mathds 1}\, =
    \,\frac{1}{\sqrt{6}}N_c {\mathds 1},
\end{eqnarray}
a todos los \'ordenes en la expansi\'on $1/N_c$. Por
otra parte, la combinaci\'on sabor octete axial de piones se
acopla a la corriente axial, octete de bariones,
\begin{equation}\label{acopla-axialOctete}
    A^{kc}=\left\langle {\cal B^\prime} \bigg| \left(\bar{q}\gamma^k\gamma_5 \frac{\lambda^c}{2}
    q\right)_{\rm QCD} \bigg| {\cal B}\right\rangle,
\end{equation}
mientras que la combinaci\'on singulete de piones se acopla con la
corriente axial, contribuci\'on de sabor singulete
\begin{equation}\label{acopla-axialsingulete}
    A^{k}=\left\langle {\cal B}^\prime \bigg| \left(\bar{q}\gamma^k\gamma_5 \frac{I}{\sqrt{6}}
    q\right)_{\rm QCD} \bigg| {\cal B}\right\rangle,
\end{equation}
donde el sub{\'\i}ndice QCD en las Ecs.
(\ref{acopla-vectorialOctete}) y (\ref{acopla-axialOctete}) indica
que los campos de quarks $q$ y $\bar{q}$ son campos de quarks de
QCD y no los operadores de creaci\'on y aniquilaci\'on de la
representaci\'on de quarks.

Resumiendo tenemos que el Lagrangiano quiral de bariones describe las interacciones de los piones y bariones en t\'erminos de operadores de bariones del \'algebra SU(2) y SU(3).

La corriente axial de bariones se analiza en el contexto de la
expansi\'on $1/N_c$ para bariones de sabor singulete, sabor
octete y sabor {\bf 27} como el las Refs. \cite{RFM2006,DashenPRD51}. El operador de
co\-rrien\-te axial de bariones $A^{kc}$ es un objeto de
esp{\'\i}n entero ($J=1$) y se transforma como un {\bf octete}
bajo SU(3), adem\'as es impar bajo inversi\'on temporal. Este
operador est\'a dado de la siguiente forma \cite{JenkinsPRD53,DashenPRD51}
\begin{equation}{\label{corrienteAxial}}
A^{kc} = a_1 G^{kc} + \sum^{N_c}_{n=2,3}b_n \frac{1}{N^{n-1}_{c}}{\cal D}^{kc}_{n} + \sum^{N_c}_{3,5} c_n \frac{1}{N^{n-1}_{c}} {\cal O}^{kc}_{n},
\end{equation}
donde los operadores ${\cal D}^{kc}_{n}$ son diagonales con
elementos de matriz no cero y s\'olo act\'uan entre estados con el
mismo esp{\'\i}n, los ${\cal O}^{kc}_{n}$ son operadores fuera de
la diagonal con elementos de matriz no cero y solo act\'uan entre
estados con esp{\'\i}n diferente. Los primeros t\'erminos en la
expansi\'on son los siguientes
\begin{equation}\label{D2}
{\cal D}^{kc}_{2} = J^{k}T^{c},
\end{equation}
\begin{equation}\label{D3}
{\cal D}^{kc}_{3} = \{J^{k},\{J^{r},G^{rc}\}\},
\end{equation}
\begin{equation}\label{O3}
{\cal O}^{kc}_{3} = \{J^{2},G^{kc}\} - \frac{1}{2}
\{J^{k},\{J^{r},G^{rc}\}\}.
\end{equation}
Los t\'erminos a orden m\'as alto se obtienen como
\[
{\cal D}^{kc}_{n} = \{J^{2},{\cal D}^{kc}_{n-2}\}\hspace{0.8cm}  {\rm y}
\hspace{0.8cm} {\cal O}^{kc}_{n} = \{J^{2},{\cal O}^{kc}_{n-2}\},
\hspace{0.6cm} {\rm para}\hspace{0.6cm} n\geq 4.
\]
Los operadores ${\cal O}^{kc}_{2m}$  $(m=1,2, \ldots)$ est\'an
prohibidos en la expansi\'on  debido a que ellos son pares bajo
inversi\'on temporal. Los coeficientes desconocidos $a_{1}$,
$b_{n}$, y $c_{n}$ en la Ec. (\ref{corrienteAxial}) tiene
expansiones en potencias $1/N_{c}$ comenzando con orden 1. Para el
valor $N_{c}=3$ la serie puede truncarse como
\begin{equation}\label{Akc3}
A^{kc} = a_1 G^{kc} + b_2 \frac{1}{N_c} {\cal D}_2 + b_3 \frac{1}{N_c^2} {\cal D}_3 + c_3 \frac{1}{N_c^2}{\cal O}_3.
\end{equation}
Los elementos de matriz de las componentes espaciales de $A^{kc}$ entre estados de SU(6) dan los valores ordinarios de
los acoplamientos axial vector. Para los bariones del octete, los acoplamientos axial son $g_A$, como convencionalmente se definen
en ex\-pe\-ri\-men\-tos de decaimientos de bariones, con una normalizaci\'on tal que $g_A \approx $1.27, $g_{_V} = 1$ para
el decaimiento $\beta$ del neutr\'on y $g$ para decuplete octete.

Por otra parte, los coeficientes $a_1$, $b_2$, $b_3$ y $c_3$ en la Ec. (\ref{Akc3}), no los predice la teor{\'\i}a y las expresiones finales de la corriente axial ya renormalizada contiene a estos par\'ametros. Posteriormente, realizando los ajustes con los efectos de correcciones a un loop, de los datos sobre los decaimientos semilept\'onicos de bariones, podemos extraer los valores num\'ericos para los par\'ametros ya mencionados anteriormente. Esta es la forma como se extraen los par\'ametros b\'asicos $a_1$, $b_2$, $b_3$ y $c_3$ del Lagrangiano quiral $1/N_c$ y ellos se presentan en el Cap{\'\i}tulo 4.


\chapter{Renormalizaci\'on de la Corriente Axial Vector de Bariones}
En el contexto de la expansi\'on $1/N_c$, la corriente axial de bariones $A^{kc}$ dada en la Ec. (\ref{Akc3}), es un operador y los elementos de matriz de las componentes espaciales de $A^{kc}$ entre estados bari\'onicos, dan como resultado los valores usuales de los acoplamientos axial vector $g_A$ \cite{RFM2006}. El formalismo para el desarrollo del c\'alculo, consiste en utilizar una expansi\'on combinada en $m_q/\Lambda_\chi$ y en $1/N_c$, es decir, sobre el doble l{\'\i}mite, con rotura de simetr{\'\i}a quiral $m_q \rightarrow 0$ y el l{\'\i}mite $N_c \rightarrow \infty$, como en la Ref. \cite{RFM2006}. La contribuci\'on m\'as importante de este trabajo, es considerar inserciones de masa mediante el operador ${\cal M}$, el cual depende del esp{\'\i}n \cite{JenkinsPRD53} como lo podemos ver en la Ec. (\ref{ec:3masa-barion}). Ahora, en este cap{\'\i}tulo presentamos la renormalizaci\'on de la corriente axial vector de bariones, para lo cual se requiere hacer la  reducci\'on de una gran cantidad de productos de operadores que contribuyen a dicha renormalizaci\'on. Espec{\'\i}ficamente para el caso no degenerado, estos operadores est\'an contenidos en las expresiones para la contribuci\'on de sabor singulete, sabor octete y sabor ${\bf 27}$, manteniendo $N_c$ arbitrario y los cuales se presentan en las secciones 3.2 y 3.3.

Los procesos de decaimientos semilept\'onicos de bariones (DSB), $B_i\rightarrow B_j e^- \bar{\nu}_e$, del octete de bariones con esp{\'\i}n $\frac{1}{2}$ y paridad positiva ($J^{^P}=\frac{1}{2}^{+}$), son particularmente interesantes, debido a la relaci\'on que existe entre las interacciones d\'ebiles y fuertes junto con la matriz de mezcla de Cabibbo-Kobayashi-Maskawa (CKM).
En la des\-crip\-ci\'on de los DSB, el modelo de Cabibbo \cite{cabmodel} ha sido muy exitoso. Este modelo establece que las corrientes hadr\'onicas vectoriales y axiales pertenecen a octetes de una simetr{\'\i}a exacta de sabor SU(3). El poder predictivo de este modelo permite conocer todos los factores de forma de los posibles decaimientos semilept\'onicos del octete de ba\-rio\-nes en t\'erminos de los coeficientes de Clebsch-Gordan y de algunos par\'ametros libres (factores de forma reducidos).

La extra\~neza $S$ es un n\'umero cu\'antico conservado en las interacciones fuertes y electromagn\'eticas, mientras que no necesariamente se conserva en las transiciones d\'ebiles. Todos los DSB que se han observado en los experimentos, se pueden clasificar en dos grandes grupos, de acuerdo al n\'umero de extra\~neza, como se muestra en la Tabla \ref{t:DSB}. Cabe mencionar, que las transiciones con $|\Delta S| > 1$ no han sido observadas. La corriente hadr\'onica d\'ebil cargada se descompone en las componentes $\Delta S = 0$ y $|\Delta S| = 1$, expresados como una suma de t\'erminos vector y axial vector y el n\'umero cu\'antico de la corriente debe satisfacer las reglas de selecci\'on dadas sobre las dos clases de decaimientos. Por otra parte, tambi\'en presentamos en la Tabla \ref{t:DSB}
los decaimientos fuertes del decuplete de bariones de esp{\'\i}n $\frac{3}{2}$ hacia el octete de bariones y piones.
\begin{table}[h]
\begin{center}
\begin{tabular}{llr}
  \hline
  \hline
  \multicolumn{2}{c}{Procesos semilept\'onicos}& Procesos fuertes\\
  \hline
   \hspace{0.5cm} $\Delta S = 0$ & \hspace{0.5cm}$|\Delta S| = 1$& $\Delta S = 0$\\
  \hline
                \hspace{0.5cm}$n \hspace{0.2cm}\rightarrow \hspace{0.2cm} pe^-{\bar\nu_e}$ & \hspace{0.5cm}$\Lambda \hspace{0.2cm}\rightarrow \hspace{0.2cm}pe^-{\bar \nu}_e$ & $\Delta \rightarrow N \pi$ \\
                \hspace{0.5cm}$\Sigma^{+} \hspace{0.2cm}\rightarrow \hspace{0.2cm}\Lambda e^+ \nu_e$ & \hspace{0.5cm}$\Sigma^- \hspace{0.2cm}\rightarrow \hspace{0.2cm}ne^-{\bar\nu}_e$ & $\Sigma^{*}\rightarrow \Lambda\pi$ \\
                \hspace{0.5cm}$\Sigma^{-} \hspace{0.2cm}\rightarrow \hspace{0.2cm}\Lambda e^- {\bar \nu_e}$ & \hspace{0.5cm}$\Xi^- \hspace{0.2cm}\rightarrow \hspace{0.2cm}\Lambda e^-{\bar \nu}_e$ & $\Sigma^{*}\rightarrow \Sigma \pi$\\
                & \hspace{0.5cm}$\Xi^- \hspace{0.2cm}\rightarrow \hspace{0.2cm}\Sigma^0 e^-{\bar \nu}_e$ & $\Xi^{*}\rightarrow \Xi\pi$ \\
                & \hspace{0.5cm}$\Xi^0 \hspace{0.2cm} \rightarrow \hspace{0.2cm}\Sigma^+ e^-{\bar \nu}_e$\\
  \hline
  \hline
\end{tabular}
\caption{Decaimientos $\beta$ de bariones y decaimientos a piones. \label{t:DSB}}
\end{center}
\end{table}

\section{La corriente axial de bariones a nivel \'arbol}
El c\'alculo del operador de corriente axial vector para bariones $A^{kc}$ Ec. (\ref{Akc3}) al orden mas bajo, a nivel \'arbol, consiste en hacer la sustituci\'on de los elementos de matriz que se encuentran listados en la Tabla \ref{tabla2}, 
En el ejemplo siguiente presentamos $A_{\small{\acute{A}rbol}}^{kc}$, para la transici\'on neutr\'on-prot\'on. 
El resultado se obtiene en funci\'on de los par\'ametros b\'asicos $a_1$, $b_2$ y $b_3$ de la teor{\'\i}a.
\begin{eqnarray}
\left(g_A^{np}\right)_{\small{\acute{A}rbol}}&=&
       \left\langle \, p \, | \,A^{kc}_{\small{\acute{A}rbol}} \, | \, n \, \right\rangle \nonumber \\
       &=& a_1 \left\langle p | G^{kc} | n \right\rangle + \frac{b_2}{N_c} \left\langle p |{\cal D}_2^{kc} |n \right\rangle +
           \frac{b_3}{N_c^2} \left\langle p | {\cal D}_3^{kc} | n \right\rangle +
           \frac{c_3}{N_c^2} \left\langle p | {\cal O}_3^{kc} | n \right\rangle\nonumber\\
       &=& a_1 \frac{5}{6} + b_2 \frac{1}{6}  + b_3 \frac{5}{18}, \label{ga-arbol}
\end{eqnarray}
En la Tabla \ref{tabla2} se listan los valores de los elementos de matriz de los operadores de bariones $G^{kc}$, ${\cal D}_{2}^{kc}$ ${\cal D}_{3}^{kc}$ y ${\cal O}_3^{kc}$ para ocho decaimientos semilept\'onicos de bariones y para cuatro decaimientos fuertes de bariones a piones.

El c\'alculo del operador $A^{kc}$ se complica en la siguientes secciones de este cap{\'\i}tulo, cuando consideramos correcciones a un loop y con rotura de simetr{\'\i}a de sabor SU(3) perturvativa.
\begingroup
\begin{table}[ht]
\begin{center}
\begin{tabular}{lccccccccccc}
\hline\hline
$B_1 B_2$ & $np$& $\Sigma^\pm \Lambda$& $\Lambda p$ & $\Sigma^- n$ & $\Xi^-\Lambda$ & $\Xi^-\Sigma^0$ & $\Xi^0\Sigma^+$ & $\Delta N $& $\Sigma^{*}\Lambda$ & $\Sigma^{*}\Sigma$ & $\Xi^{*}\Xi$\\
\hline
$\langle G^{kc} \rangle$ & $\frac{5}{6}$ & $\frac{1}{\sqrt{6}}$ & $-\frac12\sqrt{\frac{3}{2}}$ & $\frac{1}{6}$ & $\frac{1}{2 \sqrt{6}}$ & $\frac{5}{6 \sqrt{2}}$ & $\frac{5}{6}$ & $-1$ & $-1$ & $-1$ & $-1$\\
$\langle \mathcal{D}_2^{kc} \rangle$ & $\frac{1}{2}$ & $0$ & $-\frac12\sqrt{\frac{3}{2}}$ & $-\frac{1}{2}$ & $\frac12\sqrt{\frac{3}{2}}$ & $\frac{1}{2 \sqrt{2}}$ & $\frac{1}{2}$ &0&0&0&0\\
$\langle \mathcal{D}_3^{kc} \rangle$ & $\frac{5}{2}$ & $\sqrt{\frac{3}{2}}$ & $-\frac32\sqrt{\frac{3}{2}}$ & $\frac{1}{2}$ & $\frac12\sqrt{\frac{3}{2}}$ & $\frac{5}{2 \sqrt{2}}$ & $\frac{5}{2}$ & $0$ &0&0&0\\
$\langle {\cal O}_3^{kc} \rangle$ & 0&0&0&0&0&0&0&$-\frac{9}{2}$&$-\frac{9}{2}$&$-\frac{9}{2}$&$-\frac{9}{2}$\\
\hline\hline
\end{tabular}
\caption{\label{tabla2} Elementos de matriz de los operadores de
bariones: Nivel \'arbol y contribuci\'on
singulete.}
\end{center}
\end{table}
\endgroup

\section{Correcciones a un loop a la corriente axial de bariones}
La corriente axial $A^{kc}$ se renormaliza por los diagramas a un loop Fig. \ref{Fig:diagramas}. Estos diagramas de loop dependen de la raz\'on $\Delta/m_{\Pi}$, donde $\Delta\equiv M_T - M_B$ es la diferencia de masas decuplete-octete y $m_\Pi$ es la masa del mes\'on $\pi$, $K$ y $\eta$. Los diagramas \ref{Fig:diagramas}(a-c) son de orden $N_c$ veces el v\'ertice a nivel \'arbol y la Fig. \ref{Fig:diagramas}(d) es de orden $1/N_c$ veces el v\'ertice a nivel \'arbol. Las cancelaciones en el l{\'\i}mite de $N_c$ grande ocurren entre los diagramas a un loop Figs. \ref{Fig:diagramas}(a-c).

\subsection{Correcciones a un loop: Diagramas Fig. 2.1(a-c)}

La correcci\'on a la corriente axial $\delta A^{kc}$ se obtiene de
la suma de los diagramas de las Figs. 2.1(a-c), la expresi\'on
matem\'atica es \cite{RFM2000},
\begin{eqnarray}\label{eq:deltaAkc}
\delta
A^{kc}&=&\frac{1}{2}\left[A^{ja},\left[A^{jb},A^{kc}\right]\right]
                \Pi^{ab}_{(1)} - \frac{1}{2}\left\{A^{ja},\left[A^{kc},
                \left[{\cal M}, A^{jb}\right]\right]\right\}\Pi^{ab}_{(2)}\nonumber\\
      &&+\frac{1}{6} \biggl(\left[A^{ja},\left[\left[{\cal M},
                \left[{\cal
                M},A^{jb}\right]\right],A^{kc}\right]\right] - \frac{1}{2} \left[\left[{\cal M},A^{ja}\right],\left[\left[{\cal M},
                  A^{jb}\right],A^{kc}\right]\right]\biggr)
                  \Pi^{ab}_{(3)}+\ldots\nonumber\\
\end{eqnarray}
En la Ec. (\ref{eq:deltaAkc}) el tensor sim\'etrico $\Pi^{ab}_{(n)}$ contiene integrales a un loop Ec. (\ref{delta}) con intercambio de un mes\'on: Un mes\'on de sabor $a$ se emite y un mes\'on de sabor $b$ se absorbe. $\Pi^{ab}_{(n)}$ se descompone en
representaciones de sabor singulete {\bf 1}, sabor octete {\bf 8}
y sabor {\bf 27},
\begin{equation}
    \Pi^{ab}_{(n)}=F^{(n)}_{\bf 1}\delta^{ab} + F^{(n)}_{\bf 8}d^{ab8}+F^{(n)}_{\bf 27}\left[\delta^{a8}\delta^{b8}-
    \frac{1}{8}\delta^{ab} - \frac{3}{5} d^{ab8}d^{888}\right]. \label{eq:Pi}
\end{equation}
Las combinaciones lineales de sabor singulete $F_{\bf 1}$, octete $F_{\bf 8}$ y $F_{\bf 27}$ de las integrales sobre loop est\'an dadas en las Ecs. (\ref{eq:Fn1}) - (\ref{eq:Fn27}). En el l{\'\i}mite de degeneraci\'on se obtiene
\begin{eqnarray}
  F^{(1)}(m_\Pi,0,\mu) &=& -\frac{m_\Pi^2}{16 \pi^2 f^2}{\rm ln} \frac{m_\Pi^2}{\mu^2}, \label{ec:20a}\\
  F^{(2)}(m_\Pi,0,\mu) &=& -\frac{1}{8 \pi f^2} m_\Pi, \label{ec:20b}\\
  F^{(3)}(m_\Pi,0,\mu) &=& \frac{1}{4 \pi^2 f^2}{\rm ln}\frac{m_\Pi^2}{\mu^2} \label{ec:20c}.
\end{eqnarray}

\subsubsection{Esquema de conteo de potencias}
Para facilitar el c\'alculo, en primer lugar, podemos hacer uso del esquema de conteo de potencias $1/N_c$ \cite{RFM2006,RFM2000}, el cual establece que, para bariones con espines de orden uno,
\begin{equation}
T^a \sim N_c, \qquad G^{ia} \sim N_c, \qquad J^i \sim 1.
\label{eq:crules}
\end{equation}
Esto es equivalente a afirmar que los factores de $J^i/N_c$ est\'an $1/N_c$ suprimidos en relaci\'on con los factores de $T^a/N_c$ y $G^{ia}/N_c$. Con seguridad se puede implementar esta regla de conteo $1/N_c$ para restringir los estados de bariones al orden m\'as bajo, es decir, aquellos que constituyen la
representaci\'on $\bf{56}$ dimensional de SU(6).

En segundo lugar, tambi\'en debemos tener en cuenta lo siguiente: Un n\'umero par o impar de inserciones del operador de masa de
bariones, en la Ec. (\ref{eq:deltaAkc}), da estructuras con un orden diferente en $N_c$. Esta $N_c$-dependencia se determin\'o en
Ref.~\cite{RFM2000} a trav\'es de un an\'alisis detallado. B\'asicamente uno necesita contar potencias de $J$
porque de la supresi\'on $1/N_c$ el factor $J/N_c$ se introduce.

Por ejemplo, en $A^{kc}$ y $\mathcal{M}$ el operador de esp{\'\i}n
$J$ aparece un m{\'\i}nimo de 0 y 2 veces, respectivamente.

Sea $r$ el n\'umero de $J$'s de $A^{kc}$ y $\mathcal{M}$ m\'as all\'a
de estos valores m{\'\i}nimos en una estructura dada.

En la Ec.~(\ref{eq:deltaAkc}), las contribuciones, sin inserci\'on de masa son las siguientes:
\begin{eqnarray*}
   && \mathcal{O}(N_c^0)\hspace{.5cm} \mbox{para} \hspace{.5cm} r = 0, 1\\
   && \mathcal{O}(N_c^{2-r})\hspace{.5cm} \mbox{para} \hspace{.5cm} r \geq 2.
\end{eqnarray*}
Con una inserci\'on de masa, son listadas a continuaci\'on:
\begin{eqnarray*}
 && \mathcal{O}(N_c^0) \hspace{.5cm} \mbox{para} \hspace{.5cm} r = 0, 1 \\
 &&\mathcal{O}(N_c^{1-r}) \hspace{.5cm} \mbox{para} \hspace{.5cm} r\geq 2.
\end{eqnarray*}
Con dos inserciones de masa, son al siguiente orden
\[
\mathcal{O}(N_c^{-r})
\]
Observemos que estas potencias en $N_c$ incluyen
la supresi\'on $1/N_c$ debido al factor de $1/f^2$ el cual
acompa\~na la integral de loop \cite{RFM2000}.
Analizaremos brevemente las implicaciones que este esquema de conteo de potencias tiene en los diferentes sumandos en Ec. (\ref{eq:deltaAkc}).
\subsubsection{A. Diagramas Fig. 2.1(a-c): L{\'\i}mite degenerado}
El primer t\'ermino de la Ec. (\ref{eq:deltaAkc}) es
\begin{equation}{\label{AAA}}
\delta A_{\rm deg}^{kc} =\frac12 \left[A^{ja},\left[A^{jb},A^{kc}\right]\right]\Pi_{(1)}^{ab}
\end{equation}
este t\'ermino corresponde al l{\'\i}mite degenerado $\Delta/m_\Pi\to 0$ y
ha sido analizado en Ref. \cite{RFM2006}. Uno esperar{\'\i}a que
el doble conmutador fuera de orden $\mathcal{O}(N_c^3)$, es decir,
un factor $N_c$ por cada $A^{kc}$. Sin embargo existen
cancelaciones a $N_c$ grande entre los diagramas de Feynman de las
Figs. 2.1(a-c), siempre y cuando todos los estados
bari\'onicos en un multiplete completo de la simetr{\'\i}a
esp{\'\i}n sabor SU(6), para $N_c$ grande, sean incluidos en la suma
sobre estados intermediarios y las razones del acoplamiento axial
predichas por esta simetr{\'\i}a esp{\'\i}n sabor sean utilizados
\cite{RFM2000}. Se puede demostrar anal{\'\i}ticamente que este
doble conmutador es ${\cal O}(N_c)$ \cite{RFM2006}.

Utilizando la regla de conteo mencionada anteriormente se listan en la Tabla \ref{tAAA} los diferentes t\'erminos del producto $AAA$ Ec. (\ref{AAA}) es decir, $GGG$, $GG{\cal D}_2$, $G{\cal D}_2{\cal D}_2$, $GG{\cal D}_3$ y $GG{\cal O}_3$, contribuyen al mismo orden al doble conmutador y da correcciones de orden $1/N_c$ a los resultados a nivel \'arbol, que es de orden $N_c$. Al siguiente orden de importancia los t\'erminos con $\mathcal{D}_2\mathcal{D}_2\mathcal{D}_2$,
$G\mathcal{D}_2\mathcal{D}_3$ y $G\mathcal{D}_2\mathcal{O}_3$ hace correcciones de orden $1/N_c^2$ a los valores a nivel \'arbol.
\begin{table}[h]
\begin{center}
\begin{tabular}{cc}
\hline
\hline
\multicolumn{2}{c}{T\'erminos con $r = 0,1,2$}\\
$\mathcal{O}(1/N_c)$ al valor nivel arbol. & $GGG$, $GG\mathcal{D}_2$, $G\mathcal{D}_2\mathcal{D}_2$, $GG\mathcal{D}_3$ y $GG\mathcal{O}_3$\\
\multicolumn{2}{c}{T\'erminos con $r = 3$}\\
$\mathcal{O}(1/N^2_c)$ al valor nivel arbol. & $\mathcal{D}_2\mathcal{D}_2\mathcal{D}_2$, $G\mathcal{D}_2\mathcal{D}_3$, y $G\mathcal{D}_2\mathcal{O}_3$\\
\hline
\hline
\end{tabular}
\caption{T\'erminos que contribuyen al doble conmutador del producto $AAA$. \label{tAAA}}
\end{center}
\end{table}

\subsubsection{B. Diagramas Fig. 2.1(a-c): L{\'\i}mite no degenerado}
Los siguientes sumandos de la Ec. (\ref{eq:deltaAkc}) contienen
inserciones de masa y son los t\'erminos de inter\'es en este
trabajo,
\begin{eqnarray}\label{Akc-nodegenerado2}
&&\hspace{-1cm}\delta A^{kc}_{\Delta} =-\,\frac{1}{2}\left\{A^{ja},\left[A^{kc},
         \left[{\cal M},
         A^{jb}\right]\right]\right\}\Pi_{(2)}^{ab}\nonumber \\
               &&+\,\frac{1}{6} \biggl(\left[A^{ja},\left[\left[{\cal M},\left[{\cal M},A^{jb}\right]\right],A^{kc}\right]\right]-\frac{1}{2} \left[\left[{\cal
M},A^{ja}\right],\left[\left[{\cal M},
         A^{jb}\right],A^{kc}\right]\right]\biggr)\Pi_{(3)}^{ab} + \cdots, \nonumber\\
\end{eqnarray}
como hemos mencionado lo que nos interesa calcular es la contribuci\'on debida a los t\'erminos de Ec. (\ref{Akc-nodegenerado2}). Estas correcciones contienen inserciones de masa y forman el caso m\'as realista, el l{\'\i}mite no degenerado cuando $\Delta \equiv M_T - M_B \neq 0$.

Aunque el operador de masa bari\'onica $\mathcal{M}$ entra
expl{\'\i}citamente en la expresi\'on anterior, uno se queda s\'olo
con el operador de desdoblamiento de masa hiperfina
$\mathcal{M}_{\textrm{hiperfina}}$, Ec. (\ref{ec:3masa-barion}),
porque el t\'ermino independiente de esp{\'\i}n en $\mathcal{M}$
es proporcional al operador identidad por lo que queda fuera de
los conmutadores que lo contienen. De acuerdo a las reglas de
conteo de potencias $N_c$, para $r = 0,1$ los t\'erminos en el
producto $AAA\mathcal{M}$, es decir, $GGGJ^2$ y
$GG\mathcal{D}_2J^2$, producen las correcciones de orden $1/N_c$ a
nivel \'arbol, mientras que, al siguiente orden de importancia,
para $r = 2$, las contribuciones surgen de
$G\mathcal{D}_2\mathcal{D}_2J^2$, $GG\mathcal{D}_3J^2$ y
$GG\mathcal{O}_3J^2$, como se muestra en la Tabla \ref{tAAAM}

\begin{table}[h]
\begin{center}
\begin{tabular}{cc}
\hline
\hline
\multicolumn{2}{c}{T\'erminos con $r = 0,1$}\\
$\mathcal{O}(1/N_c)$ al valor nivel arbol. & $GGGJ^2$, $GG\mathcal{D}_2J^2$\\
\multicolumn{2}{c}{T\'erminos con $r = 2$}\\
$\mathcal{O}(1/N^2_c)$ al valor nivel arbol. & $G\mathcal{D}_2\mathcal{D}_2J^2$, $GG\mathcal{D}_3J^2$, y $GG\mathcal{O}_3J^2$\\
\hline
\hline
\end{tabular}
\caption{T\'erminos contenidos en el producto $AAA\mathcal{M}$. \label{tAAAM}}
\end{center}
\end{table}

Mediante la regla de conteo mencionada anteriormente, ahora corresponde al producto $AAA\mathcal{M}\mathcal{M}$ con t\'erminos, como
$GGGJ^2J^2$ y $GG\mathcal{D}_2J^2J^2$ producir\'a la correcci\'on de \'ordenes de $1/N_c$ y $1/N_c^2$ a nivel
\'arbol, respectivamente, como se muestra en la Tabla \ref{tAAAMM}. Adem\'as, como parte interesante de informaci\'on que tambi\'en podemos extraer es, que las
correcciones dominantes $1/N_c$ de los desdoblamientos de masa del bari\'on se deben a m\'ultiples inserciones del operador $J^2$ en
lugar de las contribuciones de potencias de $J^2$. Por ejemplo, dos inserciones de $J^2$, como en $GGGJ^2J^2$ son m\'as grandes
(por una potencia de $N_c$) que una inserci\'on de $J^4$ como en $GGGJ^4$.

\begin{table}[h]
\begin{center}
\begin{tabular}{cc}
\hline
\hline
\multicolumn{2}{c}{T\'erminos con $r = 0$}\\
$\mathcal{O}(1/N_c)$ al valor nivel arbol. & $GGGJ^2J^2$, \\
\multicolumn{2}{c}{T\'erminos con $r = 1$}\\
$\mathcal{O}(1/N^2_c)$ al valor nivel arbol. & $GG\mathcal{D}_2J^2J^2$\\
\hline
\hline
\end{tabular}
\caption{T\'erminos contenidos en el producto $AAA\mathcal{MM}$. \label{tAAAMM}}
\end{center}
\end{table}

Para evaluar los elementos de matriz entre estados
bari\'onicos utilizamos las identidades de la Tabla \ref{identi},
$a_1\,G^{kc}$ es el t\'ermino a orden l{\'\i}der en $N_c$ para la
corriente axial $A^{kc}$, as{\'\i} tenemos que el
anticonmutador-conmutador correspondiente al caso no degenerado
con una inserci\'on de masa es el siguiente
\begin{equation}\label{GGG}
\frac{a_1^3}{N_c}
m_2\left\{G^{ia},\left[G^{kc},\left[J^2,G^{ia}\right]\right]\right\},
\end{equation}
el desarrollo para la reducci\'on de operadores de la
estructura anticonmutador-conmutador Ec. (\ref{GGG}) se realiza en
el Ap\'endice B de esta tesis, el resultado manteniendo $N_c$ y
$N_f$ arbitrarios es el siguiente
\begin{equation}\label{correccionloop}
\left\{G^{ia},\left[G^{kc},\left[J^2,G^{ia}\right]\right]\right\}
               = - \frac{1}{2}(N_f -2)G^{kc} + \frac{1}{2}(N_c +
               N_f){\cal D}_2^{kc}-\frac{1}{2}D_3^{kc}-{\cal
               O}_3^{kc}.
\end{equation}
Una vez que tenemos identificadas todas las diferentes
contribuciones que se requieren al orden de aproximaci\'on
implementada aqu{\'\i}, se realiza el c\'alculo de \'estas, siguiendo el
procedimiento como en Ref.\cite{RFM2006}. La correcci\'on $\delta
A^{kc}$ en t\'erminos de las diferentes re\-pre\-sen\-taciones de sabor
singulete, octete y {\bf 27} de SU(3) Ec.(\ref{Akc-nodegenerado2}),
son complicadas estructuras conmutador y/o anticonmutador y contienen operadores de quarks de $n$
cuerpos, los cuales act\'uan sobre estados bari\'onicos.
Estos operadores son polinomios en los ge\-ne\-ra\-do\-res de un
cuerpo, de esp{\'\i}n $J^k$, de sabor $T^c$ y de esp{\'\i}n sabor
$G^{kc}$ de SU(2$N_f$) \cite{DashenPRD51}. Los generadores de la
simetr{\'\i}a esp{\'\i}n sabor, expresados en t\'erminos de
ope\-ra\-do\-res bos\'onicos de quarks de creaci\'on
$q_\alpha^\dagger$ y de aniquilaci\'on $q^\alpha$, se encuentran escritos en la Ec. (\ref{ec:op}).

No todos los productos de operadores de los generadores de
esp{\'\i}n sabor son li\-neal\-men\-te independientes. Existen
identidades entre operadores \cite{DashenPRD51, Manohar}  que
permiten eliminar ciertas combinaciones en la expansi\'on general.
La reducci\'on aunque muy larga y tediosa en vista de la
considerable cantidad de teor{\'\i}a de grupo involucrada, es sin
embargo, realizable dado que la base de operadores es completa
e independiente \cite{DashenPRD49,{DashenPRD51}}.

Ahora obtenemos la correcci\'on a un loop para $\delta A^{kc}$ a orden relativo ${\cal O} (1/N_c^2)$
\begin{equation}{\label{deltaAkc1-8-27}}
\delta A^{kc} = \delta A^{kc}_{\bf 1} + \delta A^{kc}_{{\bf 8}} + \delta A^{kc}_{\bf 27}
\end{equation}
donde
\begin{equation}
\delta A_{\mathbf{1}}^{kc} = \sum_{i=1}^{7} s_i S_i^{kc}, \label{eq:das}
\end{equation}
\begin{equation}
\delta A_{\mathbf{8}}^{kc} = \sum_{i=1}^{30} o_i O_i^{kc} \label{eq:dao}
\end{equation}
y
\begin{equation}
\delta A_{\mathbf{27}}^{kc} = \sum_{i=1}^{61} t_i T_i^{kc}, \label{eq:dat}
\end{equation}
donde el sub{\'\i}ndice {\bf 1, 8 y 27} en Ec. (\ref{deltaAkc1-8-27}) se refiere a las diferentes representaciones de sabor SU(3).

Para la contribuci\'on de sabor singulete, obtenemos la base de operadores $S_1^{kc}$ al $S_7^{kc}$ al orden de aproximaci\'on usado aqu{\'\i}, estos operadores se encuentran dados de la siguiente forma
\begin{eqnarray}
   && S_1^{kc}= G^{kc}, \hspace{0.5cm} S_2^{kc}= {\cal D}_2^{kc},\hspace{0.5cm} S_3^{kc}= {\cal D}_3^{kc}, \nonumber\\
   && S_4^{kc}= {\cal O}_3^{kc}, \hspace{0.5cm} S_5^{kc}= {\cal D}_4^{kc},\hspace{0.5cm} S_6^{kc}= {\cal D}_5^{kc}, \nonumber\\
   && S_7^{kc}= {\cal O}_5^{kc},
\end{eqnarray}
y los diferentes coeficientes considerados en Ec. (\ref{eq:das}) son los siguientes

\begin{eqnarray}
s_{1} & = & \left[\frac{23}{24} a_1^3 - \frac{N_c+3}{3N_c} a_1^2b_2 + \frac{N_c^2+6N_c-54}{12N_c^2} a_1b_2^2 - \frac{N_c^2+6N_c+2}{2N_c^2} a_1^2b_3
          -\frac{N_c^2+6N_c-3}{2N_c^2} \right.\nonumber\\
          & & \left. \times a_1^2c_3 - \frac{6(N_c+3)}{N_c^3} a_1b_2b_3 \right] F_{\mathbf{1}}^{(1)}
          + \left[\frac14 a_1^3 - \frac{N_c+3}{N_c} a_1^2b_2 - \frac{N_c^2+6N_c+6}{N_c^2} a_1^2b_3 \right] \frac{\Delta}{N_c} F_{\mathbf{1}}^{(2)} \nonumber\\
          & & + \frac{1}{12} (N_c^2+6N_c-3) a_1^3 \frac{\Delta^2}{N_c^2} F_{\mathbf{1}}^{(3)}, \label{eq:s1}
\end{eqnarray}
\begin{eqnarray}
s_{2} & = & \left[ \frac{101}{24N_c} a_1^2b_2 + \frac{2(N_c+3)}{3N_c^2} a_1b_2^2 + \frac{N_c^2+6N_c-18}{12N_c^3} b_2^3 - \frac{3(N_c+3)}{2N_c^2} a_1^2b_3 -
            \frac{N_c+3}{4N_c^2} a_1^2c_3\right.\nonumber \\
& & + \left. \frac{N_c^2+6N_c+2}{2N_c^3} a_1b_2b_3 - \frac{3(N_c^2+6N_c-24)}{4N_c^3} a_1b_2c_3 \right] F_{\mathbf{1}}^{(1)}\nonumber\\
&&+ \left[ - \frac14 (N_c+3) a_1^3 - \frac{N_c^2+6N_c-29}{4N_c} a_1^2b_2 - \frac{5(N_c+3)}{N_c^2} a_1^2b_3 - \frac{3(N_c+3)}{2N_c^2}a_1^2c_3 \right] \frac{\Delta}{N_c} F_{\mathbf{1}}^{(2)} \nonumber\\
& & + \left[ - \frac{11}{24}(N_c+3) a_1^3 - \frac{3(N_c^2+6N_c-16)}{8N_c} a_1^2b_2 \right] \frac{\Delta^2}{N_c^2} F_{\mathbf{1}}^{(3)},
\end{eqnarray}
\begin{eqnarray}
s_{3} & = & \left[ \frac{11}{8N_c^2} a_1b_2^2 + \frac{51}{8N_c^2} a_1^2b_3 + \frac{1} {N_c^2}a_1^2c_3 + \frac{17(Nc+3)}{6N_c^3} a_1b_2b_3 - \frac{9(N_c+3)}{4N_c^3} a_1b_2c_3 \right] F_{\mathbf{1}}^{(1)} \nonumber \\
& & \mbox{} + \left[ \frac14 a_1^3 - \frac{N_c+3}{4N_c} a_1^2b_2 - \frac{2N_c^2+12N_c-53}{4N_c^2} a_1^2b_3 + \frac{9}{4N_c^2} a_1^2c_3 \right] \frac{\Delta}{N_c} F_{\mathbf{1}}^{(2)}\nonumber\\
& & + \left[ \frac12 a_1^3 - \frac{19(N_c+3)}{24N_c} a_1^2b_2 \right] \frac{\Delta^2}{N_c^2} F_{\mathbf{1}}^{(3)},
\end{eqnarray}
\begin{eqnarray}
s_{4} & = & \left[\frac{3}{4N_c^2} a_1b_2^2 + \frac{7}{6N_c^2} a_1^2b_3 + \frac{167}{24N_c^2} a_1^2c_3 + \frac{5(N_c+3)}{3N_c^3} a_1b_2b_3 - \frac{N_c+3}{3N_c^3} a_1b_2c_3 \right] F_{\mathbf{1}}^{(1)} \nonumber \\
& & +\left[ \frac12 a_1^3 - \frac{5}{2N_c^2} a_1b_2^2 + \frac{1}{N_c^2} a_1^2b_3 - \frac{2N_c^2+12N_c-37}{4N_c^2} a_1^2c_3 \right] \frac{\Delta}{N_c} F_{\mathbf{1}}^{(2)}\nonumber\\
& & + \left[\frac{2}{3} a_1^3 - \frac{N_c+3}{3N_c} a_1^2b_2 \right] \frac{\Delta^2}{N_c^2} F_{\mathbf{1}}^{(3)},
\end{eqnarray}
\begin{eqnarray}
s_{5} &=& \left[ \frac{5}{4N_c^3} b_2^3 + \frac{11}{6N_c^3} a_1b_2b_3 + \frac{19}{2N_c^3} a_1b_2c_3\right] F_{\mathbf{1}}^{(1)}\nonumber\\
       && \mbox{} + \left[\frac{1}{N_c} a_1^2b_2 - \frac{N_c+3}{2N_c^2} a_1^2b_3 - \frac{N_c+3}{2N_c^2} a_1^2c_3 \right]
            \frac{\Delta}{N_c} F_{\mathbf{1}}^{(2)}\nonumber\\
       && \mbox{}+ \frac{49}{12N_c} a_1^2b_2 \frac{\Delta^2}{N_c^2} F_{\mathbf{1}}^{(3)},
\end{eqnarray}
\begin{equation}
\hspace{-2cm} s_{6} = \left[ \frac{3}{2N_c^2} a_1^2b_3 + \frac{1}{2N_c^2} a_1^2c_3 \right] \frac{\Delta}{N_c} F_{\mathbf{1}}^{(2)},
\end{equation}
\begin{equation}
s_{7} = \left[ \frac{5}{2N_c^2} a_1^2c_3 \right] \frac{\Delta}{N_c} F_{\mathbf{1}}^{(2)}. \label{eq:s7}
\end{equation}
Las contribuciones de sabor $\mathbf{8}$ y $\mathbf{27}$ se encuentran listadas en el Ap\'endice D.

En las Ecs.~(\ref{eq:s1} - \ref{eq:s7}) hemos definido $\Delta=m_2$, como
consecuencia de la co\-rres\-pon\-den\-cia entre los par\'ametros del
octete y decuplete del Lagrangiano quiral $1/N_c$ para $N_c=3$,
Ref. \cite{JenkinsPRD53}, es decir,
\begin{equation}
M_B=3 m_0 + \frac14 m_2, \qquad M_T = 3m_0+\frac54 m_2.
\end{equation}

Por otra parte, $a_1$, $b_2$, $b_3$ y $c_3$ son los coeficientes indeterminados no predichos por la teor{\'\i}a.

Los elementos de matriz correspondientes se presentan en la Tabla \ref{tabla2}, para los ocho procesos entre bariones del octete de esp{\'\i}n $\frac{1}{2}$.

\subsection{Correcciones a un loop: Diagrama Fig. 2.1(d)}
La correcci\'on a un loop a la corriente axial vector para el diagrama de la Fig. 2.1(d) est\'a dada por la expresi\'on
\begin{equation}\label{eq:Akc(d)}
    \delta A^{kc} = -\frac{1}{2} \left[T^{a},\left[T^{b}, A^{kc}\right]\right]\Pi^{ab},
\end{equation}
donde $\Pi^{ab}$ es un tensor sim\'etrico con una estructura similar al de la Ec. (\ref{eq:Pi}),
\begin{equation}\label{Pi(d)}
  \Pi^{ab} = I_{\bf 1} \delta^{ab} + I_{\bf 8} d^{ab8} + I_{\bf 27} \left[ \delta^{a8} \delta^{b8} - \frac{1}{8} \delta^{ab} - \frac{3}{5} d^{ab8} d^{888} \right],
\end{equation}
nuevamente, los tensores de sabor singulete, octete y 27 en Ec. (\ref{Pi(d)}) son proporcionales a  sabor singulete $I_{\bf 1}$, sabor octete $I_{\bf 8}$
y sabor {\bf 27} $I_{27}$, combinaciones lineales de las integrales a un loop $I(m_\pi, \mu)$, $I(m_K, \mu)$ y $I(m_\eta, \mu)$, como sigue
\begin{equation}\label{Ec:I(d)}
I(m_\Pi, \mu) = \frac{i}{f^2} \int \frac{d^4 k}{(2\pi)^4} \frac{1}{k^2 - m_\Pi^2} = \frac{m_\Pi^2}{16 \pi^2 f^2} \left[{\rm ln}\frac{m_\Pi^2}{\mu^2} - 1 \right].
\end{equation}
Estos forman las combinaciones lineales Ecs. (\ref{eq:I1})-(\ref{eq:I27})

El c\'alculo del doble conmutador da como resultado en Ref. \cite{RFM2006}, las siguientes contribuciones de sabor para $N_f = 3$:
\begin{enumerate}
  \item Contribuci\'on de sabor singulete
  \begin{equation}\label{Ta1}
    [T^a, [T^a, A^{kc}]] = 3A^{kc},
  \end{equation}
  \item Contribuci\'on de sabor octete
  \begin{equation}\label{Ta8}
    d^{ab8}[T^a, [T^a, A^{kc}]] = \frac{3}{2}d^{c8e}A^{ke},
  \end{equation}
  \item Contribuci\'on de sabor {\bf 27}
  \begin{equation}\label{Ta27}
    [T^8, [T^8, A^{kc}]] = f^{c8e}f^{8eg}A^{kg},
  \end{equation}
\end{enumerate}

Los dobles conmutadores en Ecs. (\ref{Ta1}) - (\ref{Ta27}) son proporcionales a $A^{kc}$ y son ${\cal O}(N_c)$; as{\'\i} la correcci\'on a un loop de la Fig. 2.1(d) es a lo m\'as ${\cal O}(1)$ ya que $f^2$ se escala como $N_c$. En consecuencia, estas correcciones son del mismo orden que la obtenidas de las Figs. 2.1(a-c), es decir, esto es de orden relativo $1/N_c$ a la contribuci\'on a nivel \'arbol y no implica ninguna cancelaci\'on entre estados octete y decuplete.

\subsection{Correcci\'on total a un loop}

Finalmente la correcci\'on total a la corriente axial vector de bariones $A^{kc}$ considerando cada uno de los diagramas a un loop de la Fig. 2 (a-d) contiene todos los t\'erminos de la Ec. (\ref{eq:AkcTotal}),

\begin{eqnarray}\label{eq:AkcTotal}
\delta
A^{kc}&=&\frac{1}{2}\left[A^{ja},\left[A^{jb},A^{kc}\right]\right]
                \Pi^{ab}_{(1)}-\frac{1}{2}\left\{A^{ja},\left[A^{kc},
                \left[{\cal M}, A^{jb}\right]\right]\right\}\Pi^{ab}_{(2)}\nonumber\\
      &&+\frac{1}{6} \biggl(\left[A^{ja},\left[\left[{\cal M},
                \left[{\cal
                M},A^{jb}\right]\right],A^{kc}\right]\right] - \frac{1}{2} \left[\left[{\cal M},A^{ja}\right],\left[\left[{\cal M},
                  A^{jb}\right],A^{kc}\right]\right]\biggr)
                  \Pi^{ab}_{(3)}\nonumber \\
      &&-\frac{1}{2} \left[T^{a},\left[T^{b}, A^{kc}\right]\right]\Pi^{ab} + \ldots
\end{eqnarray}

Cabe mencionar que el diagrama a un loop de la Fig. 2(d), aunque esta considerado en la renormalizaci\'on de la corriente axial, este no depende de la raz\'on $\Delta / m_\Pi$. La contribuci\'on del diagrama Fig. 2(d), si est\'a tomada en cuenta en el presente an\'alisis.

El inter\'es por calcular todas estas estructuras de conmutadores y anticonmutadores Ec. (\ref{eq:AkcTotal}) se basa en el hecho de que los c\'alculos realizados se pueden comparar con las predicciones te\'oricas dentro de otros enfoques y con los datos experimentales reportados \cite{PDG}. Espec{\'\i}ficamente, una comparaci\'on directa puede llevarse a cabo con el acoplamiento axial $g_{A}$ el cual se ha obtenido en el marco de teor{\'\i}a de perturbaciones quirales para bariones pesados.
Hasta aqu{\'\i}, hemos llevado a cabo la renormalizaci\'on de la corriente axial vector tomando en cuenta la contribuci\'on del octete y el decuplete de bariones donde el Lagrangiano quiral $1/N_c$ de bariones posee una simetr{\'\i}a contra{\'\i}da esp{\'\i}n sabor SU(6).

\section{Correcciones con rotura de simetr{\'\i}a SU(3) perturbativa}

Una parte importante que debemos considerar en el presente an\'alisis es el tema de rotura de simetr{\'\i}a SU(3) perturbativa para el operador de corriente axial $A^{kc}$. Como hab{\'\i}amos mencionado anteriormente $A^{kc}$ es un objeto de esp{\'\i}n $1$ y se transforma como sabor octete bajo SU(3). La rotura de simetr{\'\i}a tambi\'en se transforma como un octete bajo SU(3).

Si ignoramos la rotura de isosp{\'\i}n e incluimos rotura de simetr{\'\i}a SU(3) perturbativa a primer orden, entonces $A^{kc}$ contiene partes que se transforman de acuerdo a todas las representaciones de SU(3) contenidas en el producto tensorial $(1, {\bf 8} \otimes {\bf 8}) = (1,{\bf 1}) \oplus (1,{\bf 8}_S) \oplus (1, {\bf 8}_A) \oplus (1, {\bf 10} + \overline{\bf 10}) \oplus(1, {\bf 27})$, es decir,

\begin{equation}\label{eq:delta8y10}
\delta A^{kc}_{{\rm SB}} = \delta A_{{\rm SB},{\bf 1}}^{kc} + \delta A_{{\rm SB},{\bf 8}}^{kc} + \delta A_{{\rm SB}, {\bf 10} + \overline{\bf 10}}^{kc} + \delta A_{{\rm SB},{\bf 27}}^{kc}
\end{equation}
donde el sub{\'\i}ndice ${\rm SB}$, significa rotura de simetr{\'\i}a SU(3), y los n\'umeros la representaci\'on de sabor, singulete {\bf 1}, octete {\bf 8}, {\bf 10} + $\overline{\bf 10}$ y {\bf 27} respectivamente. En principio, $\delta A^{kc}_{{\rm SB}}$ es de orden ${\cal O}(\epsilon N_c)$
siguiendo el desarrollo como en las Refs. \cite{DashenPRD51, Jen2012, Dai}, para cons\-truir los operadores que tenemos en Ec. (\ref{eq:delta8y10}) a orden relativo ${1}/{N^2_c}$.


\subsection{Contribuci\'on de sabor singulete} 
La expansi\'on $1/N_c$ para el operador $(1, {\bf 1})$, a orden relativo $1/N_c$, tiene dos t\'erminos los cuales est\'an contenidos en la ecuaci\'on siguiente

\begin{equation}\label{eq:(1,1)}
\delta A^{kc}_{{\rm SB},{\bf 1}} = c_{1}^{1,{\bf 1}}\delta^{c{\bf 8}}J^k + c_{3}^{1,{\bf 1}} \delta^{c {\bf 8}} \left\{ J^{2}, J^k \right\}
\end{equation}
donde los super{\'\i}ndices adjuntos a los coeficientes $c_i^{1,{\bf 1}}$ indican la representaci\'on. T\'erminos de orden m\'as alto pueden obtenerse
anticonmutando los operadores con $J^2 /N_c^2$. La contribuci\'on de la Ec. (\ref{eq:(1,1)}) s\'olo es importante para el operador de momento magn\'etico \cite{RMFPRD80(2009), Jen2012}.

\subsection{Contribuci\'on de sabor octete} 
La expansi\'on $1/N_c$ para el operador $(1, {\bf 8})$ tiene una estructura similar como $A^{kc}$ en la Ec. (\ref{Akc3}). As{\'\i} la correcci\'on de la rotura $(1, {\bf 8})$ es la siguiente,
\begin{equation}\label{eq:(1,8)}
\delta A^{kc}_{{\rm{SB}},{\bf 8}} = c_{1}^{1,{\bf 8}}d^{ce{\bf 8}}G^{ke} + b_{2}^{1,{\bf 8}} \frac{1}{N_c} d^{ce{\bf 8}}{\cal D}_2^{ke} + b_{3}^{1,{\bf 8}} \frac{1}{N^2_c} d^{ce{\bf 8}}{\cal D}_3^{ke} + c_{3}^{1,{\bf 8}} \frac{1}{N^2_c} d^{ce{\bf 8}}{\cal O}_3^{ke}.
\end{equation}
Las reglas de inversi\'on temporal proh{\'\i}ben el reemplazo de los s{\'\i}mbolos $d$ por $f$.  Existe otra serie para el operador $(1, {\bf 8})$; este inicia con el t\'ermino
\begin{equation}\label{eq:c2(1,8)}
  c_2^{1, {\bf 8}} \frac{1}{N_c} f^{ce8} \epsilon^{ijk} \left\{J^i, G^{je}\right\},
\end{equation}
y t\'erminos de orden m\'as alto pueden ser construidos anticonmutando el operador principal con $J^2/N_c^2$, podemos observar que
\begin{equation}\label{eq:fepsilon}
f^{ce8}e^{ijk}\left\{J^i, G^{je}\right\} = [J^2, [T^8,G^{kc}]].
\end{equation}
El lado derecho de la Ec. (\ref{eq:fepsilon}) muestra que el operador s\'olo contribuye a procesos con cambios de esp{\'\i}n y extrañeza.
Estos procesos no han sido observados, por ello, la serie (\ref{eq:c2(1,8)}) ser\'a excluida.

\subsection{Contribuci\'on de sabor ${\bf 10}+\overline{\bf 10}$}
Hasta orden relativo $1/N_c^2$, las series para el operador $(1, {\bf 10} + \overline{\bf 10})$ contienen operadores a dos-cuerpos y tres-cuerpos, esto es,
\begin{equation}
   \left\{ G^{kc}, T^8 \right\} - \left\{ G^{k8}, T^c \right\}, \label{eq:(10+10)a}\\
\end{equation}
\begin{equation}
   \left\{ G^{kc}, \left\{ J^r, G^{r8} \right\}  \right\} - \left\{ G^{k8}, \left\{ J^r, G^{rc} \right\}\right\}, \label{eq:(10+10)b}
\end{equation}
los cuales requieren substracciones de los operadores sabor octete \cite{DashenPRD51}. Las series para los t\'erminos de rotura de simetr{\'\i}a $(1, {\bf 10} + {\bf 10})$ podemos escribirlos como

\begin{eqnarray}
  &&\delta A^{kc}_{{\rm SB},{\bf 10} + {\overline{\bf 10}}} = c_2^{1,{\bf 10} + \overline{\bf 10}} \frac{1}{N_c} \biggl( \left\{G^{kc}, T^8\right\} - \left\{G^{k8}, T^c \right\} - \frac{1}{3} f^{ce8} f^{egh} \bigl( \left\{G^{kg}, T^h\right\} \nonumber \\
  && - \left\{G^{kh}, T^g \right\}\bigr) \biggr)
+ c_3^{1,{\bf 10}+\overline{{\bf 10}}} \frac{1}{N_c^2} \biggl( \left\{G^{kc}, \left\{J^r, G^{r8}\right\}\right\} - \left\{G^{k8}, \left\{J^{r}, G^{rc}\right\}\right\} \nonumber \\
   &&- \frac{1}{3} f^{ce8} f^{egh} \bigl( \left\{G^{kg}, \left\{J^r, G^{rh}\right\}\right\} - \left\{G^{kh}, \left\{J^r, G^{rg}\right\}\right\}  \bigl) \biggr). \label{eq:(1,10)}
\end{eqnarray}
Reducciones adicionales implican que
\begin{equation}\label{eq:10+10}
  \frac{1}{3}f^{ce8}f^{egh} \left( \left\{G^{kg}, T^h\right\} - \left\{G^{kh}, T^g\right\}\right) = \frac{2}{3}\left[J^2 , \left[T^8, G^{kc} \right] \right],
\end{equation}
y
\begin{eqnarray}\label{eq:10}
  &&\frac{1}{3}f^{ce8}f^{egh} \left( \left\{G^{kg}, \left\{J^r, G^{rh}\right\}\right\} - \left\{G^{kh}, \left\{J^r, G^{rg}\right\}\right\}\right) = \frac{1}{3}(N_c + N_f)\left[J^2 , \left[T^8, G^{kc} \right] \right],\nonumber\\
\end{eqnarray}
as{\'\i} los t\'erminos substra{\'\i}dos en Ec. (\ref{eq:(1,10)}) pueden ser absorbidos en las series ya existentes para $(1,{\bf 1})$ y $(1, {\bf 8})$, Eqs. (\ref{eq:(1,1)}) y (\ref{eq:(1,8)}) respectivamente.

\subsection{Contribuci\'on de sabor 27} 

Finalmente a orden relativo $1/N_c^2$, las series para el operador $(1, {\bf 27})$ contiene tres t\'erminos: un operador de dos-cuerpos y dos operadores de tres-cuerpos, los cuales son los siguientes,
\begin{equation}
\left\{G^{kc}, T^8\right\} + \left\{G^{k8}, T^c\right\}, \label{eq:27a}
\end{equation}
\begin{equation}
\left\{J^k, \left\{T^{c}, T^8\right\}\right\}  \hspace{0.5cm}\mbox{y} \hspace{0.5cm}
\left\{G^{kc}, \left\{J^r, G^{r8} \right\}\right\} + \left\{G^{k8}, \left\{J^r, G^{rc}\right\} \right\} \label{eq:27c}.
\end{equation}
Estos operadores requieren substracciones de las partes de sabor singulete y sabor octete \cite{DashenPRD51}. La serie de rotura de simetr{\'\i}a $(1,{\bf 27})$ es la siguiente,
\begin{eqnarray}
  &&\delta A^{kc}_{{\rm SB},{\bf 27}} = c_2^{1,{\bf 27}} \frac{1}{N_c}
                                  \biggl( \left\{G^{kc},T^8 \right\} + \left\{G^{k8},T^c\right\} - \frac{2}{N_f^2 - 1}\delta^{c8} \left\{G^{ke},T^e \right\} \nonumber\\
                               &&-\,\frac{2N_f}{N_f^2-4}d^{ce8}d^{egh} \left\{G^{kg},T^h\right\}\biggr)
                                  + c_3^{1,{\bf 27}} \frac{1}{N^2_c} \biggl( \left\{J^{k},\left\{T^c,T^8\right\} \right\}
                                   \nonumber\\
                               &&-\, \frac{1}{N_f^2 - 1}\delta^{c8} \left\{J^{k},\left\{T^e,T^e\right\} \right\}    -\,\,\frac{N_f}{N_f^2 - 4}d^{ce8}d^{egh}\left\{J^{k},\left\{T^g,T^h\right\}\right\}\biggr)  \nonumber\\
                               &&+\,\bar{c}_3^{1,{\bf 27}}\frac{1}{N_c^2} \Biggl(\left\{G^{kc},\left\{J^r,G^{r8}\right\} \right\} + \left\{G^{k8},\left\{J^r,G^{rc}\right\}\right\} - \frac{2}{N_f^2 - 1}\delta^{c8}\left\{G^{ke},\left\{J^r,G^{re}\right\}\right\}\nonumber \\
                               &&-\, \frac{2N_f}{N_f^2 - 4}d^{ce8}d^{egh}\left\{G^{kg},\left\{J^r, G^{rh}\right\}\right\}\Biggr).\label{eq:D27}
\end{eqnarray}
Nuevamente, reducciones adicionales dan las expresiones siguientes,
\begin{eqnarray}
&& \hspace{-3cm}\frac{2}{N_f^2 - 1}\delta^{c8}\left\{G^{ke},T^e\right\} + \frac{2N_f}{N_f^2 -4}d^{ce8}d^{egh}\left\{G^{kg},T^h\right\}= \nonumber\\
&& \frac{2(N_c+N_f)}{N_f+2}d^{c8e}G^{ke} +\,\,\frac{2(N_c + N_f)}{N_f(N_f + 1)}\delta^{c8}J^k +\,\, \frac{2}{N_f + 2}d^{c8e}{\cal D}_2^{ke},
\end{eqnarray}
\begin{eqnarray}
  &&\hspace{-1cm} \frac{1}{N_f^2 - 1}\delta^{c8}\left\{ J^k, \left\{ T^e, T^e \right\}\right\}+\frac{N_f}{N_f^2 - 4}d^{ce8}d^{egh}\left\{J^k,\left\{ T^g,T^h\right\}\right\}=\nonumber\\
  &&\hspace{1cm}\frac{N_ c(N_c + 2N_f)(N_f - 2)}{N_f(N_f^2 -1)}\delta^{c8}J^k + \frac{2(N_c + N_f)(N_f -4)}{N_f^2 -4}d^{c8e}{\cal D}_2^{ke}+\frac{2N_f}{N_f^2 -4}d^{c8e}{\cal D}_3^{ke} \nonumber \\
  &&\hspace{1cm} + \frac{2}{N_f^2 - 1}\delta^{c8}\left\{J^2,J^k\right\},
\end{eqnarray}
y
\begin{eqnarray}
&&\hspace{-1.5cm}\frac{2}{N^2_f-1}\delta^{c8}\left\{G^{ke},\left\{J^r, G^{re}\right\}\right\} +
\frac{2N_f}{N^2_f-4}d^{ce8}d^{egh}\left\{G^{kg},\left\{J^r, G^{rh}\right\}\right\}=
\frac{2N_f}{N_f+2}d^{c8e}G^{ke} \nonumber\\
&& +\,\,\frac{(N_c+2)(N_c+2N_f-2)}{2(N^2_f-1)}\delta^{c8}J^k + \frac{N_f(N_c + N_f)}{N_f^2-4}d^{c8e}{\cal D}_2^{ke} + \frac{N_f-4}{N_f^2-4}d^{c8e}{\cal D}_3^{ke}\nonumber\\
&& +\,\,\frac{2}{N_f+2}d^{c8e}{\cal O}_3^{ke}+\frac{N_f-2}{N_f(N_f^2-1)}\delta^{c8}\left\{J^2,J^k\right\},
\end{eqnarray}
como se esperaba, la substracci\'on de las contribuciones de sabor singulete y sabor octete en la expansi\'on $1/N_c$ Ec. (\ref{eq:D27}), contienen operadores ya definidos en las series (\ref{eq:(1,1)}) y (\ref{eq:(1,8)}) de este modo, en la expansi\'on $1/N_c$ (\ref{eq:D27}) s\'olo mantenemos los t\'erminos principales (\ref{eq:27a} - \ref{eq:27c}).

\section{Correcci\'on total de la corriente axial vector}
Las correcciones a la corriente axial vector para bariones se pueden obtener de manera similar, a orden de un loop y con rotura de simetr{\'\i}a SU(3)
perturbativo. Las correcciones a un loop, $\delta A_{1\rm L}^{kc}$, surge de los diagramas Fig. 2(a-d) como se muestra en la Ec. (\ref{eq:AkcTotal}). Por otro lado, la correcci\'on con rotura de simetr{\'\i}a SU(3) perturbativa $\delta A^{kc}_{\rm SB}$ viene de la Ec.(\ref{eq:delta8y10}). De esta manera, obtenemos la expresi\'on para la correcci\'on total a la corriente axial de bariones dada de la siguiente forma
\begin{equation}\label{eq:AkcSB}
A^{kc} + \delta A^{kc} = A^{kc} + \delta A_{1\rm L}^{kc} + \delta A^{kc}_{\rm SB}.
\end{equation}
Los elementos de matriz de las componentes espaciales de $A^{kc} + \delta A^{kc}$ entre los estados de simetr{\'\i}a SU(6) dan los valores comunes de los acoplamientos axial vector. Para el octete de bariones, los acoplamientos axial vector son $g_A$. Tal como est\'an definidos en los experimentos, en decaimientos semilept\'onicos de bariones, $g_A \approx 1.27$ para el decaimiento de neutr\'on. Para los bariones del decuplete los acoplamientos axial vector son $g$, los cuales son extra{\'\i}dos de las anchuras de los decaimientos fuertes del decuplete de bariones hacia el octete de bariones y piones.


\chapter{Resultados Num\'ericos\label{c:ccuatro}}

\section{Aspectos te\'oricos de los decaimientos semi\-lep\-t\'o\-ni\-cos de bariones}

A manera de notaci\'on, definamos al decaimiento semilept\'onico
de bariones (DSB) como
\begin{equation}
B_1\to B_2 + \ell + \bar{\nu}_l, \label{eq:ec1}
\end{equation}
donde $B_1$ y $B_2$ son los bariones inicial y final, $\ell$ es el
lept\'on cargado y $\bar{\nu_l}$ su correspondiente antineutrino
(o neutrino, dependiendo del decaimiento).

\subsection{Propiedades del elemento de matriz de bariones \label{sec:ff}} Los
elementos de matriz de la corriente lept\'onica d\'ebil se
obtienen f\'acilmente para los diferentes operadores involucrados en el c\'alculo. La co\-rrien\-te
hadr\'onica d\'ebil, por otra parte, se construye a partir de las
cantidades independientes disponibles. Para DSB se dispone de tres
vectores polares: $\bar{u}_{B_2}{\gamma_\mu} u_{B_1}$,
$\bar{u}_{B_2} \sigma_{\mu \nu} q_\nu u_{B_1}$ y $\bar{u}_{B_2}
q_\mu u_{B_1}$; y de tres vectores axiales:
$\bar{u}_{B_2}{\gamma_\mu} \gamma_5 u_{B_1}$,
$\bar{u}_{B_2}{\sigma}_{\mu \nu} q_\nu\gamma_5 u_{B_1}$ y
$\bar{u}_{B_2} \gamma_5 q_\mu u_{B_1}$, siendo $q$ el 4-mo\-mento
transferido. Entonces la forma m\'as general de la corriente hadr\'onica
d\'ebil puede expresarse de la siguiente forma
\begin{eqnarray}
\langle B_2 |J_\mu^h|B_1 \rangle & = & V_{\rm CKM} \,
\bar{u}_{B_2} (p_2) \left[ f_1(q^2) \gamma_\mu +
\frac{f_2(q^2)}{M_1}
\sigma_{\mu\nu}q^\nu + \frac{f_3(q^2)}{M_1} q_\mu \right. \nonumber \\
&   & \mbox{} + \left. \left(g_1(q^2) \gamma_\mu +
\frac{g_2(q^2)}{M_1} \sigma_{\mu\nu}q^\nu + \frac{g_3(q^2)}{M_1}
q_\mu \right) \gamma_5 \right] u_{B_1}(p_1),
\label{eq:bme}
\end{eqnarray}
donde $u_{B_1}$, $p_1$, $M_1$ $[\bar{u}_{B_2}, p_2, M_2]$, son los
espinores de Dirac, el 4-momento y la masa del bari\'on inicial
[final] y $q=p_1-p_2$. En este trabajo adoptamos la convenci\'on
de matrices $\gamma$ utilizada en la Ref. \cite{AGarcia}.

Cada modo de decaimiento semilept\'onico depende de seis funciones
independientes entre s{\'\i}, los llamados {\it factores de
forma}. \'Estos son $f_1$, $f_2$, $f_3$, $g_1$, $g_2$, $g_3$, los
cuales son funciones del invariante $q^2$ y son a su vez escalares
de Lorentz. $f_1(q^2)$ y $g_1(q^2)$ son los factores de forma
vector y axial vector l{\'\i}deres, $f_2(q^2)$ y $g_2(q^2)$
son los factores de forma magn\'etico d\'ebil y el\'ectrico, y
$f_3(q^2)$, $g_3(q^2)$ son los factores de forma inducido escalar
y pseudoescalar, respectivamente. La invariancia bajo inversi\'on
temporal requiere que los factores de forma sean reales. El factor
$V_{\rm CKM}$ se extrae de los elementos de matriz
como un preludio a la referencia del \'angulo de Cabibbo. $f_3(q^2)$ y $g_3(q^2)$ para
modos electr\'onicos tienen una contribuci\'on despreciable sobre la
raz\'on de decaimiento debido a la peque\~nez del factor
$(m/M_1)^2$ que los acompa\~na; por tanto, a un alto grado de
precisi\'on, los modos electr\'onicos de DSB se describen en
t\'erminos de cuatro y no seis factores de forma. En contraste,
para modos mu\'onicos aunque el factor $(m/M_1)^2$ sea peque\~no,
$f_3(q^2)$ y $g_3(q^2)$ pueden contribuir significativamente a la
raz\'on de transici\'on y por tanto estos t\'erminos deben ser
retenidos. Por conveniencia se introduce la definici\'on
$f_i\equiv f_i(0)$ y $g_i\equiv g_i(0)$ con $i= 1,2,3$.

La descripci\'on te\'orica de los DSB est\'a basada fuertemente en
el Modelo de Cabibbo \cite{cabmodel}. Entre otros aspectos, el
modelo predice que todos los factores de forma ser\'an una
funci\'on de solamente tres par\'ametros independientes no
predichos por la teor{\'\i}a. Para tener una mejor claridad,
revisemos brevemente los postulados
de Cabibbo.

{\bf Postulado 1}. Los componentes de la corriente hadr\'onica
d\'ebil pertenecen a una representaci\'on autoconjugada de SU(3)
simple. Los componentes de la corriente son los miembros cargados
de un octete de SU(3) y los n\'umeros cu\'anticos de isosp{\'\i}n
y extra\~neza portados por la corriente hadr\'onica d\'ebil
corresponden a aquellos de los mesones cargados del octete
$J^{P}=0^-$:
\[
\begin{array}{*{20}c}
   {\Delta {\sf S} = \;\;\;0} & {\left\{ \begin{array}{l}
 \Delta {\sf Q} = \Delta {\sf I}_3  =  + 1 \\
 \Delta {\sf Q} = \Delta {\sf I}_3  =  - 1 \\
 \end{array} \right.} & \begin{array}{l}
 \left| {\Delta {\sf I}} \right| = 1 \\
 \left| {\Delta {\sf I}} \right| = 1 \\
 \end{array} & \begin{array}{l}
 \pi^+   \\
 \pi^-   \\
 \end{array}  \\
   {\Delta {\sf S} =  + 1} & {\Delta {\sf S} = \Delta {\sf Q} =  + 1} & {\left| {\Delta {\sf I}} \right| = {\raise0.7ex\hbox{$1$}
   \!\mathord{\left/
 {\vphantom {1 2}}\right.\kern-\nulldelimiterspace}
\!\lower0.7ex\hbox{$2$}}} & {K^ +  }  \\
   {\Delta {\sf S} =  - 1} & {\Delta {\sf S} = \Delta {\sf Q} =  - 1} & {\left| {\Delta {\sf I}} \right| = {\raise0.7ex\hbox{$1$}
   \!\mathord{\left/
 {\vphantom {1 2}}\right.\kern-\nulldelimiterspace}
\!\lower0.7ex\hbox{$2$}}} & {K^ -  }  \\
\end{array}
\]
La autoconjugaci\'on implica que las corrientes con $\Delta {\sf
Q}=1$ y $\Delta {\sf Q}=-1$ pertenecen al mismo octete.

Los estados bari\'onicos $J^{P}=\frac{1}{2}^+$ se identifican en
t\'erminos de los componentes del octete SU(3). Los elementos de
matriz de un operador --por s{\'\i} mismo perteneciente a un
octete-- entre dos estados de octete es una combinaci\'on lineal
de dos elementos de matriz reducidos, a causa de los octetes
sim\'etricos y antisim\'etricos que aparecen en el producto
directo de dos octetes en SU(3) $[{\bf 8} \times {\bf 8} ={\bf
1}\oplus{\bf 8}_a\oplus {\bf 8}_b \oplus {\bf
10}\oplus\overline{{\bf 10}}\oplus {\bf 27}]$.

Espec{\'\i}ficamente, cualquiera de los seis factores de forma
$f_{m}(q^2)$, $g_{m}(q^2)$ mencionados anteriormente est\'a dado
por
\begin{eqnarray}
\left\{ {\begin{array}{*{20}l}
{f_m(q^2)  = r_1F_m (q^2 ) + r_2D_m (q^2 )} \\[3mm]
{g_m(q^2)  = r_1F_{m + 3} (q^2 ) + r_2D_{m + 3} (q^2 )}, \\
\end{array}} \right. \qquad \quad (m =1,2,3) \label{eq:ffr}
\end{eqnarray}
donde $F_i(q^2)$ y $D_i(q^2)$, $i=1,\ldots,6$, son funciones de
$q^2$, diferentes para cada uno de los seis factores de forma. Las
constantes $r_1$ y $r_2$ en la (\ref{eq:ffr}) son los coeficientes
generalizados de Glebsch-Gordan, cuyos valores est\'an dados en la
Tabla~\ref{table:cg} para varias transiciones.

\begin{table}[ht]
\begin{center}
\begin{tabular}{|l|c|c|}
\hline
Transici\'on \quad & \quad $r_1$ \quad & \quad $r_2$ \quad \\[0.9mm]
\hline
&&\\[-2.8mm]
$n\to p$ & $1$ & $1$ \\[0.6mm]
$\Sigma^\pm\to \Lambda$ & $0$ & $\sqrt{\frac{2}{3}}$ \\[0.6mm]
$\Sigma^-\to \Sigma^0$ & $\sqrt{2}$ & $0$ \\[0.6mm]
$\Xi^-\to \Xi^0$ & $-1$ & $1$ \\[0.6mm]
$\Lambda\to p$ & $-\sqrt{\frac{3}{2}}$ & $-\sqrt{\frac{3}{2}}$ \\[0.6mm]
$\Sigma^-\to n$ & $-1$ & $1$ \\[0.6mm]
$\Xi^-\to \Lambda$ & $\sqrt{\frac{3}{2}}$ & $-\sqrt{\frac{3}{2}}$ \\[0.6mm]
$\Xi^-\to \Sigma^0$ & $\frac{1}{\sqrt{2}}$ & $\frac{1}{\sqrt{2}}$ \\[0.6mm]
$\Xi^0\to \Sigma^+$ & $1$ & $1$ \\[0.6mm]
\hline
\end{tabular}
\end{center}
\caption{Coeficientes de Clebsch-Gordan en DSB.\label{table:cg}}
\end{table}

El primer postulado de la teor{\'\i}a de Cabibbo, al hacer de cada
factor de forma una combinaci\'on lineal de dos funciones en $q^2$
(independientes entre s{\'\i}), ha reducido el n\'umero de
funciones independientes a doce. La Ec.~(\ref{eq:ffr}) se basa en
una simetr{\'\i}a SU(3) exacta, y posteriormente discutiremos las
consecuencias de la rotura de esta simetr{\'\i}a. Como la
separaci\'on de masa para los bariones es peque\~na ($\sim 15\%$)
comparada con la de mesones, uno puede esperar que las
predicciones de SU(3) exacta se cumplan mejor en el caso de
bariones.

{\bf Postulado 2}. Universalidad. La corriente lept\'onica est\'a acoplada a una sola co\-rrien\-te hadr\'onica simple de longitud
unitaria:
\begin{equation}
{J^h}_\mu= \cos\theta_C{J^h}_\mu(\Delta {\sf S}=0)+\sin\theta_C{J^h}_\mu(|\Delta {\sf S}|=1)
\end{equation}
con la constante de acoplamiento $G=G_\mu$. La constante $V_{\rm
CKM}$ de la Ec.~(\ref{eq:bme}) es igual a $\cos\theta_C$ para
transiciones $\Delta {\sf S}=0$ y $\sin \theta_C$ para
transiciones $|\Delta {\sf S}|=1$.

{\bf Postulado 3}. Hip\'otesis generalizada de la corriente
vectorial conservada (CVC). La hip\'otesis CVC establece que la
parte vectorial ${J_\mu}^\pm$ de la corriente d\'ebil $\Delta {\sf
S}=0$ es una corriente conservada como la corriente
electromagn\'etica. Adem\'as CVC implica que ${J_\mu}^\pm$ forma
un isotriplete con la parte isovector de la corriente
electromagn\'etica. CVC ge\-ne\-ra\-li\-za\-da supone que la parte
vectorial de las corrientes d\'ebiles $\Delta {\sf S}=0$ y
$|\Delta {\sf S}|=1$ y de la corriente electromagn\'etica son
miembros del mismo octete de SU(3). Los factores de forma
electromagn\'eticos y los factores de forma vectoriales d\'ebiles
est\'an por tanto directamente relacionados. Observamos que,
hist\'oricamente, la hip\'otesis CVC fue un paso importante hacia
la unificaci\'on de las fuerzas electromagn\'eticas y d\'ebiles.

Los factores de forma electromagn\'eticos del prot\'on y el
neutr\'on est\'an dados por las relaciones
\begin{equation}
f^{p}_{m}(q^2) = F_{m}(q^2) + \frac{1}{3}D_{m}(q^2), \qquad \qquad
f^{n}_{m}(q^2) = -\frac{2}{3}D_{m}(q^2).
\end{equation}
Entonces las seis funciones $F_{m}(q^2)$ y $D_{m}(q^2)$ para los
factores de forma vectoriales en la Ec.~(\ref{eq:ffr}) est\'an
completamente determinadas por las expresiones
\begin{equation}
F_{m}(q^2) = f^{p}_{m}(q^2)+\frac{1}{2}f^{n}_{m}(q^2), \qquad
\qquad D_{m}(q^2)=-\frac{2}{3}f^{n}_{m}(q^2).
\end{equation}
Para $q^2=0$, el factor de forma electromagn\'etico $f_1(0)$ es
igual a la carga el\'ectrica del bari\'on; por consiguiente
$F_1(0)=1$ y $D_1(0)=0$.

El factor de forma electromagn\'etico $f_2(q^2)$ est\'a
relacionado con el momento magn\'etico an\'omalo de los nucleones
$\mu$. Definiendo $f_2^\prime(q^2)=[(M_1+M_2)/M_1]f_2(q^2)$,
tenemos ${f^\prime}^{p,n}(0)=\mu_{p,n}$, lo cual conduce a
$F_2^\prime(0)=\mu_p+\frac{1}{2}\mu_n$ y
$D_2^\prime(0)=-\frac{3}{2}\mu_{p,n}$. El factor de forma $f_2$ es
frecuentemente llamado factor de forma magn\'etico d\'ebil.

Finalmente la conservaci\'on de la corriente electromagn\'etica
requiere que $f^{p,n}_3(q^2)$ sea igual a cero, implicando que
$F_3(q^2)=D_3(q^2)=0$. Entonces el factor de forma $f_3(q^2)$ es
nulo para todos los decaimientos semilept\'onicos.

Dentro de este esquema, los factores de forma vectoriales para DSB
est\'an determinados por la hip\'otesis de CVC generalizada y
est\'an relacionados, en $q^2=0$, con la carga el\'ectrica y el
momento magn\'etico an\'omalo de los nucleones. En el l{\'\i}mite
de simetr{\'\i}a exacta SU(3), la dependencia en $q^2$ de los
factores de forma est\'a tambi\'en dada por la dependencia de
$q^2$ de los factores de forma electromagn\'eticos de los
nucleones.

{\bf Postulado 4}. Ausencia de corrientes de segunda clase. Aun
queda, a este nivel, la determinaci\'on de los tres factores de
forma axial vectoriales $g_1$, $g_2$ y $g_3$, cada uno de ellos
siendo una combinaci\'on lineal de dos funciones desconocidas para
los distintos decaimientos. Para reducir aun m\'as el n\'umero de
factores de forma, se necesita una hip\'otesis suplementaria.

Consideremos las propiedades de las operaciones combinadas de
conjugaci\'on de carga $C$ y rotaci\'on por $180^{\circ}$
alrededor del eje $I_2$ del espacio de isosp{\'\i}n; esta
operaci\'on combinada es llamada {\bf $G$-paridad} ($G_\lambda$). Cuando se aplica a
la matriz de transici\'on  del decaimiento beta del neutr\'on,
\'esta intercambia el prot\'on y el neutr\'on dos veces y regresa
a la matriz inicial, hasta un posible cambio de signo. Bajo
$G$-paridad, las corrientes d\'ebiles vectorial $(f_1)$ y axial
$(g_1)$ se transforman como $G_\lambda$ Ec. (\ref{eq:fcl})
\begin{eqnarray}
\begin{array}{l}
V_\mu\to V_\mu, \\[0.6mm]
 A_\mu\to -A_\mu.
\end{array} \label{eq:fcl}
\end{eqnarray}
Las corrientes que se transforman como la Ec.~(\ref{eq:fcl}) son
llamadas {\it corrientes de primera clase}; aquellas se que
transforman con un signo opuesto son llamadas {\it corrientes de
segunda clase}. Los t\'erminos $f_1$, $f_2$, $g_1$ y $g_3$
corresponden a corrientes de primera clase, mientras que los
t\'erminos $f_3$ y $g_2$ son de segunda clase. Dado que las
interacciones fuertes son invariantes bajo $C$ y transformaciones
en el espacio de isosp{\'\i}n, las propiedades de las corrientes
anteriormente mencionadas son inalterables debido a los efectos de
la interacci\'on fuerte.

Notemos que a nivel de quarks hay solamente corrientes de primera
clase para las transiciones $d\to u$, ocurriendo en el decaimiento
beta del neutr\'on, y que el t\'ermino $f_3$ ya ha sido eliminado
por la hip\'otesis CVC generalizada.

El mismo razonamiento aplica al doblete de isosp{\'\i}n
$(\Xi^-,\Xi^0)$ para el decaimiento $\Xi^-\to \Xi^0\l\bar{\nu}_l$.
La ausencia de corrientes de segunda clase conduce a la Ec. (\ref{eq:g2np})
\begin{equation}\label{eq:g2np}
g_2^{n,p} = F_5(q^2)+D_5(q^2)=0, \quad \qquad
g_2^{\Xi^-,\Xi^0}=D_5(q^2)-F_5(q^2)=0,
\end{equation}
as{\'\i} que
\begin{equation}
F_5(q^2)=D_5(q^2)=0.
\end{equation}
Entonces todos los factores de forma pseudotensoriales $g_2$ son
nulos en todos los decaimientos, hasta efectos de rotura de
simetr{\'\i}a. En lenguaje moderno, se establece que el factor de
forma $g_2$ para elementos de matriz diagonales de corrientes
herm{\'\i}ticas (por ejemplo $\langle B|\overline{u}\gamma^\mu
\gamma_5 u-\overline{d}\gamma^\mu\gamma_5 d|B\rangle$) se anula
por hermiticidad e invariancia bajo inversi\'on temporal. Por
tanto, SU(3) implica que $g_2=0$ en el  l{\'\i}mite de
simetr{\'\i}a.

Adicionalmente, en la expresi\'on para la raz\'on de decaimiento
diferencial, todos los t\'erminos que involucran a los factores de
forma $f_3$ y $g_3$ est\'an multiplicados por un factor
$(m/M_1)^2$. Como los experimentos precisos han sido desarrollados
solamente sobre los decaimientos electr\'onicos de bariones,
$g_3(q^2)$ puede ser ignorado.

De esta forma, el l{\'\i}mite de simetr{\'\i}a  SU(3) exacta
establece que, para modos electr\'onicos,  s\'olo cuatro factores
de forma en DSB se requieren para determinar los elementos de la
matriz de transici\'on: $f_1(q^2)$ y $f_2(q^2)$ est\'an bien
determinados, mientras que $g_1(q^2)$ son una combinaci\'on lineal
de dos funciones desconocidas $F(q^2)=F_{4}(q^2)$ y
$D(q^2)=D_{4}(q^2)$ y $g_2(q^2)=0$. Los valores en $q^2=0$ de los
factores de forma para todos los decaimientos semilept\'onicos de
bariones est\'an dados en la Tabla~\ref{t:ffsu3}.

\begin{table}[h]
\begin{center}
\begin{tabular}{|c|c|c|c|c|}
\hline
&&&& \\[-3.0mm]
Transici\'on & $V_{\rm CKM}$ & $f_1(0)$ & $f_2(0)$ & $g_1(0)$ \\[2mm]
\hline
$n\to pe^-\bar{\nu}_e$ & $V_{ud}$ & $1$ & $\mu_p-\mu_n$ & $F+D$ \\[2mm]
$\Sigma^\pm\to \Lambda e^\pm\nu_e$ & $V_{ud}$ & $0$ & $-\sqrt{3/2}\mu_n$ & $\sqrt{2/3}D$ \\[2mm]
$\Sigma^-\to \Sigma^0e^-\bar{\nu}_e$ & $V_{ud}$ & $\sqrt{2}$ & $\sqrt{2}[\mu_p+(\mu_n/2)]$ & $\sqrt{2}F$ \\[2mm]
$\Lambda\to pl^-\bar{\nu}_l$ & $V_{us}$ & $-\sqrt{3/2}$ & $-\sqrt{3/2}\mu_p$ & $-\sqrt{3/2}(F+D/3)$ \\[2mm]
$\Sigma^-\to nl^-\bar{\nu}_l$ & $V_{us}$ & $-1$ & $-(\mu_p+2\mu_n)$ & $-(F-D$ \\[2mm]
$\Xi^-\to \Lambda l^-\bar{\nu}_l$ & $V_{us}$ & $\sqrt{3/2}$ & $\sqrt{3/2}(\mu_p+\mu_n)$ & $\sqrt{3/2}(F-D/3)$ \\[2mm]
$\Xi^-\to \Sigma^0l^-\bar{\nu}_l$ & $V_{us}$ & $\frac{1}{\sqrt{2}}$ & $\frac{1}{\sqrt{2}}(\mu_p-\mu_n)$ & $(F+D)/\sqrt{2}$ \\[2mm]
$\Xi^0\to \Sigma^+l^-\bar{\nu}_l$ & $V_{us}$ & $1$ & $(\mu_p-\mu_n)$ & $F+D$ \\[2mm]
$\Xi^-\to \Xi^0l^-\bar{\nu}_l$ & $V_{ud}$ & $1$ & $(\mu_p+2\mu_n)$ & $F-D$ \\[2mm]
\hline
\end{tabular}
\end{center}
\caption{Par\'ametros de la matriz d\'ebil de
bariones.\label{t:ffsu3}}
\end{table}
Todos los decaimientos est\'an por tanto descritos por tres
par\'ametros: el \'angulo de Cabibbo $\theta_C$ y dos constantes
de acoplamiento $F$ y $D$. En lo sucesivo tomaremos la
convenci\'on de que el signo de $g_1/f_1=F+D$ es positivo para el
decaimiento del neutr\'on, lo cual fija todos los otros signos. En
algunas otras referencias (por ejemplo, en el Particle Data Group
\cite{PDG}) se toma la convenci\'on opuesta. La determinaci\'on
del \'angulo de Cabibbo requiere la medici\'on de razones de
decaimiento, mientras que $F$ y $D$ pueden ser determinados
tambi\'en estudiando la distribuci\'on de algunas variables
cinem\'aticas.

Los resultados dados en la Tabla \ref{t:ffsu3} son v\'alidos en
$q^2=0$ e ignoran los efectos de rotura de la simetr{\'\i}a
de sabor SU(3). Dicha rotura puede introducir modificaciones
notables a dichos valores.

\section{Raz\'on diferencial de decaimiento \label{sec:razon}}
La amplitud de transici\'on para procesos semilept\'onicos de
bariones puede ser cons\-tru{\'\i}\-da a partir del producto de
los elementos de matriz de las corrientes hadr\'onica y
lept\'onica \cite{AGarcia}. De esta amplitud, la raz\'on
diferencial de decaimiento en DSB, denotada aqu{\'\i} por
$d\Gamma$, puede ser derivada usando t\'ecnicas est\'andares. Para
el decaimiento en tres cuerpos Ec.~(\ref{eq:ec1}), diferentes
elecciones de las cinco variables relevantes en el estado final
conducen a expresiones apropiadas para $d\Gamma$. En la
Ref.~\cite{AGarcia} se presentan expresiones detalladas para
$d\Gamma$ en el sistema en reposo de $A$ $(B)$ cuando dicho
bari\'on est\'a polarizado a lo largo de la direcci\'on $s_1$
$(s_2)$, con el lept\'on cargado y el neutrino emitidos dentro de
los \'angulos s\'olidos $d\Omega_{\ell}$ y $d\Omega_\nu$,
respectivamente. Similarmente en las
Refs.~\cite{RFM1997,PRD63(2001)} $d\Gamma$ ha sido obtenida, en el
sistema de reposo de $A$, dejando a las energ{\'\i}as del
electr\'on y bari\'on final como las variables relevantes, junto
con algunas variables angulares.

En todos los casos anteriores la raz\'on de decaimiento
diferencial puede escribirse, en la forma m\'as general como
\begin{equation}
d\Gamma = G^2 d\Phi_3 \left[A_0^\prime - A_0^{\prime \prime} \,
{\hat {\mathbf s}} \cdot {\hat {\mathbf p}} \right],
\end{equation}
donde $d\Phi_3$ es un elemento del espacio fase apropiado de tres
cuerpos y $A_0^\prime$ and $A_0^{\prime \prime}$ dependen de las
variables cinem\'aticas y son funciones cuadr\'aticas de los
factores de forma. El producto escalar ${\hat {\mathbf s}} \cdot
{\hat {\mathbf p}}$, donde ${\hat {\mathbf s}}$ denota el
esp{\'\i}n ya sea de $B_1$ o de $B_2$ y ${\hat {\mathbf p}}= {\hat
{\mathbf l}}, {\hat {\mathbf p}_2}, {\hat {\mathbf p}_\nu}$,
representa la correlaci\'on angular entre dicho esp{\'\i}n y el
momento de la part{\'\i}cula correspondiente
\cite{RFM1997,PRD63(2001)}.

\section{Observables integrados\label{sec:obs}}

Cuando los experimentos en DSB tienen baja estad{\'\i}stica en
general no es posible de\-sa\-rro\-llar un an\'alisis detallado de la
raz\'on de decaimiento diferencial $d\Gamma$. Ante esto es
necesario producir observables integrados; entre \'estos se
encuentran la raz\'on de decaimiento total $R$ y las correlaciones
angulares y los coeficientes de asimetr{\'\i}a. La definici\'on de
estos observables implica solamente cinem\'atica y no supone
alg\'un modelo te\'orico en particular. Por ejemplo, el
coeficiente de correlaci\'on angular del lept\'on cargado-neutrino
est\'a definido como
\begin{equation}
\alpha_{\ell\nu} = 2 \frac{N(\Theta_{\ell\nu} < \pi/2) -
N(\Theta_{\ell\nu} > \pi/2)}{N(\Theta_{\ell\nu} < \pi/2) +
N(\Theta_{\ell\nu}> \pi/2)},
\end{equation}
donde $N(\Theta_{\ell\nu} < \pi/2)$ $[N(\Theta_{\ell\nu} >
\pi/2)]$ es el n\'umero de pares lept\'on cargado-neutrino
emitidos en direcciones que forman un \'angulo entre ellos menor
[mayor] que $\pi/2$. Expresiones similares pueden ser derivadas
para los coeficientes de asimetr{\'\i}a del lept\'on cargado
$\alpha_\ell$, del neutrino $\alpha_\nu$, y del bari\'on emitido
$\alpha_{B}$, siendo esta vez $\Theta_\ell$, $\Theta_\nu$ y
$\Theta_{B}$ los \'angulos entre las direcciones de $\ell$, $\nu$
y $B$ y la polarizaci\'on de $B_1$, respectivamente. Cuando la
polarizaci\'on del bari\'on emitido es observada, es posible
definir dos coeficientes de asimetr{\'\i}a $A$ y $B$
\cite{AGarcia}. Si la masa del lept\'on cargado puede ser
despreciada es bastante sencillo calcular expresiones te\'oricas
aproximadas de estos observables. Esto ha sido hecho en la Ref.\
\cite{AGarcia} para varios decaimientos. Para la raz\'on total de
decaimiento no corregida se tiene por ejemplo
\begin{eqnarray}
R^0 & = & G^2 \frac{(\Delta M)^5}{60\pi^3} \left[\left(1-\frac32
\beta + \frac67 \beta^2\right)f_1^2 + \frac47 \beta^2 f_2^2
+ \left(3 - \frac92 \beta + \frac{12}{7} \beta^2 \right) g_1^2 \right. \nonumber  \\
&  & \mbox{\hglue1.0truecm} + \left. \frac{12}{7} \beta^2 g_2^2 +
\frac67 \beta^2 f_1f_2 + (-4\beta+6\beta^2)g_1g_2 \right],
\label{eq:gam}
\end{eqnarray}
donde $\beta = (M_1-M_2)/M_1$ y el super{\'\i}ndice $0$ en un
observable dado es usado como un indicador de que no se han
incorporado correcciones radiativas en dicho observable. En la
Ec.~(\ref{eq:gam}) aunque los factores de forma se han supuesto
constantes, su dependencia en $q^2$ no siempre puede ser
despreciada ya que podr{\'\i}a dar una contribuci\'on
significativa. Para obtener expresiones corregidas a orden
${\mathcal O}(q^2)$, la dependencia de $q^2$ de $f_2$ y $ g_2$
puede ser ignorada debido a que ya contribuyen a orden ${\mathcal
O}(q)$ a la raz\'on de decaimiento. Para $f_1(q^2)$ y $g_1(q^2)$,
sin embargo, es suficiente una expansi\'on lineal en $q^2$ debido
a que potencias m\'as altas dan lugar a contribuciones
despreciables en la raz\'on de decaimiento, no mayores que una
fracci\'on porcentual. Entonces
\begin{equation}
f_1(q^2) = f_1(0) + \frac{q^2}{M_1^2} \lambda_1^f, \qquad \qquad
g_1(q^2) = g_1(0) + \frac{q^2}{M_1^2} \lambda_1^g,
\end{equation}
donde los par\'ametros de pendiente $\lambda_1^f$ y $\lambda_1^g$
son ambos de orden uno \cite{AGarcia}. Una parametrizaci\'on
dipolar para los factores de forma del tipo $f(q^2) =
f(0)/(1-q^2/M^2)^2$ conduce a
\begin{equation}
\lambda_1^f = \frac{2M_1^2 f_1}{M_V^2}, \qquad \qquad \lambda_1^g
= \frac{2M_1^2 g_1}{M_A^2}, \label{eq:slop}
\end{equation}
donde $M_V= 0.97$ GeV y $M_A=1.11$ GeV para procesos $|\Delta {\sf
S}|=1$ y $M_V= 0.84$ GeV y $M_A=0.96$ GeV para procesos $|\Delta
{\sf S}|=0$ \cite{AGarcia}. Para f\'ormulas m\'as precisas o
cuando la masa del lept\'on cargado no pueda despreciarse es
necesario integrar num\'ericamente sobre las variables
cinem\'aticas las expresiones para $d\Gamma$ y los coeficientes
angulares dados en trabajos previos
\cite{AGarcia,RFM1997,PRD63(2001)}. Respecto a esto, en la
referencia \cite{AGarcia} se proporcionan f\'ormulas num\'ericas
completas para la raz\'on de decaimiento y los coeficientes
angulares para 16 modos $e$ y 10 modos $\mu$ en DSB. Estas
f\'ormulas, sin embargo, tienen m\'as de 20 a\~nos de haber sido
publicadas y los datos experimentales recientes de las masas de
los bariones \cite{PDG} introducen modificaciones, las cuales
necesitan ser tomadas en cuenta. Una actualizaci\'on a estas
f\'ormulas se encuentra en las Refs.~\cite{tesisra, RFM2004}.

\section{Correcciones radiativas\label{sec:cr}}

Los experimentos en DSB se han vuelto suficientemente sensitivos
para requerir correcciones radiativas a los observables
integrados. Sin embargo, el c\'alculo de las correcciones
radiativas de procesos que involucran hadrones ha sido un problema
te\'orico abierto por muchos a\~nos. A pesar del notable progreso
logrado en el entendimiento de las interacciones fundamentales con
el Modelo Est\'andar \cite{PDG}, los c\'alculos a primeros
principios de correcciones radiativas no son posibles
todav{\'\i}a. Estas correcciones est\'an comprometidas con
dependencia de modelo y los an\'alisis experimentales que los
utilizan se vuelven a su vez dependientes de modelo. Incluso si la
dependencia de modelo que surge de las correcciones radiativas
virtuales no puede ser eliminada, un an\'alisis del decaimiento
beta del neutr\'on posteriormente extendido a DSB \cite{sirlin}
muestra que a \'ordenes $(\alpha/\pi)(q/M_1)^0$ y
$(\alpha/\pi)(q/M_1)$ dicha dependencia de modelo se puede
englobar en algunas constantes, las cuales pueden ser absorbidas
en los factores de forma originalmente definidos en el elemento de
matriz de la corriente hadr\'onica. Adicionalmente el teorema de
Low en la versi\'on de Chew \cite{low} puede ser usado para
mostrar que a esos dos \'ordenes de aproximaci\'on las
correcciones radiativas de bremsstrahlung dependen tanto de los
factores de forma no radiativos como de los multipolos
electromagn\'eticos est\'aticos de las part{\'\i}culas
involucradas en el decaimiento solamente, as{\'\i} que en este
sector no existe dependencia de modelo en las correcciones
radiativas. Por tanto a estos \'ordenes de aproximaci\'on
te\-ne\-mos \'unicamente expresiones generales que pueden ser
usados en an\'alisis que no est\'an comprometidos con alg\'un
modelo en particular \cite{AGarcia,RFM1997,PRD63(2001)}. Por
supuesto, a \'ordenes m\'as altos esto no necesariamente ser\'a
cierto y podr{\'\i}amos esperar una fuerte dependencia de modelo.

Las correcciones radiativas a orden $(\alpha/\pi)(q/M_1)^0$ a las
razones de transici\'on $R$ y a los coeficientes angulares y de
asimetr{\'\i}a en DSB referidos anteriormente, $\alpha_k$, han
sido calculadas en la Ref.~\cite{AGarcia}. En ese trabajo se
mostr\'o que a este orden de aproximaci\'on $\alpha_k$ para los
modos $e$ y $\mu$ no se ven afectadas por correcciones radiativas,
as{\'\i} que a una buena aproximaci\'on $\alpha_k \simeq
\alpha_k^0$. En contraste, la raz\'on total de decaimiento
adquiere correcciones de la forma $R = R^0 [1 + (\alpha/\pi)
\Phi]$, donde $R^0$ es la raz\'on de decaimiento no corregida y
$\Phi$ proviene de correcciones radiativas independientes de
modelo. La funci\'on $\Phi$ se obtiene de las Ecs.\ (5.25) y
(5.28) de la Ref.~\cite{AGarcia}; sus valores num\'ericos para
varios decaimientos de inter\'es en este trabajo est\'an listados
en la Tabla~\ref{table:phirad}

\begin{table}
\begin{center}
\begin{tabular}{|lc|r|}
\hline\hline
& & \\[-3mm]
Proceso &\qquad & $\Phi$ \mbox{\hglue0.42truecm} \\[2mm] \hline
$n \to p e^-\bar{\nu}_e$ & & $0.0486$ \\[1mm]
$\Sigma^+ \to \Lambda e^+ \nu_e$ & & $0.0015$ \\[1mm]
$\Sigma^- \to \Lambda e^- \bar{\nu}_e$ & & $0.0012$ \\[1mm]
$\Lambda \to p e^- \bar{\nu}_e$ & & $0.0207$ \\[1mm]
$\Sigma^- \to n e^- \bar{\nu}_e$ & & $-0.0025$ \\[1mm]
$\Xi^- \to \Lambda e^- \bar{\nu}_e$ & & $0.0015$ \\[1mm]
$\Xi^- \to \Sigma^0 e^- \bar{\nu}_e$ & & $-0.0000$ \\[1mm]
$\Xi^0 \to \Sigma^+ e^- \bar{\nu}_e$ & & $0.0226$ \\[1mm]
\hline\hline
\end{tabular}
\caption{Valores de $\Phi$ para correcciones radiativas en algunos
DSB.\label{table:phirad}}
\end{center}
\end{table}

Por otra parte, dado que la dependencia de modelo de las
correcciones radiativas no puede ser calculada rigurosamente, la
Ref.~\cite{AGarcia} propone parametrizarla a trav\'es de una
constante de acoplamiento d\'ebil modificada de la forma $G\equiv
G(1+C)$, donde $C\sim 0.0234$. Este valor de $C$ puede dar una
contribuci\'on notable a la raz\'on de decaimiento total.
Adoptaremos esta parametrizaci\'on en el presente an\'alisis.

\section{Datos experimentales acerca de DSB}

La informaci\'on experimental disponible en decaimientos
semilept\'onicos de bariones \cite{PDG} est\'a constituida por
las razones totales de decaimiento $R$, los coeficientes de
correlaci\'on angular $\alpha_{e\nu}$, los coeficientes de
asimetr{\'\i}a angular $\alpha_e$, $\alpha_\nu$, $\alpha_B$, $A$ y
$B$ y por los cocientes $g_A/g_V$. Esta informaci\'on est\'a
listada en la Tabla 4.4. Los datos
contenidos en esta tabla proviene de diferentes fuentes. Por
ejemplo, la raz\'on de transici\'on puede obtenerse f\'acilmente a
partir del tiempo de vida media del bari\'on y de la anchura de
decaimiento del proceso en cuesti\'on; estos datos est\'an
contenidos en la Ref.~\cite{PDG}. Para algunos de los
coeficientes de correlaci\'on angular y de asimetr{\'\i}a se
consultaron, hasta donde fue posible, los trabajos que contienen
los resultados experimentales originales
\cite{wise,bour,hsueh,dwor} y para los coeficientes restantes se
utiliz\'o la Ref.~\cite{AGarcia}.

\begin{figure}[h]
\begin{center}
\includegraphics[scale=0.96]{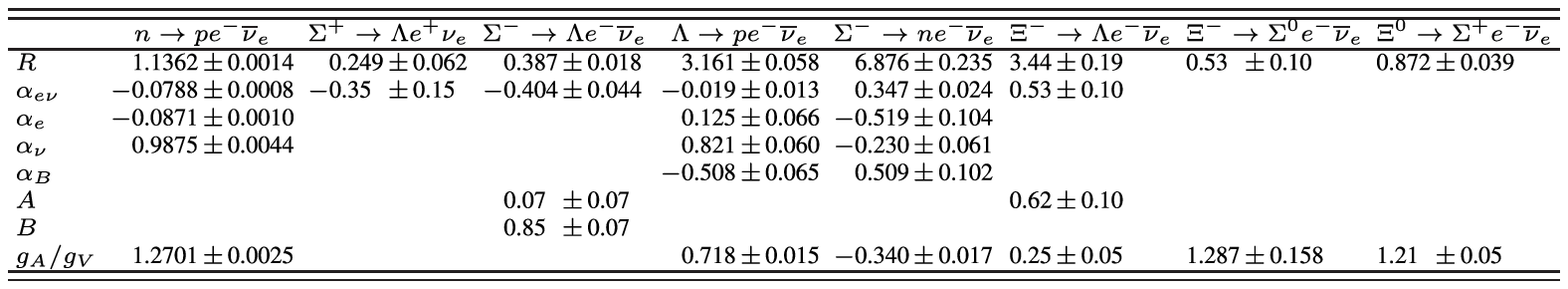}
\end{center}
Tabla 4.4: Datos experimentales de ocho decaimientos semilept\'onicos de bariones observados. Las unidades
de $R$ son 10$^{-3}$ s$^{-1}$ para el decaimiento del neutr\'on y $10^6$ s$^{-1}$ para los decaimientos restantes.
\end{figure}

Es una pr\'actica com\'un en el an\'alisis de datos de DSB clasificar a la informaci\'on experimental en dos grupos. El primero est\'a constituido por las razones de decaimiento totales, los coeficientes de correlaci\'on angular y los coeficientes de asimetr{\'\i}a angular. El segundo grupo lo conforman las razones de transici\'on totales y los cocientes $g_A/g_V$. El primer conjunto es aparentemente m\'as abundante en informaci\'on que el segundo dado que el cociente $g_A/g_V$ no es una medida independiente y se determina a partir de otros observables. En el presente trabajo utilizaremos ambos conjuntos en la comparaci\'on entre teor{\'\i}a y experimento.

\section{Los Ajustes con los datos experimentales}
En esta secci\'on realizamos una comparaci\'on detallada de la complicada expresi\'on en Ec. (\ref{eq:AkcSB}) con los datos experimentales disponibles \cite{PDG} sobre decaimientos semilept\'onicos de bariones, estos se encuentran listados en la Tabla 4.4 en la forma de la raz\'on total de decaimiento $R$, los coeficientes de correlaci\'on angular $\alpha_{e\nu}$ y los coeficientes de asimetr{\'\i}a angular $\alpha_e$, $\alpha_\nu$, $\alpha_B$, $A$ y $B$ junto con las razones $g_A/g_V$. Con el fin de obtener informaci\'on acerca de los par\'ametros libres de la teor{\'\i}a, es decir, de los par\'ametros b\'asicos $a_1$, $b_2$,  $b_3$ y $c_3$ del operador axial vector de bariones $A^{kc}$~Ec. (\ref{Akc3}).

El an\'alisis num\'erico se puede realizar de dos formas. Primero estudiamos los efectos de las correcciones a un loop, solo comparando las expresiones te\'oricas con los datos disponibles sobre decaimientos semilept\'onicos de bariones, para posteriormente incorporar los efectos de ambas correcciones a un loop y con rotura de simetr{\'\i}a perturbativa en el an\'alisis, usamos los datos experimentales sobre decaimientos semilept\'onicos y los decaimientos fuertes del decuplete de bariones. Los resultados nos permiten comparar, de igual forma, con otros c\'alculos realizados.

\subsection{Ajustes de los datos sobre decaimientos semilept\'onicos de bariones: Efectos de correcciones a un loop}
Los ajustes que presentamos a continuaci\'on han sido publicados en la Ref. \cite{PRD2012}. El ajuste m\'as sencillo que podemos realizar es un ajuste con simetr{\'\i}a exacta SU(3), el cual involucra solo dos par\'ametros, $a_1$ y $b_2$; esto es equivalente a un ajuste con solo $F$ y $D$, a este nivel tenemos lo siguiente,

\begin{equation}\label{eq:FitsDyF}
  D = \frac{1}{2}a_1, \hspace{1cm} F=\frac{1}{3} a_1 + \frac{1}{6} b_2,
\end{equation}
utilizando las razones de decaimiento y las razones $g_A/g_V$ los valores del mejor ajuste, los presentamos en la Tabla \ref{tab:fit1},
\begingroup
\setcounter{table}{4}
\begin{table}[ht]
\begin{center}
\begin{tabular}{lr}
\hline \hline
Par\'ametros & Valores\\
\hline
$a_1$ & $1.61 \pm 0.01$\\
$b_2$ & $-0.40 \pm 0.06$\\
$D$ & $0.81 \pm 0.01$\\
$F$ & $0.47 \pm 0.01$\\
\hline\hline
\end{tabular}
\caption{\label{tab:fit1} Ajuste 1, con simetr{\'\i}a exacta SU(3) y con una $\chi^2 = 53.85$ para 12 grados de libertad.}
\end{center}
\end{table}
\endgroup
El siguiente ajuste consiste en ignorar la diferencia de masa bari\'onica $\Delta$, lo cual es equivalente a considerar el l{\'\i}mite degenerado $\Delta/m_\Pi \rightarrow 0$. En realidad, un ajuste bajo esta suposici\'on ya se realiz\'o en la Ref.~\cite{RFM2006}.
Usando las razones de decaimiento y las razones $g_A/g_V$ de la Tabla 4.4, encontramos
los valores listados en la Tabla \ref{tab:fit2}. En lo sucesivo, los errores citados de los par\'ametros mejor ajustados deber\'an ser del ajuste $\chi^2 $ \'unicamente, y no incluyen las incertidumbres te\'oricas. Una inspecci\'on detallada de la salida del ajuste revela que, con excepci\'on de $c_3$, los valores de los par\'ametros obtenidos son como se esperaban de la expansi\'on $1/N_c$, es decir, son aproximadamente de orden 1. Para $c_3$ la situaci\'on es radicalmente diferente, ya que est\'a muy lejos de cualquier expectativa coherente. Sorprendentemente, los efectos de las correcciones del loop cambian notablemente los valores de $a_1$ y $b_1$ con respecto al caso discutido anteriormente de simetr{\'\i}a SU(3). Realmente cuando consideramos correcciones del loop con ambos octete y decuplete de bariones aparecen dos coeficientes m\'as $b_3$ y $c_3$, los cuales est\'an directamente relacionados a los acoplamientos ${\cal C}$ y ${\cal H}$. Para $N_c = 3$ las ecuaciones son, \cite{JenkinsPRD53}
\begin{eqnarray}
  D &=& \frac{1}{2}a_1 +\frac{1}{6}b_3, \nonumber \\
  F &=& \frac{1}{3}a_1 +\frac{1}{6}b_2 + \frac{1}{9}b_3, \nonumber \\
   && \\
  {\cal C} &=& - a_1 -\frac{1}{2}c_3,\nonumber\\
  {\cal H} &=& -\frac{3}{2}a_1 - \frac{3}{2}b_2 -\frac{5}{2}b_3. \nonumber
\end{eqnarray}

\begingroup
\begin{table}[ht]
\begin{center}
\begin{tabular}{lr}
\hline \hline
Par\'ametros & Valores\\
\hline
$a_1$ & $ 0.28 \pm 0.07$\\
$b_2$ & $-0.67 \pm 0.04$\\
$b_3$ & $ 4.02 \pm 0.26$\\
$c_3$ & $ -13.95 \pm 2.92$\\
$D$ & $0.81 \pm 0.01$\\
$F$ & $0.43 \pm 0.01$\\
${\cal C}$ & $6.70 \pm 1.10$ \\
${\cal H}$ & $-9.47\pm 0.50$  \\
\hline\hline
\end{tabular}
\caption{\label{tab:fit2} Ajuste 2. Correcciones a un loop con ambos octete y decuplete de bariones, con $\chi^2 = 39.33$ para 10 grados de libertad.}
\end{center}
\end{table}
\endgroup

Como podemos observar, en la Tabla 4.7 tenemos las diferentes contribuciones de sabor SU(3) para $g_A$ y estas, siguen el patr\'on dictado por la expansi\'on $1/N_c$. Las correcciones singulete $-$las m\'as importantes$-$ hablando rigurosamente, est\'an suprimidas $1/N_c$ con respecto al valor de nivel \'arbol.
Supresiones posteriores de las contribuciones octete y ~${\bf 27}$ tambi\'en son notables. Por lo tanto, a pesar del valor alto de $\chi^2$, el ajuste en el caso degenerado si da predicciones de $g_A$ las cuales son consistentes con lo esperado. Sin embargo, el precio que se paga depende de los valores m\'as altos de los par\'ametros de la teor{\'\i}a, lo cual no es completamente satisfactorio.

\begin{figure}[ht]
\begin{center}
\includegraphics[scale=1]{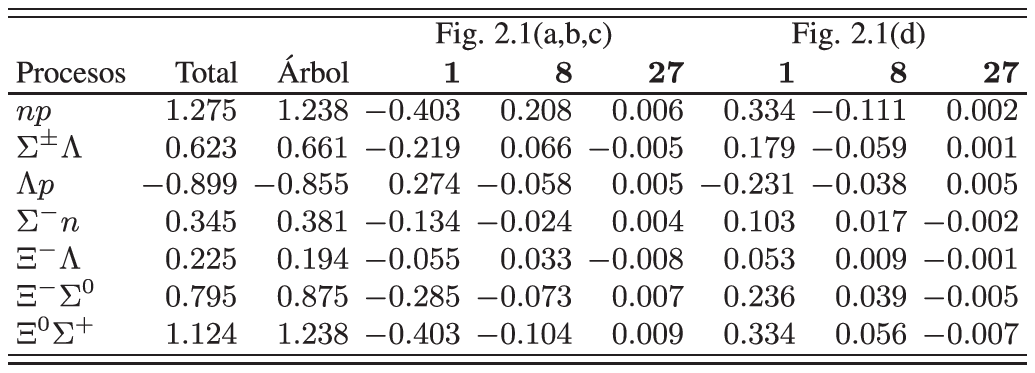}
\end{center}
Tabla 4.7: Valores de $g_A$ de algunos decaimientos semilept\'onicos de bariones para $\Delta$ nulo. Presentamos las contribuciones de las diferentes representaciones de SU(3). Se utilizan las razones de decaimiento y las razones $g_A/g_V$ en el ajuste.
\end{figure}

Como se mencion\'o anteriormente, un ajuste similar se llev\'o a cabo en la Ref.~\cite{RFM2006}. Nuestro an\'alisis aqu{\'\i} difiere del anterior en dos aspectos. Primero, en la Ref.~\cite{RFM2006} se utiliz\'o la informaci\'on experimental accesible en ese momento \cite{part2}. Los valores de $V_{ud}$ y $V_{us}$, sin embargo, se han actualizado con la informaci\'on experimental sobre los procesos $n \rightarrow p$ y $\Xi^0 \to \Sigma^+$ \cite{PDG}. Estas mejoras introducen diferencias perceptibles en el an\'alisis actual. Segundo, en la Ref.~\cite{RFM2006} se realiz\'o un ajuste limitado con el fin de conseguir $c_3$ del acoplamiento mes\'on-bari\'on $|\mathcal{C}|=1.6$. Ahora obtenemos $c_3$ a partir de los datos solamente en las mismas condiciones que los dem\'as par\'ametros de $a_1$, $b_2$ y  $b_3$. Por lo tanto podemos decir que nuestros resultados num\'ericos actuales sustituyen a los de la Ref.~\cite{RFM2006}.
En el presente an\'alisis cuantificamos los efectos de $\Delta$ no nula para $g_A$, para ello hacemos la evaluaci\'on de los efectos de una diferencia de masa $\Delta$ no nula octete decuplete. Al igual que en el ajuste anterior, se utiliza la informaci\'on experimental sobre las razones de decaimiento y las razones $g_A/g_V$ con el fin de determinar los par\'ametros $a_1$, $b_2$, $b_3$ y $c_3$. Los valores del mejor ajuste obtenidos esta vez son listados en la Tabla \ref{tab:fit3},
\setcounter{table}{7}
\begingroup
\begin{table}[ht]
\begin{center}
\begin{tabular}{lr}
\hline \hline
Par\'ametros & Valores\\
\hline
$a_1$ & $ -0.35 \pm 0.02$\\
$b_2$ & $ -2.40 \pm 0.16$\\
$b_3$ & $ 6.53 \pm 0.16$\\
$c_3$ & $ 5.86 \pm 0.29$\\
$D$ & $0.91 \pm 0.02$\\
$F$ & $0.21 \pm 0.02$\\
${\cal C}$ & $ -2.58 \pm 0.14$ \\
${\cal H}$ & $ -12.2 \pm 0.16$  \\
\hline\hline
\end{tabular}
\caption{\label{tab:fit3} Ajuste 3. Efectos de una diferencia de masa octete-decuplete $\Delta$, con una $\chi^2 = 17.80$ para 10 grados de libertad.}
\end{center}
\end{table}
\endgroup
Aunque los valores de $b_3$ y $c_3$ son ligeramente superiores a lo esperado, podemos decir que existe una notable mejora de los par\'ametros del ajuste, en este caso con respecto al anterior. Adem\'as, $\chi^2$ reduce considerablemente su valor a 1.78/dof, donde dof se refiere a los grados de libertad, lo que indica un ajuste mucho mejor.

Ahora para los acoplamientos axial vector de bariones. Los valores predichos para $g_A$ son listados en la Tabla 4.9. Observamos que hay total consistencia en estas predicciones. Las supresiones $1/N_c$, dadas por la expansi\'on $1/N_c$, son evidentes en todas las contribuciones de sabor para $g_A$.
\begin{figure}[ht]
\begin{center}
\includegraphics[scale=.91]{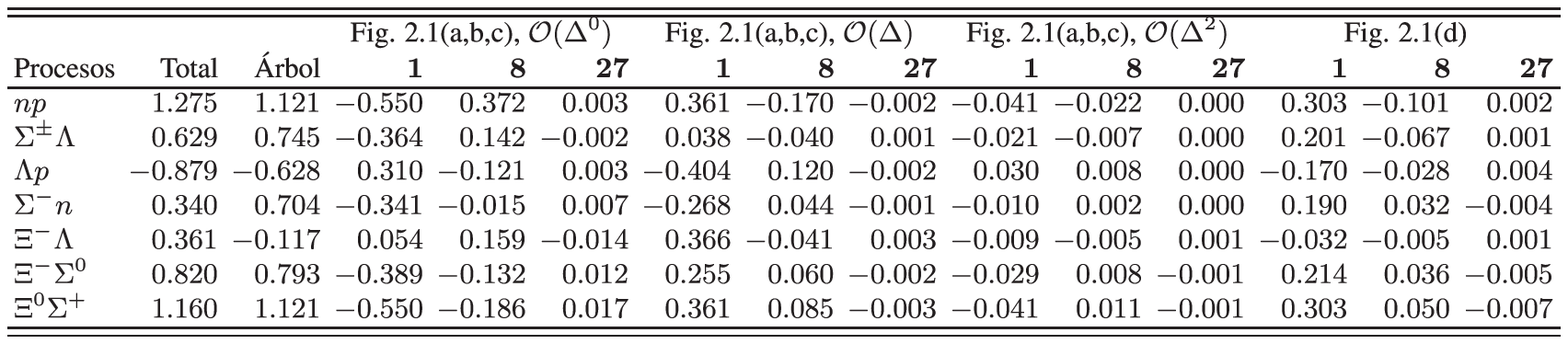}
\end{center}
Tabla 4.9: Valores de $g_A$ de algunos decaimientos semilept\'onicos de bariones observados con $\Delta$ no nulo. Presentamos las
 contribuciones de las diferentes representaciones de SU(3). Para el ajuste se utilizan las razones de decaimiento y las razones $g_A/g_V$.
\end{figure}
Si bien el sabor singulete es el m\'as significativo, los sabores octete y {\bf 27} presentan supresiones en relaci\'on con el valor a nivel \'arbol como se esperaba. Sin embargo, debemos se\~nalar que las entradas del proceso $\Xi^- \rightarrow \Lambda$ muestran preocupantes desviaciones de los valores esperados. Este comportamiento se ha observado sistem\'aticamente en otros an\'alisis \cite{RFM2006, RFM(1998), RFM2004}.

Ahora, en la Tabla 4.10 proporcionamos los observables obtenidos con los pa\-r\'a\-me\-tros del mejor ajuste con el fin de compararlos con los valores experimentales que se muestran en la Tabla 4.4. Las desviaciones m\'as importantes entre la teor{\'\i}a y experimento se derivan de las razones de decaimiento de los procesos de $\Xi^- \rightarrow \Lambda$, $\Lambda \rightarrow p$ y $\Xi^0 \rightarrow \Sigma^+$, cuyas contribuciones a la $\chi^2$ total ascienden a $\chi^2_{\Xi^- \Lambda} = 5.31$, $\chi^2_{\Lambda p} = 2.37$ y $\chi^2_{\Xi^0 \Sigma^+} = 1.78$, respectivamente y las razones $g_A/g_V$ de $n \rightarrow p$, el contribuye con $\chi^{2}_{np} = 3.87$ a $\chi^{2}$.

\begin{figure}[ht]
\begin{center}
\includegraphics[scale=.95]{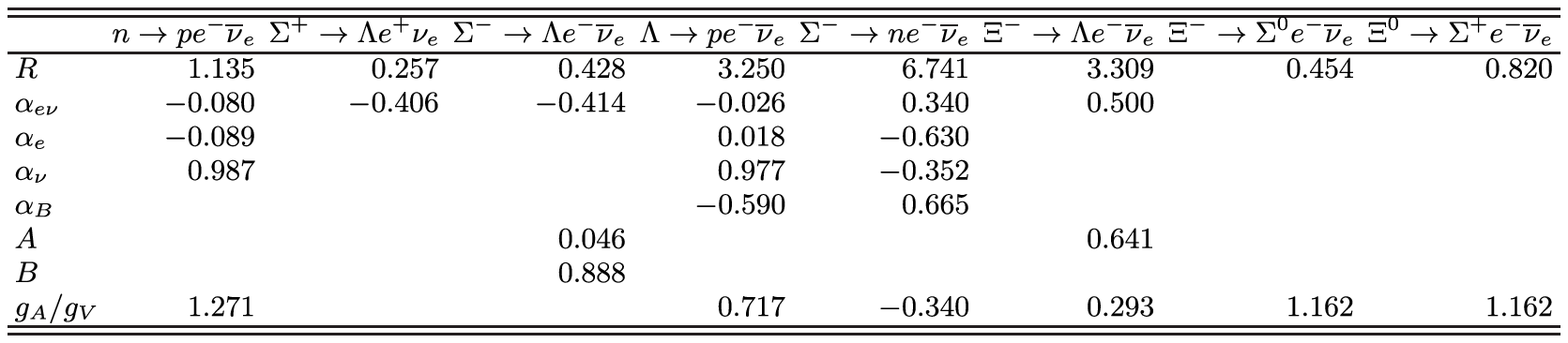}
\end{center}
Tabla 4.10: Valores predichos de algunos observables para ocho decaimientos semilept\'onicos de bariones observados. Las unidades de $R$ son $10^{-3}s^{-1}$ para el decaimiento del neutr\'on y $10^6s^{-1}$ para los otros decaimientos.
\end{figure}

Ahora podemos rehacer el an\'alisis con el fin de utilizar el otro conjunto de datos experimentales que se discuti\'o anteriormente, es decir, el constituido por las razones de decaimiento y los coeficientes de correlaci\'on angular y de asimetr{\'\i}a angular. En este momento tenemos a nuestra disposici\'on 8 razones de decaimiento y 17 coeficientes.

Al igual que en el caso anterior, realizamos la comparaci\'on entre la teor{\'\i}a y el experi\-mento en el l{\'\i}mite de $\Delta$ nulo.
El ajuste produce $a_1 = 0.30 \pm 0.06$, $b_2 = -0.65 \pm 0.03$, $b_3 = 3.92 \pm 0.24$, y $c_3 = -13.79 \pm 2.17$, con una $\chi^2 = 62.62$ para 23 grados de libertad. Observamos que los valores de los par\'ametros del ajuste no cambian sustancialmente con respecto a sus an\'alogos cuando las razones de decaimiento y las razones $g_A/g_V$ se utilizan. Sin embargo, las diferencias, aunque peque\~nas, si son perceptibles. Las predicciones para $g_A$ se enumeran en la Tabla 4.11. Las diferentes contribuciones de sabor se dan de la misma manera como en la Tabla 4.7.

\begin{figure}[ht]
\begin{center}
\includegraphics[scale=1]{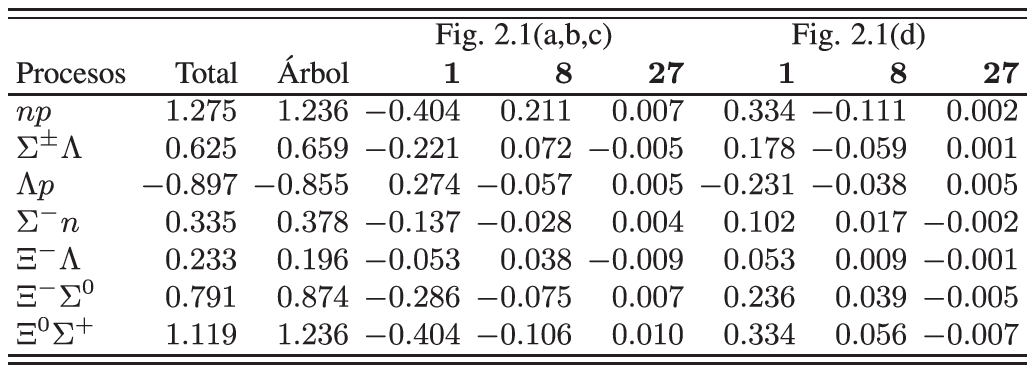}
\end{center}
Tabla 4.11: Valores predichos de $g_A$ de algunos decaimientos semilept\'onicos de bariones observados para $\Delta$ nulo. Presentamos las contribuciones de las diferentes representaciones de SU(3). Se utilizan razones de decaimiento y coeficientes de correlaci\'on angular y de asimetr{\'\i}a angular en el ajuste.
\end{figure}

Cuando consideramos $\Delta$ no nula, el ajuste da $a_1 = -0.36 \pm 0.02$, $b_2 =-2.50 \pm 0.15$, $b_3 = 6.64 \pm 0.15$, y $c_3 = 5.81 \pm 0.25$, con una
$\chi^2= 36.10$ para 23 grados de libertad. Las predicciones para $g_A$ se presentan en la Tabla 4.12. Existen peque\~nas pero perceptibles diferencias entre las entradas de esta Tabla 4.12 y las de la Tabla 4.9.
\begin{figure}[h]
\begin{center}
\includegraphics[scale=.93]{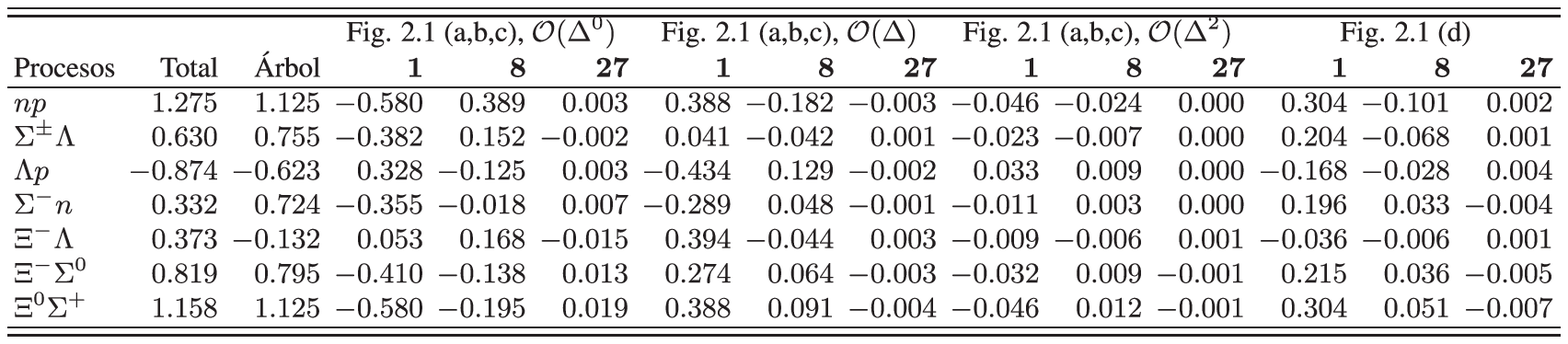}
\end{center}
Tabla 4.12: Valores de $g_A$ de algunos decaimientos semi\-lept\'o\-ni\-cos de bariones observados con $\Delta$ no nulo. Se presentan las diferentes contribuciones de SU(3). En el ajuste se utilizan los coeficientes de correlaci\'on angular y los coeficientes de asimetr{\'\i}a angular.
\end{figure}

Se concluye la comparaci\'on, proporcionando en la Tabla 4.13 los observables obtenidos, con el mejor ajuste de los par\'ametros, ahora
con el prop\'osito de compararlos con los va\-lo\-res experimentales dados en la Tabla 4.4. La m\'as alta desviaci\'on entre la teor{\'\i}a y el experimento proceden de $\alpha_\nu$ de $\Lambda \rightarrow p$ ($\chi^2_{\Lambda p} = 6.66)$ y $\Sigma^- \rightarrow n$ ($\chi^2_{\Sigma^-n} = 4.10)$ y de $\alpha_e$ en el proceso $n\rightarrow p $ ($\chi^2_{np} = 4.23$), que en su conjunto ascender\'a a la mitad del total de $\chi^2$.

\begin{figure}[ht]
\begin{center}
\includegraphics[scale=.95]{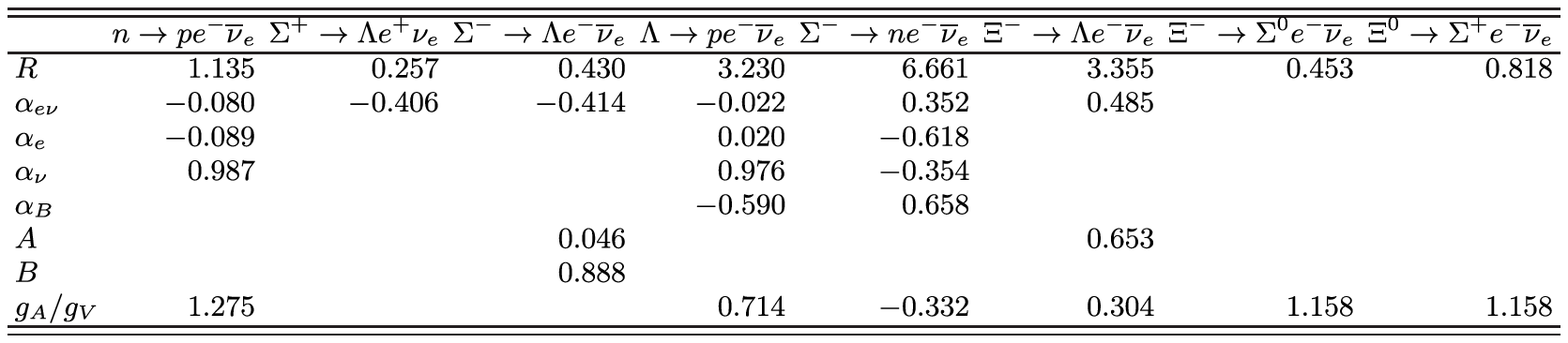}
\end{center}
Tabla 4.13: Valores predichos de algunos observables para ocho decaimientos semilept\'onicos de bariones observados. Las unidades de $R$ son $10^{-3}s^{-1}$ para el decaimiento del neutr\'on y $10^6s^{-1}$ para los otros decaimientos.
\end{figure}

La motivaci\'on de ir m\'as all\'a en este an\'alisis y dedicar un tiempo considerable en hacer la reducci\'on de todos estos
productos de operadores esp{\'\i}n sabor $J^k$, $T^c$ y $G^{kc}$ del grupo SU(6), fue para demostrar expl{\'\i}citamente como estas cancelaciones ocurren en $N_c$ grande. El hecho de que existen cancelaciones en $N_c$ grande ha sido puntualizado en la Ref. \cite{RFM2000}.

Adicionalmente, se han obtenido co\-rreccio\-nes a orden relativo en $N_c$, para los aco\-pla\-mien\-tos $g_A$ de ocho decaimientos semilept\'onicos de bariones observados, considerando el caso degenerado $\Delta \rightarrow 0$, el ejercicio  consiste en mantener $N_c$ como par\'ametro libre y generar las diferentes contribuciones de sabor singulete octete y {\bf 27}. En particular los resultados han sido publicados en Proceedings of Science ICHEP 2010 \cite{Mary} y en XXV International Symposium on Lepton Photon Interactions at High Energies 2011~\cite{Mary2011}.

\subsection{Ajuste de datos sobre los decaimientos $\beta$ y fuertes de bariones: inclusi\'on de ambas correcciones, quiral y con rotura parturbativa}
Los elementos de matriz fuera de la diagonal de $A^{kc} + \delta A^{kc}$ que involucran bariones del octete y decuplete pueden ser obtenidos a trav\'es de las transiciones del decuplete de ba\-rio\-nes al octete de bariones y piones. La informaci\'on experimental disponible sobre los decaimientos fuertes $ \Delta\rightarrow N \pi, \Sigma^* \rightarrow \Lambda \pi, \Sigma^* \rightarrow\Sigma\pi, \Xi^* \rightarrow \Xi\pi$ puede ser encontrada en Ref. \cite{PDG} en forma de anchuras.

El formalismo para obtener las anchuras de los decaimientos fuertes del decuplete de bariones en teor{\'\i}a de perturbaciones quirales
fue introducido originalmente por  Peccei \cite{peccei} y posteriormente implementada en el an\'alisis de Ref \cite{Dai}. En este formalismo la anchura de un $B'$ del decuplete de bariones decayendo en un $B$ del octete de bariones y un pi\'on esta dado por

\begin{equation}\label{eq:anchura}
  \Gamma_{B'}= \frac{g^2 C(B, B')^2(E_B + M_B)|{\bf q}|^3}{24 \pi f^2 M_{B'}}
\end{equation}
donde $g$ es el acoplamiento axial vector para este decaimiento y $C(B,B')$ es un coeficiente de Clebsch-Gordan presentados en la Tabla 4.14, $M_{B'}$ y $M_B$ son las masas de los bariones del decuplete y octete respectivamente, $f$ es la constante de decaimiento del pi\'on y $E_B$ y {\bf q} son la energ{\'\i}a del octete de bariones y el tri-momento en el sistema de reposo de $B'$, respectivamente.

\begin{table}[ht]
  \begin{center}
  \begin{tabular}{ccc}
    \hline\hline
    Decaimientos & $C(B, B\prime)$ & $g$ \\
    \hline
    $\Delta \rightarrow N \pi$ & 1 & $-2.04 \pm 0.01$ \\
    $\Sigma^{*} \rightarrow \Lambda\pi$ & $1/\sqrt{2}$ & $-1.69 \pm 0.02$  \\
    $\Sigma^{*} \rightarrow \Sigma\pi$ & $1/\sqrt{3}$  & $-1.59 \pm 0.10$  \\
    $\Xi^{*} \rightarrow \Xi\pi$  & $1/\sqrt{2}$ & $-1.46 \pm 0.04$ \\
    \hline\hline
  \end{tabular}
  \end{center}
  Tabla 4.14: Coeficientes de Clebsch-Gordan $C(B, B')$ y acoplamientos axial vector $g$ para los decaimientos fuertes.
\end{table}
Con ayuda de la Ec. (\ref{eq:anchura}), los acoplamientos axial vector $g$ pueden ser determinados para cada decaimiento y son listados en la Tabla 4.14. Notemos que los coeficientes de Clebsch-Gordan se han elegido de tal forma que los acoplamientos $g$ est\'en todos en el l{\'\i}mite de simetr{\'\i}a exacta SU(3) \cite{Dai}.

En esta etapa contamos con cuatro piezas extra de informaci\'on experimental. Con el fin de realizar una comparaci\'on mas pr\'actica entre la teor{\'\i}a y el experimento, realizamos un ajuste global usando la informaci\'on experimental $g_A$ y $g$, esta vez incluyendo ambas correcciones quirales y con rotura de simetr{\'\i}a perturbativa en juego.

La rotura de simetr{\'\i}a perturbativa implica varios par\'ametros libres extra en el an\'alisis. Podemos mantener la rotura de simetr{\'\i}a perturbativa hasta un cierto orden cercano al principal, es decir, podemos considerar solo correcciones a orden ${\cal O}(N_c^{0})$ a $A^{kc}$ en la expansi\'on Ec. (\ref{eq:delta8y10}), de lo contrario se pierde poder predictivo. Tambi\'en incluimos rotura de simetr{\'\i}a SU(3) solo en el sector de extrañeza cero. Bajo esta suposici\'on, $A^{kc} + \delta_{{\rm SB}}^{kc}$ toma la forma simplificada de Ec. (\ref{eq:SB})
\begin{eqnarray}
  A^{kc} + \delta_{{\rm SB}}^{kc} &=& a_1 G^{kc} + b_2\frac{1}{N_c}{\cal D}_2^{kc} +  b_3\frac{1}{N_c^2}{\cal D}_3^{kc} + c_3\frac{1}{N_c^2}{\cal O}_3^{kc} \nonumber\\
                            &+&  W^a \left[ d_1 d^{c8e}G^{ke} + d_2\frac{1}{N_c}d^{c8e}{\cal D}_2^{kc} + d_3 \frac{1}{N_c} \left( \left\{G^{kc}, T^8\right\} -\left\{G^{k8}, T^c\right\}\right) \right.\nonumber\\
                            &+&\left. d_4 \frac{1}{N_c}\left(\left\{ g^{kc}, T^8\right\} + \left\{G^{k8}, T^c\right\}\right) \right],\label{eq:SB}
\end{eqnarray}
donde $W^a = 1$ para $a= 4, 5$ y nulo para $a= 1,2$.

Con el fin de realizar el ajuste  en el l{\'\i}mite $\Delta \rightarrow 0$ en forma consistente, debemos establecer $b_3 = 0$ en Ec. (\ref{eq:SB}) y eliminar de $\delta A_{1 {\rm L}}^{kc}$ todos los t\'erminos de orden $1/N_c^{3}$, ya que estos t\'erminos son proporcionales a $b_2^2$, $a_1, b_2, b_3$ y $a_1b_2c_3$. De acuerdo al t\'ermino $c_3$ en Ec. (\ref{eq:SB}) evitaremos mezclar los efectos de rotura de simetr{\'\i}a con correcciones $1/N_c$ en los acoplamientos $D$, $F$, ${\cal C}$ y ${\cal H}$ \cite{Dai}.

\setcounter{table}{14}
\begingroup
\begin{table}[ht]
\begin{center}
\begin{tabular}{lrrrrr}
\hline \hline
Par\'ametros & Valores con $\Delta = 0$  & Valores con $\Delta \neq 0$\\
\hline
$a_1$ & $1.00 \pm 0.02$ & $ 0.64 \pm 0.22$\\
$b_2$ & $0.73 \pm 0.06$ & $ 0.21 \pm 0.25$\\
$b_3$ & $0.0 $          & $ 1.35 \pm 0.06$\\
$c_3$ & $0.82 \pm 0.04$ & $ 1.90 \pm 0.41$\\
$d_1$ & $-0.67 \pm 0.04$ & $ -0.44 \pm 0.12 $ \\
$d_2$ & $6.66 \pm 0.41$ &  $ 6.48 \pm 0.37$  \\
$d_3$ & $0.09 \pm 0.03$ &  $0.04 \pm 0.03$   \\
$d_4$ & $-0.01 \pm 0.06$ &  $ 0.08 \pm 0.07$   \\
$D$ & $0.50 \pm 0.01 $ & $0.54 \pm 0.03$\\
$F$ & $0.45 \pm 0.01$ & $0.40 \pm 0.03$\\
${\cal C}$ & $-1.41 \pm 0.01 $ & $ -1.59 \pm 0.05$ \\
${\cal H}$ & $-2.59 \pm 0.07$ & $ -4.64 \pm 1.30$  \\
\hline\hline\\
\end{tabular}
\caption{Ajustes para los datos sobre decaimientos $\beta$ y fuertes: Incluyendo ambas correcciones quiral y rotura de simetr{\'\i}a perturbativa.\label{tab:Ajuste}}
\end{center}
\end{table}
\endgroup
El ajuste, ahora, en el l{\'\i}mite de $\Delta$ nulo, usando los datos experimentales, sobre $g_A$ (o alternativamente $g_A/g_V$), el ajuste da los valores listados en la columna Ajuste 4 de la Tabla \ref{tab:Ajuste}.
La contribuci\'on m\'as alta para $\chi^2$ viene de $g_A$ del proceso $\Xi^0 \Sigma^+(\chi_{\Xi^0\Sigma^+}^2)= 4.88$ y $g$ de los procesos $\Sigma^*\Sigma$ y $\Xi^*\Xi (\chi^2_{\Sigma^*\Sigma}= 3.15$ y $\chi^2_{\Xi^*\Xi}=3.33)$, las diferentes contribuciones SU(3) con rotura de simetr{\'\i}a para $g_A$ y $g$ se listan en la Tabla 4.16.
\begin{figure}[ht]
\begin{center}
\includegraphics[scale=1]{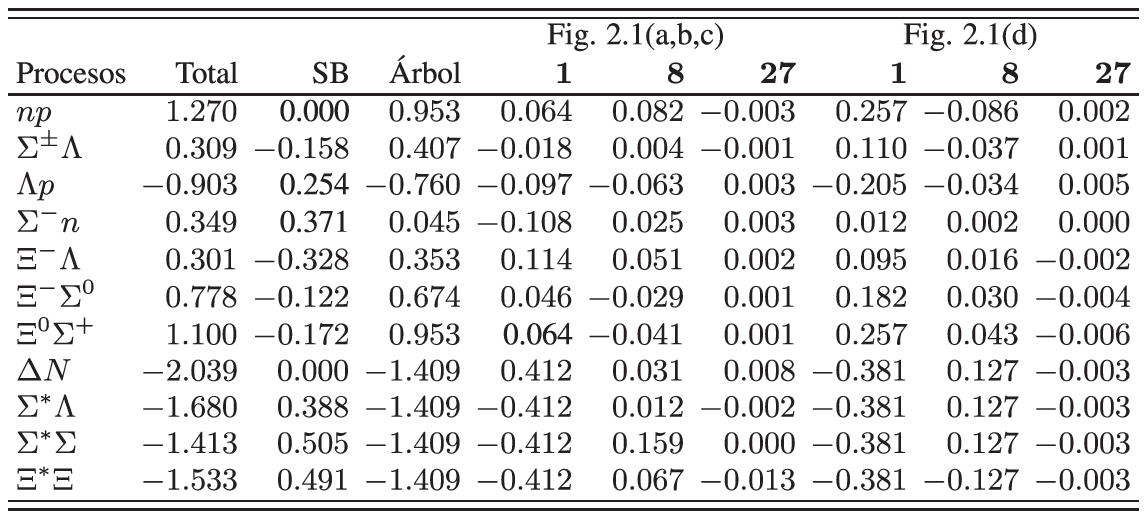}
\end{center}
Tabla 4.16: Valores predichos a los acoplamientos axial vector para $\Delta$ nulo. Presentamos las diferentes contribuciones de SU(3).
\end{figure}
Encontramos muy buena correspondencia entre lo predicho y los acoplamientos $g_A$ y $g$ observados. Tambi\'en vemos que dictan un patr\'on para las diferentes partes de rotura de SU(3) que est\'an en concordancia con las predicciones.

Cuando rehacemos el ajuste en el l{\'\i}mite no nulo, encontramos los valores listado en la columna Ajuste 5, de la Tabla 4.15 con $\chi^2= 2.28$ para dos grados de libertad. La contribucion m\'as alta viene de $g_A$ del proceso $\Xi^- \Lambda (\chi^2_{\Xi^- \Lambda} = 1.58)$, las diferentes contribuciones para $g_A$ y $g$ est\'an listadas en la Tabla 4.17.

\begin{figure}[ht]
\begin{center}
\includegraphics[scale=.85]{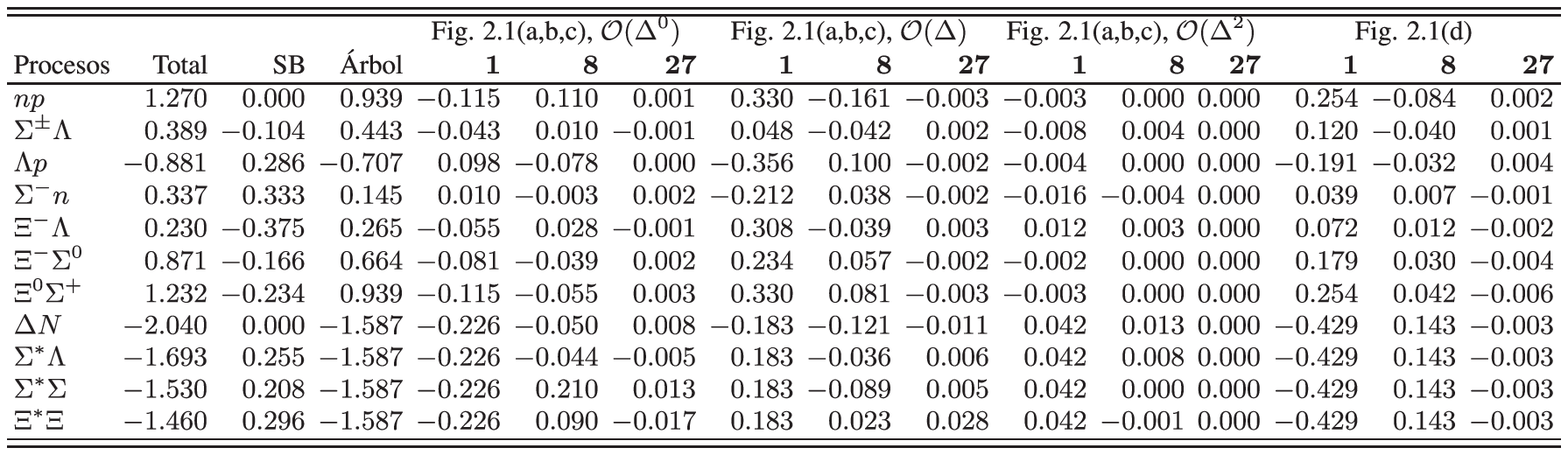}
\end{center}
Tabla 4.17: Valores predichos a los acoplamientos axial vector para $\Delta$ no nulo. Presentamos las diferentes contribuciones de SU(3).
\end{figure}

\chapter{Conclusiones}

El presente trabajo de tesis ha sido desarrollado para calcular la renormalizaci\'on de la co\-rrien\-te axial vector a orden de un loop en teor{\'\i}a de perturbaciones quirales para ba\-rio\-nes y el l{\'\i}mite de $N_c$ grande, tomando en cuenta la diferencia de masa entre el octete y el decuplete de bariones. En esta aproximaci\'on, la correcci\'on para la corriente axial vector, se da por una serie infinita, cada uno de los t\'erminos representa una complicada estructura de conmutadores y/o anticonmutadores implicados en la corriente axial vector de bariones $A^{kc}$ y del operador de masa ${\cal M}$. Hemos considerado los tres primeros t\'erminos de esta expansi\'on, es decir, el l{\'\i}mite degenerado ($AAA$), el siguiente orden de correcci\'on ($AAA{\cal M}$) y finalmente al orden de correcci\'on ($AAA{\cal MM}$) respectivamente. Hemos evaluado expl{\'\i}citamente estas expresiones $-$Individualmente para contribuciones de sabor singulete, sabor {\bf 8} y sabor {\bf 27}$-$ a orden relativo $1/N_c^2$ al valor de nivel \'arbol. En el l{\'\i}mite $N_c$ grande ocurren cancelaciones entre los diferentes diagramas de Feynman en el l{\'\i}mite degenerado, lo cual es una consecuencia de la simetr{\'\i}a contra{\'\i}da esp{\'\i}n sabor SU(6).

El orden del c\'alculo en el presente trabajo nos ha permitido llevar a cabo diferentes ajustes. M\'as precisamente, ajustando nuestras expresiones anal{\'\i}ticas con los datos experimentales sobre los decaimientos semilept\'onicos de bariones, somos capaces de extraer los par\'ametros b\'asicos $a_1$, $b_2$, $b_3$, $c_3$ del Lagrangiano quiral $1/N_c$ de bariones, as{\'\i} como los acoplamientos axial vector $g_A$ para bariones del octete. En una primera aproximaci\'on hemos ignorado la diferencia de masa $\Delta = M_T - M_B$ entre bariones del octete y del decuplete. Este an\'alisis por lo tanto sigue los lineamientos del ajuste como en Ref.~\cite{RFM2006}. El segundo enfoque, el m\'as realista, incorpora los efectos de una diferencia de masa $\Delta$ no nula y constituye la principal contribuci\'on de este trabajo.

En la primera parte del an\'alisis referente al l{\'\i}mite de degeneraci\'on $\Delta \rightarrow 0$, la comparaci\'on entre los datos experimentales y las expresiones te\'oricas se hacen a trav\'es de un ajuste de m{\'\i}nimos cuadrados, para la cual se obtiene una $\chi^2 / \mathrm{grados \, de \, libertad = 3.95}$. En la segunda parte, se toman en cuenta los efectos de $\Delta$, el ajuste da $\chi^2 = 1.78/\mathrm{dof}$, puede ser considerado como un mejor ajuste. Aunque en ambos casos las predicciones a los observables, es decir, las razones de decaimiento as{\'\i} como los coeficientes de correlaci\'on angular y los coeficientes de asimetr{\'\i}a angular en los decaimientos semilept\'onicos de ba\-rio\-nes est\'an en concordancia con sus hom\'ologos experimentales, el ajuste de este \'ultimo es mejor al anterior. Esto se debe a los par\'ametros de mejor ajuste de $a_1$, $b_2$, $b_3$ y $c_3$, introducidos en la definici\'on de la ecuaci\'on de la corriente axial vector (\ref{eq:deltaAkc}). Se ajustan tal como se esperaba de la expansi\'on $1/N_c$, es decir, son m\'as o menos de orden 1.

En conclusi\'on, es esencial considerar sistem\'aticamente, la diferencia de masa entre el octete y el decuplete de bariones en nuestro an\'alisis y en nuestros ajustes, con el fin de establecer las predicciones te\'oricas en teor{\'\i}a de perturbaciones quirales y $N_c$ grande con respecto a la renormalizaci\'on de la corriente axial vector las cuales son consistentes con la expansi\'on $1/N_c$ y con los resultados experimentales.

\appendix

\chapter{La Funci\'on $F(m_\Pi, \Delta, \mu)$}

\noindent La funci\'on $F(m_\Pi, \Delta, \mu)$ esta definida por
la integral a un loop para los diagramas de la Fig. 2.1(a, b, c) puede ser expresada como sigue

\begin{equation}\label{ec:intLoop}
\delta^{ij} F(m_\Pi, \Delta, \mu)=\frac{i}{f^2}\int
\frac{d^4k}{(2\pi)^4} \frac{(\mbox{\bf k}^i)(-\mbox{\bf
k}^j)}{(k^2-m_\Pi^{2})(k\cdot v -\Delta + i\epsilon)}.
\end{equation}
donde $\mu$ es el par\'ametro de escala de regularizaci\'on dimensional. La soluci\'on de esta integral toma la forma \cite{RFM2000}

\begin{eqnarray}
24\pi^2f^2 \, F(m,\Delta,\mu) & = & \Delta \left[\Delta^2 - \frac32 m^2 \right] \ln \frac{m^2}{\mu^2} - \frac83 \Delta^3 - \frac72 \Delta m^2 \nonumber \\
&  & \mbox{} + \left\{ \begin{array}{ll} \displaystyle
2(m^2-\Delta^2)^{3/2} \left[ \frac{\pi}{2} - \textrm{arctan}
\left( \frac{\Delta}{\sqrt{m^2-\Delta^2}} \right) \right], & m \ge \left|\Delta\right|, \nonumber \\[6mm]
\displaystyle - (\Delta^2-m^2)^{3/2} \left[-2i\pi + \ln \left( {
\frac{\Delta-\sqrt{\Delta^2-m^2}}{\Delta+\sqrt{\Delta^2-m^2}}}
\right) \right], & m < \left|\Delta\right|. \end{array} \right.
\end{eqnarray}
Las derivadas de la funci\'on $F(m, \Delta, \mu)$ que necesitamos para el c\'alculo, son las siguientes

\begin{eqnarray}
24\pi^2f^2 \, F^{(1)} (m,\Delta,\mu) & = & 3\left[\Delta^2-\frac12 m^2\right]\ln \frac{m^2}{\mu^2} - 6\Delta^2 - \frac{11}{2} m^2 \nonumber \\
&  & \mbox{} - \left\{ \begin{array}{ll} \displaystyle 6\Delta\sqrt{m^2-\Delta^2} \left[\frac{\pi}{2} - \textrm{arctan} \left(\frac{\Delta}{\sqrt{m^2-\Delta^2}} \right) \right], & m \geq |\Delta|,\nonumber \\[6mm]
\displaystyle 3\Delta\sqrt{\Delta^2-m^2}\left[-2i\pi+ \ln \left(
\frac{\Delta-\sqrt{\Delta^2-m^2}}{\Delta+\sqrt{\Delta^2-m^2}}
\right) \right], & m < |\Delta|,
\end{array} \right.
\end{eqnarray}
\begin{eqnarray}
24\pi^2f^2 \, F^{(2)} (m,\Delta,\mu) & = & 6\Delta \left[\ln\frac{m^2}{\mu^2}-1\right] \nonumber \\
&  & \mbox{} - \left\{ \begin{array}{ll} \displaystyle \frac{6(m^2-2\Delta^2)}{\sqrt{m^2-\Delta^2}} \left[\frac{\pi}{2} - \textrm{arctan} \left( \frac{\Delta} {\sqrt{m^2-\Delta^2}} \right) \right], & m \geq |\Delta|,\nonumber \\[6mm]
\displaystyle \frac{3(2\Delta^2-m^2)}{\sqrt{\Delta^2-m^2}}
\left[-2i\pi+ \ln
\left(\frac{\Delta-\sqrt{\Delta^2-m^2}}{\Delta+\sqrt{\Delta^2-m^2}}
\right) \right], & m < |\Delta|, \end{array}  \right.
\end{eqnarray}
\vspace*{0.5cm}
y
\vspace*{0.5cm}
\begin{eqnarray}
24\pi^2f^2 \, F^{(3)} (m,\Delta,\mu) & = & 6\ln\frac{m^2}{\mu^2} - \frac{6\Delta^2}{m^2-\Delta^2} \nonumber \\
&  & \mbox{} + \left\{ \begin{array}{ll} \displaystyle \frac{6\Delta(3m^2-2\Delta^2)}{(m^2-\Delta^2)^{3/2}} \left[\frac{\pi}{2} - \textrm{arctan} \left( \frac{\Delta}{\sqrt{m^2-\Delta^2}} \right) \right], & m \geq |\Delta|,\nonumber \\[6mm]
\displaystyle \frac{3\Delta(3m^2-2\Delta^2)}{(\Delta^2-m^2)^{3/2}}
\left[-2i\pi + \ln \left(
\frac{\Delta-\sqrt{\Delta^2-m^2}}{\Delta+\sqrt{\Delta^2-m^2}}
\right) \right], & m < |\Delta|. \end{array} \right.
\end{eqnarray}

\chapter{C\'alculo de la Estructura Conmutador/Anticonmutador}


Consideremos uno de los t\'erminos de la Ec. (\ref{Akc-nodegenerado2}), es decir, la contribuci\'on con una inserci\'on de masa. Esta expresi\'on algebraica genera estructuras anticonmutador-conmutador en t\'erminos de los operadores de esp{\'\i}n, sabor y esp{\'\i}n-sabor como se presenta en Ec. (\ref{ec.A2}),

\begin{center}
\begin{eqnarray}
&&\left\{A^{ia}, \left[A^{kc},\left[{\cal M},
A^{ia}\right]\right]\right\}= \frac{a^3}{N_c} m_2\left\{G^{ia},\left[G^{kc},\left[J^2,G^{ia}\right]\right]\right\}\nonumber\\
                 &&+\frac{a_1^2 b_2}{N_c^2}m_2\Bigl(\left\{G^{ia},\left[D_2^{kc},\left[J^2,G^{ia}\right]\right]\right\}+\left\{D_2^{ia},
                    \left[G^{kc},\left[J^2,G^{ia}\right]\right]\right\}\Bigr)\nonumber\\
                 &&+\frac{a_1 b_2^2}{N_c^3}m_2\left\{D_2^{ia}, \left[D_2^{kc},\left[J^2,G^{ia}\right]\right]\right\} \nonumber\\
                 &&+\frac{a_1^2 b_3}{N_c^3}m_2\Bigl(\left\{G^{ia},\left[D_3^{kc},\left[J^2,G^{ia}\right]\right]\right\}+
                    \left\{D_3^{ia},\left[G^{kc},\left[J^2,G^{ia}\right]\right]\right\}\Bigr)\nonumber\\
                 &&+\frac{a_1^2 c_3}{N_c^3}m_2\Bigl (\left\{G^{ia},\left[{\cal O}_3^{kc},\left[J^2,G^{ia}\right]\right]\right\}+ \left\{G^{ia},\left[G^{kc},\left[J^2,
       {\cal O}_3^{ia}\right]\right]\right\}\Bigr. \nonumber\\
       &&\Bigl. +\left\{{\cal
       O}_3^{ia},\left[G^{kc},\left[J^2,G^{ia}\right]\right]\right\}\Bigr).\label{ec.A2}
\end{eqnarray}
\end{center}
Consideremos el primer t\'ermino de Ec. (\ref{ec.A2}), este t\'ermino es a orden principal y esta formado de anticonmutador y doble conmutador,

\begin{equation}
\left\{G^{ia}, [G^{kc},[J^{2},G^{ia}]]\right\}, \label{ec:GGJG}
\end{equation}

A continuaci\'on realizamos el c\'alculo de la estructura (\ref{ec:GGJG}). Partimos del doble conmutador el cual se escribe de la siguiente forma,
\begin{equation*}
[G^{kc},[J^{2},G^{ia}]]=-[G^{ia},[G^{kc},J^{2}]]-[J^{2},[G^{ia},G^{kc}]]=[G^{ia},[J^{2},G^{kc}]]+[J^{2},[G^{kc},G^{ia}]],
\end{equation*}

\noindent donde se ha utilizado la identidad de Jacobi:
\begin{equation*}
[G^{kc},[J^{2},G^{ia}]]+[G^{ia},[G^{kc},J^{2}]]+[J^{2},[G^{ia},G^{kc}]]=0.
\end{equation*}

\begin{equation}
\left\{G^{ia},[G^{kc},[J^{2},G^{ia}]]\right\}=G^{ia}[G^{kc},[J^{2},G^{ia}]]+[G^{kc},[J^{2},G^{ia}]]G^{ia}
\end{equation}

\noindent utilizando,  $[A,[B,C]] = [B,[A,C]]-[C,[A,B]]$ tenemos,

\begin{eqnarray}
&&\left\{G^{ia},[G^{kc},[J^{2},G^{ia}]]\right\} = G^{ia}\bigl([J^{2},[G^{kc},G^{ia}]]-[G^{ia},[G^{kc},J^{2}]]\bigr) \nonumber\\
&& + \bigl([J^{2},[G^{kc},G^{ia}]] -[G^{ia},[G^{kc},J^{2}]]\bigr)G^{ia} \nonumber\\
&& = G^{ia}(J^{2}[G^{kc},G^{ia}]-[G^{kc},G^{ia}]J^{2}+G^{ia}[J^{2},G^{kc}]-[J^{2},G^{kc}]G^{ia})\nonumber \\
&& + (J^{2}[G^{kc},G^{ia}]-[G^{kc},G^{ia}]J^{2}+G^{ia}[J^{2},G^{kc}]-[J^{2},G^{kc}]G^{ia})G^{ia}\nonumber \\
&& = G^{ia}J^{2}[G^{kc},G^{ia}]-G^{ia}[G^{kc},G^{ia}]J^{2}+G^{ia}G^{ia}[J^{2},G^{kc}]\nonumber \\
&& -G^{ia}[J^{2},G^{kc}]G^{ia}+ J^{2}[G^{kc},G^{ia}]G^{ia}-[G^{kc},G^{ia}]J^{2}G^{ia}\nonumber \\
&& +G^{ia}[J^{2},G^{kc}]G^{ia}-[J^{2},G^{kc}]G^{ia}G^{ia}\nonumber \\
&& = G^{ia}J^{2}[G^{kc},G^{ia}]- G^{ia}[G^{kc},G^{ia}]J^{2}+ G^{ia}G^{ia}[J^{2},G^{kc}]\nonumber \\
&& +J^{2}[G^{kc},G^{ia}]G^{ia}-[G^{kc},G^{ia}]J^{2}G^{ia}-[J^{2},G^{kc}]G^{ia}G^{ia}\nonumber \\
&& = G^{ia}J^{2}G^{kc}G^{ia}-G^{ia}J^{2}G^{ia}G^{kc}-G^{ia}G^{kc}G^{ia}J^{2}+G^{ia}G^{ia}G^{kc}J^{2}\nonumber \\
&& +G^{ia}G^{ia}J^{2}G^{kc}-G^{ia}G^{ia}G^{kc}J^{2}+J^{2}G^{kc}G^{ia}G^{ia}-J^{2}G^{ia}G^{kc}G^{ia}\nonumber \\
&& -G^{kc}G^{ia}J^{2}G^{ia}+G^{ia}G^{kc}J^{2}G^{ia}-J^{2}G^{kc}G^{ia}G^{ia}+G^{ia}J^{2}G^{ia}G^{ia}\nonumber \\
&&= G^{ia}J^{2}G^{kc}G^{ia}-G^{ia}J^{2}G^{ia}G^{kc}-G^{ia}G^{kc}G^{ia}J^{2} +G^{ia}G^{ia}J^{2}G^{kc}\nonumber\\
&& -J^{2}G^{ia}G^{kc}G^{ia}-G^{kc}G^{ia}J^{2}G^{ia}+G^{ia}G^{kc}J^{2}G^{ia}+G^{kc}J^{2}G^{ia}G^{ia}\nonumber\\
&&= [G^{ia},J^{2}]G^{kc}G^{ia} +G^{ia}[G^{ia},J^{2}]G^{kc} \nonumber\\
&&+G^{ia}G^{kc}[J^{2},G^{ia}]+G^{kc}[J^{2},G^{ia}]G^{ia}.
\end{eqnarray}

\noindent Por otra parte,\\
$[J^{2},G^{ia}]=[J^{r}J^{r},G^{ia}]=J^{r}[J^{r},G^{ia}]+[J^{r},G^{ia}]J^{r},$\\
$J^{r}i\epsilon^{rim}G^{ma}+i\epsilon^{rim}G^{ma}J^{r}=i\epsilon^{rim}(J^{r}G^{ma}+G^{ma}J^{r}),$\\
$\Rightarrow$ $[J^{2},G^{ia}]=i\epsilon^{rim}\left\{J^{r},G^{ma}\right\}.$\\
\begin{eqnarray}\label{eq:nume}
&&\left\{G^{ia},[G^{kc},[J^{2},G^{ia}]]\right\} = -i\epsilon^{rim}\left\{J^{r},G^{ma}\right\}G^{kc}G^{ia}
+G^{ia}(-i\epsilon^{rim}\left\{J^{r},G^{ma}\right\})G^{kc}\nonumber \\
&& + G^{ia}G^{kc}i\epsilon^{rim}\left\{J^{r},G^{ma}\right\}+G^{kc}i\epsilon^{rim}\left\{J^{r},G^{ma}\right\}G^{ia}.
\end{eqnarray}\\

\noindent utilizando la identidad siguiente,
$\epsilon^{ijk}\left\{J^{i},G^{jc}\right\}=f^{abc}\left\{T^{a},G^{kb}\right\}$.\\


\begin{eqnarray}
&&\left\{G^{ia},[G^{kc},[J^{2},G^{ia}]]\right\} = if^{dea}\left\{T^{d},G^{ie}\right\}G^{kc}G^{ia}+G^{ia}if^{dea}\left\{T^{d},G^{ie}\right\}G^{kc}\nonumber \\
&& \,+
G^{ia}G^{kc}(-if^{dea}\left\{T^{d},G^{ie}\right\})+G^{kc}(-if^{dea}\left\{T^{d},G^{ie}\right\})G^{ia}.\\ \label{ec:operadores}
\end{eqnarray}
\begin{eqnarray}
\label{eq:second}
&&\left\{G^{ia},[G^{kc},[J^{2},G^{ia}]]\right\}=if^{dea}T^{d}G^{ie}G^{kc}G^{ia}+if^{dea}G^{ie}T^{d}G^{kc}G^{ia}\nonumber \\
&  & + if^{dea}G^{ia}T^{d}G^{ie}G^{kc} + if^{dea}G^{ia}G^{ie}T^{d}G^{kc}  -if^{dea}G^{ia}G^{kc}T^{d}G^{ie} \nonumber \\
&  & -if^{dea}G^{ia}G^{kc}G^{ie}T^{d} -if^{dea}G^{kc}T^{d}G^{ie}G^{ia}-if^{dea}G^{kc}G^{ie}T^{d}G^{ia}.
\end{eqnarray}

\noindent Consideremos el primer t\'ermino de la Ec. (\ref{eq:second})
\begin{eqnarray}
\label{eq:thr}
&&if^{dea}T^{d}G^{ie}G^{kc}G^{ia}=if^{dea}([T^{d},G^{ie}]+G^{ie}T^{d})G^{kc}G^{ia}\nonumber \\
&&= if^{dea}[T^{d},G^{ie}]G^{kc}G^{ia}+if^{dea}G^{ie}T^{d}G^{kc}G^{ia}      \nonumber \\
&&= if^{dea}if^{def}G^{if}G^{kc}G^{ia}+if^{dea}G^{ie}T^{d}G^{kc}G^{ia}\nonumber \\
&&= i^{2}N_{F}\delta^{af}G^{if}G^{kc}G^{ia}+if^{dea}G^{ie}T^{d}G^{kc}G^{ia}\nonumber \\
&&= - N_{F}G^{ia}G^{kc}G^{ia}+if^{dea}G^{ie}T^{d}G^{kc}G^{ia},
\end{eqnarray}

\noindent Desarrollos similares, se hacen para todos los dem\'as t\'erminos de la Ec. (\ref{eq:second}). Posteriormente
sustituyendo cada uno de los t\'erminos obtenidos, encontramos la siguiente ecuaci\'on

\begin{eqnarray}
&&\left\{G^{ia},[G^{kc},[J^{2},G^{ia}]]\right\} = -N_{F}G^{ia}G^{kc}G^{ia}+2if^{dea}G^{ie}T^{d}G^{kc}G^{ia}\nonumber \\
&& -N_{F}G^{ia}G^{ia}G^{kc}+2if^{dea}G^{ia}G^{ie}T^{d}G^{kc} +N_{F}G^{ia}G^{kc}G^{ia} \nonumber \\
&& -2if^{dea}G^{ia}G^{kc}G^{ie}T^{d}+N_{F}G^{kc}G^{ia}G^{ia}-2if^{dea}G^{kc}G^{ie}T^{d}G^{ia}, \nonumber \\
\end{eqnarray}
\noindent simplificando, obtenemos la expresi\'on siguiente,
\begin{eqnarray}
\label{eq:siete}
&&\left\{G^{ia},[G^{kc},[J^{2},G^{ia}]]\right\} = N_{F}[G^{kc},G^{ia}G^{ia}]+2if^{dea}G^{ie}T^{d}G^{kc}G^{ia}\nonumber \\
&& +2if^{dea}G^{ia}G^{ie}T^{d}G^{kc} -2if^{dea}G^{ia}G^{kc}G^{ie}T^{d} -2if^{dea}G^{kc}G^{ie}T^{d}G^{ia}.
\end{eqnarray}

\noindent Trabajaremos ahora el conmutador $[G^{kc},G^{ia}G^{ia}]$,

\begin{eqnarray}
&&[G^{kc},G^{ia}G^{ia}] = G^{kc}G^{ia}G^{ia}-G^{ia}G^{ia}G^{kc}\nonumber \\
&&= G^{kc}\left(\frac{1}{2}\left\{G^{ia},G^{ia}\right\}+\frac{1}{2}[G^{ia},G^{ia}]\right)
 -\left(\frac{1}{2}\left\{G^{ia},G^{ia}\right\}+\frac{1}{2}[G^{ia},G^{ia}]\right)G^{kc}\nonumber \\
&&= \frac{1}{2}G^{kc}\left\{G^{ia},G^{ia}\right\}-\frac{1}{2}\left\{G^{ia},G^{ia}\right\}G^{kc}.
\end{eqnarray}
utilizando la identidad
$\left\{G^{ia},G^{ia}\right\}=\frac{3}{8}N_{c}(N_{c}+2N_{F})+\left\{J^{i},J^{i}\right\}$,

\noindent obtenemos el resultado siguiente,
\begin{eqnarray}\label{ec:siete}
&& \frac{1}{2}G^{kc}\left\{G^{ia},G^{ia}\right\}-\frac{1}{2}\left\{G^{ia},G^{ia}\right\}G^{kc} = \frac{1}{2}G^{kc}
\left(\frac{3}{8}N_{c}(N_{c}+2N_{F})+\left\{J^{i},J^{i}\right\}\right)\nonumber \\
&&-\frac{1}{2}\left(\frac{3}{8}N_{c}(N_{c}+2N_{F})+\left\{J^{i},J^{i}\right\}\right)G^{kc} = \frac{1}{2}G^{kc}2J^{2}-\frac{1}{2}2J^{2}G^{kc}
\nonumber \\
&&= G^{kc}J^{2}-J^{2}G^{kc}.
\end{eqnarray}



\noindent sustituyendo obtenemos
\begin{eqnarray}
\label{eq:ocho}
&&\left\{G^{ia},[G^{kc},[J^{2},G^{ia}]]\right\} =  N_f (G^{kc}J^{2}-J^{2}G^{kc}) + 2if^{dea}G^{ie}T^{d}[G^{kc},G^{ia}] \nonumber \\
&& +2if^{dea}[G^{ie}T^{d}G^{ia},G^{kc}] + 2if^{dea}[G^{ia}G^{ie}T^{d},G^{kc}]+2if^{dea}[G^{kc},G^{ia}]G^{ie}T^{d}.
\end{eqnarray}

\noindent Continuando,
\begin{eqnarray}
&&2if^{dea}G^{ie}T^{d}[G^{kc},G^{ia}] = 2if^{dea}\left(\frac{1}{2}\left\{G^{ie},T^{d}\right\}-\frac{1}{2}if^{dea}G^{if}\right)[G^{kc},G^{ia}]\nonumber \\
&&= if^{dea}\left\{G^{ie},T^{d}\right\}[G^{kc},G^{ia}]-i^{2}f^{dea}f^{def}G^{if}[G^{kc},G^{ia}]\nonumber \\
&&= if^{dea}\left\{G^{ie},T^{d}\right\}[G^{kc},G^{ia}]+N_{F}\delta^{af}G^{if}[G^{kc},G^{ia}]\nonumber \\
&&= if^{dea}\left\{G^{ie},T^{d}\right\}[G^{kc},G^{ia}]+N_{F}G^{ia}[G^{kc},G^{ia}]\nonumber \\
&&= if^{dea}\bigl(\left\{G^{ie},T^{d}\right\}G^{kc}G^{ia}-\left\{G^{ie},T^{d}\right\}G^{ia}G^{kc}\bigr)\nonumber \\
&& \mbox{} + if^{dea}\bigl(N_{F}G^{ia}G^{kc}G^{ia}-N_{F}G^{ia}G^{ia}G^{kc}\bigr).
\end{eqnarray}


\begin{eqnarray}
&& G^{ie}T^{d} = \frac{1}{2}\left\{G^{ie},T^{d}\right\}+\frac{1}{2}[G^{ie},T^{d}] = \frac{1}{2}\left\{G^{ie},T^{d}\right\}+ \frac{1}{2}(-[T^{d},G^{ie}])\nonumber \\
&&=\frac{1}{2}\left\{G^{ie},T^{d}\right\}- \frac{1}{2}[T^{d},G^{ie}] = \frac{1}{2}\left\{G^{ie},T^{d}\right\}- \frac{1}{2}if^{def}G^{if}.
\end{eqnarray}

\begin{eqnarray}
&&\left\{G^{ia},\left[G^{kc},[J^{2},G^{ia}]\right]\right\} = N_{F}(G^{kc}J^{2}-J^{2}G^{kc})
  + if^{dea}\left\{G^{ie},T^{d}\right\}[G^{kc},G^{ia}] \nonumber \\
&& + N_{F}G^{ia}[G^{kc},G^{ia}] if^{dea}\left\{G^{ie},T^{d}\right\}[G^{ia},G^{kc}]+if^{dea}\left[\left\{G^{ie},T^{d}\right\},G^{kc}\right]G^{ia} \nonumber \\
&& + if^{dea}G^{ia}\left[\left\{G^{ie},T^{d}\right\},G^{kc}\right]+if^{dea}[G^{ia},G^{kc}]\left\{G^{ie},T^{d}\right\}.
\end{eqnarray}

\begin{eqnarray}
&&\left\{G^{ia},\left[G^{kc},[J^{2},G^{ia}]\right]\right\} =  N_f(G^{kc}J^{2}-J^{2}G^{kc})+if^{dea}\left\{G^{ie},T^{d}\right\}G^{kc}G^{ia} \nonumber \\
&& - if^{dea}G^{kc}\left\{G^{ie},T^{d}\right\}G^{ia}+if^{dea}G^{ia}\left\{G^{ie},T^{d}\right\}G^{kc}-if^{dea}G^{kc}G^{ia}\left\{G^{ie},T^{d}\right\}\nonumber \\
&&= - \frac{1}{2}(N_f -2)G^{kc} + \frac{1}{2}(N_c + N_f){\cal D}_2^{kc}-\frac{1}{2}D_3^{kc}-{\cal O}_3^{kc}.
\end{eqnarray}

\chapter{Reducci\'on de los Operadores de Bariones \label{app:reduccion}}

\section{La evaluaci\'on de la estructura con\-mu\-ta\-dor/anti\-con\-mu\-tador con una inserci\'on de masa}

\begin{center}
\begin{equation}
\left\{A^{ia}, \left[A^{kc},\left[{\cal M}, A^{ia}\right]\right]\right\},\nonumber
\end{equation}
\end{center}
la cual representa la primera contribuci\'on de la corriente axial vector de bariones renormalizada, para la diferencia de masa octete-decuplete, da los t\'erminos siguientes:\\

\textit{1.\ Contribuci\'on de sabor singulete}

\begin{equation}
\label{GJ2GG-singlet} \{G^{ia},[G^{kc},[J^2,G^{ia}]]\} =
-\frac12(N_f-2)G^{kc} + \frac12(N_c+N_f)\mathcal{D}_2^{kc} -
\frac12\mathcal{D}_3^{kc} - \mathcal{O}_3^{kc},
\end{equation}

\begin{eqnarray}
\label{GJ2GD2-singlet} &  &
\{G^{ia},[\mathcal{D}_2^{kc},[J^2,G^{ia}]]\} +
\{\mathcal{D}_2^{ia},[G^{kc},[J^2,G^{ia}]]\} =
2(N_c+N_f)G^{kc} \nonumber\\
&  & \mbox{} + \frac12[N_c(N_c+2N_f)-9N_f-2]\mathcal{D}_2^{kc} +
\frac12(N_c+N_f)\mathcal{D}_3^{kc} - 2\mathcal{D}_4^{kc},
\end{eqnarray}

\begin{eqnarray}
\{\mathcal{D}_2^{ia},[\mathcal{D}_2^{kc},[J^2,G^{ia}]]\} =(N_f+2)
\mathcal{O}_3^{kc},
\end{eqnarray}

\begin{eqnarray}
&  & \{G^{ia},[\mathcal{D}_3^{kc},[J^2,G^{ia}]]\} + \{\mathcal{D}_3^{ia},[G^{kc},[J^2,G^{ia}]]\} = 2[N_c(N_c+2N_f)+2N_f]G^{kc} \nonumber \\
&  & \mbox{} + 10(N_c+N_f)\mathcal{D}_2^{kc} + \frac12[2N_c(N_c+2N_f)-17N_f-2]\mathcal{D}_3^{kc} - (2N_f-4)\mathcal{O}_3^{kc} \nonumber \\
&  & \mbox{} + (N_c+N_f)\mathcal{D}_4^{kc} - 3\mathcal{D}_5^{kc},
\end{eqnarray}

\begin{eqnarray}
&  & \{G^{ia},[\mathcal{O}_3^{kc},[J^2,G^{ia}]]\} +
\{G^{ia},[G^{kc},[J^2,\mathcal{O}_3^{ia}]]\} +
\{\mathcal{O}_3^{ia},[G^{kc},[J^2,G^{ia}]]\} = \nonumber \\
&  & \mbox{} 3(N_c+N_f)\mathcal{D}_2^{kc} - \frac32 N_f\mathcal{D}_3^{kc} + \frac12[2N_c(N_c+2N_f)-13N_f+2]\mathcal{O}_3^{kc} + (N_c+N_f) \mathcal{D}_4^{kc} \nonumber \\
&  & \mbox{} - \mathcal{D}_5^{kc} - 5\mathcal{O}_5^{kc},
\end{eqnarray}\\

\textit{2.\ Contribuci\'on de sabor octete}
\begin{eqnarray}
\label{GJ2GG-octet}
&  & d^{ab8}\{G^{ia},[G^{kc},[J^2,G^{ib}]]\} = -\frac14(N_f-4)d^{c8e}G^{ke} + \frac{N_c(N_c+2N_f)-2N_f+4}{4N_f}\delta^{c8}J^k \nonumber \\
&  & \mbox{} + \frac14(N_c+N_f)d^{c8e}\mathcal{D}_2^{ke} +
\frac14(N_c+N_f)[J^2,[T^8,G^{kc}]]
 - \frac14 d^{c8e}\mathcal{D}_3^{ke} - \frac12 d^{c8e}\mathcal{O}_3^{ke} \nonumber \\
&  & \mbox{} - \frac12 \{G^{kc},\{J^r,G^{r8}\}\} + \frac{1}{N_f}\{G^{k8},\{J^r,G^{rc}\}\} + \frac18\{J^k,\{T^c,T^8\}\} - \frac{1}{N_f}\{J^k,\{G^{rc},G^{r8}\}\} \nonumber \\
&  & \mbox{} - \frac{1}{2 N_f}\delta^{c8}\{J^2,J^k\},
\end{eqnarray}

\begin{eqnarray}
\label{GJ2GD2-octet} &  &
d^{ab8}\left(\{G^{ia},[\mathcal{D}_2^{kc},[J^2,G^{ib}]]\} +
\{\mathcal{D}_2^{ia},[G^{kc},[J^2,G^{ib}]]\}
\right) = (N_c+N_f)d^{c8e}G^{ke} \nonumber \\
&  & \mbox{} - \frac{7N_f+4}{4}d^{c8e}\mathcal{D}_2^{ke} + \{G^{kc},T^8\} - \frac{N_f}{2}\{T^c,G^{k8}\} - \frac{N_f^2+4}{4N_f}[J^2,[T^8,G^{kc}]] \nonumber \\
&  & \mbox{} + \frac14 (N_c+N_f)d^{c8e}\mathcal{D}_3^{ke} - \frac{N_f-2}{2N_f}(N_c+N_f) \{G^{k8},\{J^r,G^{rc}\}\}
+ \frac14(N_c+N_f) \nonumber \\
&  & \times \{J^k,\{T^c,T^8\}\} \mbox{} + \frac{N_f-2}{2N_f}(N_c+N_f) \{J^k,\{G^{rc},G^{r8}\}\} - \frac12 d^{c8e}\mathcal{D}_4^{ke}  - \frac{N_f+1}{N_f}\nonumber \\
&  & \times \{\mathcal{D}_2^{kc},\{J^r,G^{r8}\}\}
\mbox{} + \frac12\{\mathcal{D}_2^{k8},\{J^r,G^{rc}\}\} - \frac{N_f-2}{2N_f}\{J^2,\{G^{k8},T^c\}\} \},
\end{eqnarray}

\begin{eqnarray}
&  &
d^{ab8}\{\mathcal{D}_2^{ia},[\mathcal{D}_2^{kc},[J^2,G^{ib}]]\} = -\, \frac{N_c+N_f}{N_f}[J^2,[T^8,G^{kc}]] + \frac{N_f+2}{2}d^{c8e}O_3^{ke} \nonumber \\
&  & +\{G^{kc},\{J^r,G^{r8}\}\} \mbox{} - \{G^{k8},\{J^r,G^{rc}\}\} + \frac{(N_c+N_f)(N_f-2)}{2N_f}\{\mathcal{D}_2^{kc},\{J^r,G^{r8}\}\} \nonumber \\
&  & \mbox{} - \frac{(N_c+N_f)(N_f-2)}{2N_f}\{J^2,\{G^{k8},T^c\}\},
\end{eqnarray}

\begin{eqnarray}
&& d^{ab8} \left( \{G^{ia},[\mathcal{D}_3^{kc},[J^2,G^{ib}]]\} + \{\mathcal{D}_3^{ia},[G^{kc},[J^2,G^{ib}]]\} \right) = 2N_f d^{c8e}G^{ke}\nonumber \\
&  & \mbox{} + \frac{5N_c(N_c+2N_f)}{N_f} \delta^{c8}J^k + 5(N_c+N_f)d^{c8e}\mathcal{D}_2^{ke} + 2(N_c+N_f)\{G^{kc},T^8\} \nonumber \\
&  & \mbox{} - (N_c+N_f)[J^2,[T^8,G^{kc}]] - \frac54 N_f d^{c8e}\mathcal{D}_3^{ke}+\frac{2(N_f+2)}{N_f}d^{c8e}\mathcal{O}_3^{ke} \nonumber \\
&  & \mbox{} - \frac{(N_f-2)^2}{N_f}\{G^{kc},\{J^r,G^{r8}\}\} - \frac{3N_f^2-2N_f-4}{N_f}\{G^{k8},\{J^r,G^{rc}\}\} + \frac52\{J^k,\{T^c,T^8\}\}\nonumber \\
&  & \mbox{} - 2(N_f+3)\{J^k,\{G^{rc},G^{r8}\}\} + \frac{N_c(N_c+2N_f)-10N_f}{2N_f}\delta^{c8}\{J^2,J^k\} \nonumber \\
&  & \mbox{} + \frac12(N_c+N_f)d^{c8e}\mathcal{D}_4^{ke} + 2(N_c+N_f)\{\mathcal{D}_2^{k8},\{J^r,G^{rc}\}\} - \frac{1}{N_f}\delta^{c8}\{J^2,\{J^2,J^k\}\}\nonumber\\
 && - \frac12 d^{c8e}\mathcal{D}_5^{ke} - \frac{2(N_f-2)}{N_f}\{J^2,\{G^{k8},\{J^r,G^{rc}\}\}\}
+ \frac14\{J^2,\{J^k,\{T^c,T^8\}\}\}  \nonumber \\
&  & \mbox{} - \frac{2}{N_f}\{J^2,\{J^k,\{G^{rc},G^{r8}\}\}\} - \frac{3N_f+2}{2N_f}\{J^k,\{\{J^r,G^{rc}\},\{J^m,G^{m8}\}\}\},
\end{eqnarray}

\begin{eqnarray}
&  & d^{ab8} \left( \{G^{ia},[\mathcal{O}_3^{kc},[J^2,G^{ib}]]\} +
\{G^{ia},[G^{kc},[J^2,\mathcal{O}_3^{ib}]]\} +
\{\mathcal{O}_3^{ia},[G^{kc},[J^2,G^{ib}]]\} \right) = \nonumber \\
&  & \frac{3N_c(N_c+2N_f)}{2N_f}\delta^{c8}J^k + \frac32(N_c+N_f)d^{c8e}\mathcal{D}_2^{ke} + \frac32(N_c+N_f)[J^2,[T^8,G^{kc}]] \nonumber \\
&  & \mbox{} - \frac{(3N_f+4)(N_f-2)}{4N_f}d^{c8e}\mathcal{D}_3^{ke} - \frac{7N_f-4}{4}d^{c8e}\mathcal{O}_3^{ke} - \frac32 N_f\{G^{kc},\{J^r,G^{r8}\}\} \nonumber \\
&  & \mbox{} + \frac32 N_f\{G^{k8},\{J^r,G^{rc}\}\} + \frac34\{J^k,\{T^c,T^8\}\} - \frac{N_f+4}{N_f}\{J^k,\{G^{rc},G^{r8}\}\} \nonumber \\
&  & \mbox{} + \frac{N_cN_f(N_c+2N_f)-2N_f(3N_f-1)+8}{2N_f^2}\delta^{c8}\{J^2,J^k\} + \frac12 (N_c+N_f)d^{c8e}\mathcal{D}_4^{ke} \nonumber\\
&  & \mbox{} - (N_c+N_f)\{\mathcal{D}_2^{k8},\{J^r,G^{rc}\}\} + (N_c+N_f)\{J^2,\{G^{kc},T^8\}\} - \frac12 d^{c8e}\mathcal{D}_5^{ke} \nonumber \\
&  & \mbox{} + \frac34(N_c+N_f)\{J^2,[J^2,[T^8,G^{kc}]]\} - \frac32 d^{c8e}\mathcal{O}_5^{ke} - \frac{1}{N_f} \delta^{c8}\{J^2,\{J^2,J^k\}\} \nonumber \\
&  & \mbox{} - \frac72\{J^2,\{G^{kc},\{J^r,G^{r8}\}\}\} + \frac{N_f+1}{N_f}\{J^2,\{G^{k8},\{J^r,G^{rc}\}\}\} \nonumber \\
&  & \mbox{} + \frac14\{J^2,\{J^k,\{T^c,T^8\}\}\} - \frac{2}{N_f}\{J^2,\{J^k,\{G^{rc},G^{r8}\}\}\} \nonumber \\
&  & \mbox{} + \frac{3N_f+2}{4N_f}\{J^k,\{\{J^r,G^{rc}\},\{J^m,G^{m8}\}\}\},
\end{eqnarray}

\vspace{2cm}
\textit{3.\ Contribuci\'on de sabor {\bf 27} }

\begin{eqnarray}
\label{GJ2GG-27}
&  & \{G^{i8},[G^{kc},[J^2,G^{i8}]]\} = - \frac{1}{N_f} \delta^{c8}\mathcal{O}_3^{k8} + \frac12 d^{c8e}\{J^k,\{G^{re},G^{r8}\}\} - \frac12 d^{c8e}\{G^{k8},\{J^r,G^{re}\}\} \nonumber \\
&  & \mbox{} - \frac14 \epsilon^{kim}f^{c8e}\{T^e,\{J^i,G^{m8}\}\}
,
\end{eqnarray}

\begin{eqnarray}
\label{GJ2GD2-27}
&  & \{G^{i8}, [\mathcal{D}_2^{kc}, [J^2,G^{i8}]] \} + \{G^{i8}, [G^{kc},[J^2, \mathcal{D}_2^{i8}]] \} + \{\mathcal{D}_2^{i8}, [G^{kc}, [J^2,G^{i8}]] \} = \nonumber \\
&  & -\frac{15}{4}f^{c8e}f^{8eg}\mathcal{D}_2^{kg} +
\frac{i}{2}f^{c8e}[G^{ke},\{J^r,G^{r8}\}] -
if^{c8e}[G^{k8},\{J^r,G^{re}\}] - \frac12 f^{c8e}f^{8eg}\mathcal{D}_4^{kg} \nonumber \\
&  & \mbox{} + \{\mathcal{D}_2^{kc},\{G^{r8},G^{r8}\}\} +
\{\mathcal{D}_2^{k8},\{G^{rc},G^{r8}\}\}
- \frac12 \{\{J^r,G^{rc}\},\{G^{k8},T^8\}\} \nonumber \\
&  & \mbox{} - \frac12 \{\{J^r,G^{r8}\},\{G^{k8},T^c\}\} +
\frac{i}{2}f^{c8e}\{J^k,[\{J^i,G^{ie}\},\{J^r,G^{r8}\}]\}.
\end{eqnarray}

\begin{eqnarray}
&  & \{\mathcal{D}_2^{i8},[\mathcal{D}_2^{kc},[J^2,G^{i8}]]\} = -\frac14 f^{c8e}f^{8eg}\mathcal{D}_3^{kg} + \frac12 f^{c8e}f^{8eg}\mathcal{O}_3^{kg} + \frac12 \epsilon^{kim}f^{c8e}\{T^e,\{J^i,G^{m8}\}\} \nonumber \\
&  & \mbox{} - \frac12 \epsilon^{kim}f^{c8e}\{T^8,\{J^i,G^{me}\}\}
+ \frac12 \{\mathcal{D}_2^{kc},\{T^8,\{J^r,G^{r8}\}\}\} - \frac12
\{J^2,\{G^{k8},\{T^c,T^8\}\}\},\nonumber\\
&&
\end{eqnarray}

\begin{eqnarray}
& & \{G^{i8},[\mathcal{D}_3^{kc},[J^2,G^{i8}]]\} + \{\mathcal{D}_3^{i8},[G^{kc},[J^2,G^{i8}]]\} = 3 f^{c8e} f^{8eg} G^{kg}
- \frac12 d^{c8e} d^{8eg} G^{kg} \nonumber \\
& & \mbox{} - \frac{1}{2N_f} d^{c88} J^k - 2 d^{c8e} d^{8eg} \mathcal{D}_3^{kg}  + \frac{4}{N_f} \delta^{c8} \mathcal{D}_3^{k8}
- \frac{4}{N_f} \delta^{88} \mathcal{D}_3^{kc} - d^{c8e} d^{8eg} \mathcal{O}_3^{kg} + f^{c8e} f^{8eg} \mathcal{O}_3^{kg} \nonumber \\
&  & + \frac{2}{N_f} \delta^{c8} \mathcal{O}_3^{k8} - \frac{2}{N_f} \delta^{88} \mathcal{O}_3^{kc} + 4 \{G^{kc},\{G^{r8},G^{r8}\}\} \mbox{} - 4 \{G^{k8},\{G^{rc},G^{r8}\}\} - \frac{4}{N_f} d^{c88} \{J^2,J^k\}\nonumber \\
& & + 6 d^{c8e} \{J^k,\{G^{re},G^{r8}\}\} - 2 d^{88e} \{J^k,\{G^{rc},G^{re}\}\} + \frac12 d^{c8e} \{G^{ke},\{J^r,G^{r8}\}\} \nonumber \\
& & \mbox{} + 2 d^{c8e} \{G^{k8},\{J^r,G^{re}\}\} - d^{88e} \{G^{kc},\{J^r,G^{re}\}\} - d^{88e} \{G^{ke},\{J^r,G^{rc}\}\} \nonumber \\
& & \mbox{} + \epsilon^{kim} f^{c8e} \{T^e,\{J^i,G^{m8}\}\}
- 2 \{\{J^r,G^{rc}\},\{G^{k8},\{J^i,G^{i8}\}\}\} \nonumber\\
& & + 2 \{J^k,\{\{J^i,G^{ic}\},\{G^{r8},G^{r8}\}\}\} \mbox{} - \frac12 d^{c8e} \{\mathcal{D}_3^{k8},\{J^r,G^{re}\}\}\nonumber \\
& & + d^{c8e} \{J^2,\{J^k,\{G^{re},G^{r8}\}\}\},
\end{eqnarray}

\begin{eqnarray}
& & \{G^{i8},[\mathcal{O}_3^{kc},[J^2,G^{i8}]]\} + \{G^{i8},[G^{kc},[J^2,\mathcal{O}_3^{i8}]]\} + \{\mathcal{O}_3^{i8},[G^{kc},[J^2,G^{i8}]]\} \nonumber \\
& & \mbox{} = -\, \frac12 d^{c8e} d^{8eg} G^{kg} - \frac{1}{2N_f} d^{c88} J^k - d^{c8e} d^{8eg} \mathcal{D}_3^{kg} - \frac12 d^{c8e} d^{8eg} \mathcal{O}_3^{kg}
- \frac32 f^{c8e} f^{8eg} \mathcal{O}_3^{kg}
\nonumber \\
& & \mbox{}- \frac{7}{N_f} \delta^{c8} \mathcal{O}_3^{k8} - \frac{1}{N_f} \delta^{88} \mathcal{O}_3^{kc} + 2 d^{c8e} \{J^k,\{G^{re},G^{r8}\}\} + \frac72 d^{c8e} \{G^{ke},\{J^r,G^{r8}\}\} \nonumber \\
& & \mbox{}- 3 d^{c8e} \{G^{k8},\{J^r,G^{re}\}\} - \frac12 d^{88e} \{G^{kc},\{J^r,G^{re}\}\}  + \frac12 d^{88e} \{G^{ke},\{J^r,G^{rc}\}\}\nonumber\\
& & \mbox{}- \frac{2}{N_f} d^{c88} \{J^2,J^k\} - \frac{3}{N_f} \delta^{c8} \mathcal{O}_5^{k8} - \epsilon^{kim} f^{c8e} \{T^e,\{J^i,G^{m8}\}\} \nonumber\\
& & \mbox{}- \{G^{kc},\{\{J^i,G^{i8}\},\{J^r,G^{r8}\}\}\} + \{\{J^r,G^{rc}\},\{G^{k8},\{J^i,G^{i8}\}\}\} \nonumber \\
& & \mbox{}- \{J^k,\{\{J^i,G^{ic}\},\{G^{r8},G^{r8}\}\}\} + 2 \{J^2,\{G^{kc},\{G^{r8},G^{r8}\}\}\} \nonumber \\
& & \mbox{} + \frac14 d^{c8e} \{\mathcal{D}_3^{k8},\{J^r,G^{re}\}\} + d^{c8e} \{J^2,\{J^k,\{G^{re},G^{r8}\}\}\} \nonumber \\
& & \mbox{} - \frac34 d^{c8e} \{J^2,\{G^{k8},\{J^r,G^{re}\}\}\} - \frac34 \epsilon^{kim} f^{c8e} \{J^2,\{T^e,\{J^i,G^{m8}\}\}\}.
\end{eqnarray}

\section{La evaluaci\'on de la estructura de conmutadores con dos inserciones de masa \label{app:evaluacion}}

La contribuci\'on al siguiente orden de la renormalizaci\'on de la corriente axial vector de bariones, para la diferencia de masas octete-decuplete de dos estructuras de operadores es la siguiente,

\[ \left[A^{ja},\left[\left[{\cal M},
         \left[{\cal M},A^{jb}\right]\right],A^{kc}\right]\right] \hspace{1cm}
\mbox{y}\hspace{1cm}
 \left[\left[{\cal M},A^{ja}\right],\left[\left[{\cal M},
         A^{jb}\right],A^{kc}\right]\right],
\]
cada una con dos inserciones de masa, explicitamente los diferentes t\'erminos son,\\

{\it 1. Contribuci\'on de sabor singulete}
\begin{equation}
[G^{ia},[[J^2,[J^2,G^{ia}]],G^{kc}]] = -\frac32 (N_c+N_f)
\mathcal{D}_2^{kc} + \frac12 (N_f+1) \mathcal{D}_3^{kc} + N_f
\mathcal{O}_3^{kc},
\end{equation}

\begin{eqnarray}
&  & [G^{ia},[[J^2,[J^2,G^{ia}]],\mathcal{D}_2^{kc}]] +  [\mathcal{D}_2^{ia},[[J^2,[J^2,G^{ia}]],G^{kc}]] = - \frac32 \left[ N_c(N_c+2N_f)-6 N_f \right] \mathcal{D}_2^{kc}  \nonumber \\
&  & \mbox{} - \frac{7}{2} (N_c+N_f) \mathcal{D}_3^{kc} - 2
(N_c+N_f) \mathcal{O}_3^{kc} + (3N_f+8) \mathcal{D}_4^{kc},
\end{eqnarray}

\begin{eqnarray}
[[J^2,G^{ia}],[[J^2,G^{ia}],G^{kc}]] & = & - \left[N_c(N_c+2 N_f)-N_f \right] G^{kc} + \frac52 (N_c+N_f) \mathcal{D}_2^{kc} \nonumber \\
&  & \mbox{} - \frac12 (N_f+1) \mathcal{D}_3^{kc} - (N_f-1) \mathcal{O}_3^{kc},
\end{eqnarray}

\begin{eqnarray}
[[J^2,G^{ia}],[[J^2,G^{ia}],\mathcal{D}_2^{kc}]] & = & \frac32 \left[N_c(N_c+2N_f)-4 N_f\right] \mathcal{D}_2^{kc} + \frac52 (N_c+N_f) \mathcal{D}_3^{kc} \nonumber \\
&  &       \mbox{} - 3 (N_f+2) \mathcal{D}_4^{kc},
\end{eqnarray}\\

{\it 2. Contribuci\'on de sabor octete}

\begin{eqnarray}
&  & d^{ab8} [G^{ia},[[J^2,[J^2,G^{ib}]],G^{kc}]] = - \frac{3N_c(N_c+2N_f)}{4N_f} \delta^{c8} J^k  - \frac34 (N_c+N_f) d^{c8e} \mathcal{D}_2^{ke} \nonumber \\
&  & \mbox{} - \frac14 (N_c+N_f) [J^2,[T^8,G^{kc}]] + \frac{N_f^2+N_f-4}{4N_f} d^{c8e} \mathcal{D}_3^{ke} + \frac{N_f^2-2}{2N_f} d^{c8e} \mathcal{O}_3^{ke} \nonumber \\
&  & \mbox{} - \frac38 \{J^k,\{T^c,T^8\}\} + \frac{(N_f+4)}{2 N_f} \{J^k,\{G^{rc},G^{r8}\}\} + \frac{1}{N_f} \{G^{kc},\{J^r,G^{r8}\}\} \nonumber \\
&  & \mbox{}  - \frac{1}{N_f} \{G^{k8},\{J^r,G^{rc}\}\} + \frac{2
N_f^2+N_f-4}{2N_f^2} \delta^{c8} \{J^2,J^k\} ,
\end{eqnarray}

\begin{eqnarray}
&  & d^{ab8} \left( [G^{ia},[[J^2,[J^2,G^{ib}]],\mathcal{D}_2^{kc}]] + [\mathcal{D}_2^{ia},[[J^2,[J^2,G^{ib}]],G^{kc}]] \right) = \frac92 N_f d^{c8e} \mathcal{D}_2^{ke} \nonumber \\
&  & \mbox{} + \frac12 (N_f-2) [J^2,[T^8,G^{kc}]] - \frac{(N_c+N_f)(5N_f+4)}{4N_f} d^{c8e} \mathcal{D}_3^{ke}- \{J^2,\{G^{kc},T^8\}\}
\nonumber \\
&  & \mbox{} - \frac{(N_c+N_f)(N_f+2)}{2N_f}d^{c8e}\mathcal{O}_3^{ke} - \frac{(N_c+N_f)(N_f-2)}{2N_f} \{G^{kc},\{J^r,G^{r8}\}\}
\nonumber \\
&  & \mbox{}+ \frac{(N_c+N_f)(N_f-2)}{2N_f}\{G^{k8},\{J^r,G^{rc}\}\} - \frac34 (N_c+N_f) \{J^k,\{T^c,T^8\}\} \nonumber \\
&  & \mbox{} - \frac{(N_c+N_f)(N_f-2)}{N_f} \{J^k,\{G^{rc},G^{r8}\}\} + \frac{(N_c+N_f)(N_f-2)}{N_f^2} \delta^{c8}\{J^2,J^k\} \nonumber \\
&  & \mbox{} + \frac12 (N_f+7)d^{c8e} \mathcal{D}_4^{ke} + \frac{N_f^2+7N_f+4}{2N_f} \{\mathcal{D}_2^{kc},\{J^r,G^{r8}\}\}
- \frac{5}{2} \{\mathcal{D}_2^{k8},\{J^r,G^{rc}\}\} \nonumber \\
&  & \mbox{} + \frac{N_f^2+2N_f-4}{2N_f} \{J^2,\{G^{k8},T^c\}\} + \frac{N_f+2}{2N_f} \{J^2,[J^2,[T^8,G^{kc}]]\},
\end{eqnarray}

\begin{eqnarray}
&  & d^{ab8} [[J^2,G^{ia}],[[J^2,G^{ib}],G^{kc}]] = \frac{N_f}{2} d^{c8e} G^{ke} + \frac{5N_c(N_c+2N_f)}{4N_f} \delta^{c8}J^k  + \frac54 (N_c+N_f) \nonumber \\
&  & \mbox{} \times d^{c8e} \mathcal{D}_2^{ke} - (N_c+N_f) \{G^{kc},T^8\} + \frac12 (N_c+N_f)[J^2,[T^8,G^{kc}]] - \frac14 (N_f+1) d^{c8e} \mathcal{D}_3^{ke}
\nonumber \\
&  & \mbox{} - \frac{N_f}{2} d^{c8e} \mathcal{O}_3^{ke} + \{G^{kc},\{J^r,G^{r8}\}\} + \frac{N_f}{2} \{G^{k8},\{J^r,G^{rc}\}\} + \frac58 \{J^k,\{T^c,T^8\}\} \nonumber \\
&  & \mbox{} - \frac12 (N_f+3) \{J^k,\{G^{rc},G^{r8}\}\} - \frac{2N_f+1}{2N_f} \delta^{c8} \{J^2,J^k\},
\end{eqnarray}

\begin{eqnarray}
&  & d^{ab8} [[J^2,G^{ia}],[[J^2,G^{ib}],\mathcal{D}_2^{kc}]] = - 3 N_f d^{c8e} \mathcal{D}_2^{ke} + \frac54 (N_c+N_f) d^{c8e} \mathcal{D}_3^{ke} \nonumber \\
&   & \mbox{} + \frac34 (N_c+N_f) \{J^k,\{T^c,T^8\}\} - \frac12 (N_f+5) d^{c8e} \mathcal{D}_4^{ke} - \frac{2N_f+7}{2} \{\mathcal{D}_2^{kc},\{J^r,G^{r8}\}\} \nonumber \\
&  & \mbox{} + \frac52 \{\mathcal{D}_2^{k8},\{J^r,G^{rc}\}\},
\end{eqnarray}\\

{\it 3. Contribuci\'on de sabor {\bf 27}}

\begin{eqnarray}
&  & [G^{i8},[[J^2,[J^2,G^{i8}]],G^{kc}]] = \frac12 d^{c8e} d^{8eg} \mathcal{D}_3^{kg} + \frac12 f^{c8e} f^{8eg} \mathcal{O}_3^{kg} + \frac12 d^{c8e} d^{8eg} \mathcal{O}_3^{kg} \nonumber \\
&  & \mbox{} + \frac{2}{N_f} \delta^{c8} \mathcal{O}_3^{k8} - d^{c8e} \{J^k,\{G^{re},G^{r8}\}\} - \frac12 d^{c8e}\{G^{ke},\{J^r,G^{r8}\}\} \nonumber \\
&  & \mbox{} + \frac12 d^{c8e}\{G^{k8},\{J^r,G^{re}\}\} + \frac{1}{N_f} d^{c88}\{J^2,J^k\},
\end{eqnarray}

\begin{eqnarray}
&  & [G^{i8},[[J^2,[J^2,G^{i8}]],\mathcal{D}_2^{kc}]] + [\mathcal{D}_2^{i8},[[J^2,[J^2,G^{i8}]],G^{kc}]] = \frac{15}{2} f^{c8e} f^{8eg} \mathcal{D}_2^{kg} \nonumber \\
&  & \mbox{} + \frac{i}{2} f^{c8e} [G^{k8},\{J^r,G^{re}\}] + 4 f^{c8e} f^{8eg}\mathcal{D}_4^{kg} + \frac{2}{N_f} \delta^{c8} \mathcal{D}_4^{k8} + \frac{2}{N_f} \delta^{88} \mathcal{D}_4^{kc} \nonumber \\
&  & \mbox{} - 2 \{\mathcal{D}_2^{kc},\{G^{r8},G^{r8}\}\} - 2 \{\mathcal{D}_2^{k8},\{G^{rc},G^{r8}\}\} + \frac12 d^{c8e} \{J^2,\{G^{ke},T^8\}\} \nonumber \\
&  & \mbox{} + \frac12 d^{88e} \{J^2,\{G^{ke},T^c\}\} + \frac12 d^{c8e} \{\mathcal{D}_2^{k8},\{J^r,G^{re}\}\} + \frac12 d^{88e} \{\mathcal{D}_2^{kc},\{J^r,G^{re}\}\} \nonumber \\
&  & \mbox{} + \frac12 \{\{J^r,G^{rc}\},\{G^{k8},T^8\}\} - \frac12 \{\{J^r,G^{r8}\},\{G^{kc},T^8\}\} \nonumber \\
&  & \mbox{} - if^{c8e} \{\{J^r,G^{re}\},[J^2,G^{k8}]\} + \frac{i}{2} f^{c8e} \{\{J^r,G^{r8}\},[J^2,G^{ke}]\} \nonumber\\
&  & \mbox{} - 3if^{c8e} \{J^k,[\{J^i,G^{ie}\},\{J^r,G^{r8}\}]\},
\end{eqnarray}

\begin{eqnarray}
& & [[J^2,G^{i8}],[[J^2,G^{i8}],G^{kc}]] = \frac34 f^{c8e} f^{8eg} G^{kg} - \frac12 d^{c8e}d^{8eg}\mathcal{D}_3^{kg}
+ \frac{1}{N_f} \delta^{c8}\mathcal{D}_3^{k8} \nonumber \\
& & \mbox{} - \frac12 d^{c8e} d^{8eg}\mathcal{O}_3^{kg} - \frac{2}{N_f} \delta^{c8}\mathcal{O}_3^{k8} - \{G^{kc},\{G^{r8},G^{r8}\}\}
- \{G^{k8},\{G^{rc},G^{r8}\}\} \nonumber \\
& & \mbox{} + \frac32 d^{c8e} \{J^k,\{G^{re},G^{r8}\}\} - \frac12 d^{88e} \{J^k,\{G^{rc},G^{re}\}\} + d^{c8e} \{G^{ke},\{J^r,G^{r8}\}\} \nonumber \\
& & \mbox{}- \frac12 d^{c8e} \{G^{k8},\{J^r,G^{re}\}\} + \frac12 d^{88e} \{G^{ke},\{J^r,G^{rc}\}\} - \frac{1}{N_f} d^{c88} \{J^2,J^k\} \nonumber \\
& & - \frac14 \epsilon^{kim} f^{c8e} \{T^e,\{J^i,G^{m8}\}\} - \frac12 f^{c8e}f^{8eg}\mathcal{O}_3^{kg}
\end{eqnarray}

\begin{eqnarray}
& & [[J^2,G^{i8}],[[J^2,G^{i8}],\mathcal{D}_2^{kc}]] = -6 f^{c8e} f^{8eg}\mathcal{D}_2^{kg} - \frac72 f^{c8e} f^{8eg} \mathcal{D}_4^{kg}
- \frac{2}{N_f} \delta^{88} \mathcal{D}_4^{kc} \nonumber \\
& & \mbox{} + 2 \{\mathcal{D}_2^{kc},\{G^{r8},G^{r8}\}\} - d^{88e} \{\mathcal{D}_2^{kc},\{J^r,G^{re}\}\} + if^{c8e} \{J^2,[G^{ke},\{J^r,G^{r8}\}]\}
\nonumber \\
& & \mbox{} - if^{c8e} \{J^2,[G^{k8},\{J^r,G^{re}\}]\} - if^{c8e} \{\{J^r,G^{re}\},[J^2,G^{k8}]\} \nonumber \\
& & + if^{c8e} \{\{J^r,G^{r8}\},[J^2,G^{ke}]\} + \frac32 if^{c8e} \{J^k,[\{J^i,G^{ie}\},\{J^r,G^{r8}\}]\}.
\end{eqnarray}


\chapter{Contribuciones de sabor 8 y 27 para $g_A$}

Las contribuciones de sabor {\bf 8} y {\bf 27} para el acoplamiento axial $g_{_A}$
pueden escribirse como

\begin{equation}
\delta A_{\mathbf{8}}^{kc} = \sum_{i=1}^{30} o_i O_i^{kc}, \label{eq:da8}
\end{equation}
donde los operadores $S_i^{kc}$ considerados a este orden son

\begin{eqnarray}
\begin{array}{ll}
O_{1}^{kc} = d^{c8e} G^{ke}, &
O_{2}^{kc} = \delta^{c8} J^k, \\
O_{3}^{kc} = d^{c8e} \mathcal{D}_2^{ke}, &
O_{4}^{kc} = \{G^{kc},T^8\}, \\
O_{5}^{kc} = \{G^{k8},T^c\}, &
O_{6}^{kc} = d^{c8e} \mathcal{D}_3^{ke}, \\
O_{7}^{kc} = d^{c8e} \mathcal{O}_3^{ke}, &
O_{8}^{kc} = \{G^{kc},\{J^r,G^{r8}\}\}, \\
O_{9}^{kc} = \{G^{k8},\{J^r,G^{rc}\}\}, &
O_{10}^{kc} = \{J^k,\{T^c,T^8\}\}, \\
O_{11}^{kc} = \{J^k,\{G^{rc},G^{r8}\}\}, &
O_{12}^{kc} = \delta^{c8} \{J^2,J^k\}, \\
O_{13}^{kc} = d^{c8e} \mathcal{D}_4^{ke}, &
O_{14}^{kc} = \{\mathcal{D}_2^{kc},\{J^r,G^{r8}\}\}, \\
O_{15}^{kc} = \{\mathcal{D}_2^{k8},\{J^r,G^{rc}\}\}, &
O_{16}^{kc} = \{J^2,\{G^{kc},T^8\}\}, \\
O_{17}^{kc} = \{J^2,\{G^{k8},T^c\}\}, &
O_{18}^{kc} = \{J^2,[ G^{kc},\{J^r,G^{r8}\}]\}, \\
O_{19}^{kc} = \{J^2,[G^{k8},\{J^r,G^{rc}\}]\}, &
O_{20}^{kc} = \{[J^2,G^{kc}],\{J^r,G^{r8}\}\}, \\
O_{21}^{kc} = \{[J^2,G^{k8}],\{J^r,G^{rc}\}\}, &
O_{22}^{kc} = \{J^k,[ \{J^m,G^{mc}\},\{J^r,G^{r8}\}]\}, \\
O_{23}^{kc} = d^{c8e} \mathcal{D}_5^{ke}, &
O_{24}^{kc} = d^{c8e} \mathcal{O}_5^{ke}, \\
O_{25}^{kc} = \{J^2,\{G^{kc},\{J^r,G^{r8}\}\}\}, &
O_{26}^{kc} = \{J^2,\{G^{k8},\{J^r,G^{rc}\}\}\}, \\
O_{27}^{kc} = \{J^2,\{J^k,\{T^c,T^8\}\}\}, &
O_{28}^{kc} = \{J^2,\{J^k,\{G^{rc},G^{r8}\}\}\}, \\
O_{29}^{kc} = \{J^k,\{\{J^r,G^{rc}\},\{J^m,G^{m8}\}\}\}, &
O_{30}^{kc} = \delta^{c8} \{J^2,\{J^2,J^k\}\}.
\end{array}
\end{eqnarray}
y los coeficientes correspondientes son
\begin{eqnarray}
o_{1} & = & \left[ \frac{11}{48}a_1^3 - \frac{N_c+3}{3N_c}a_1^2b_2 - \frac{9}{4N_c^2}a_1b_2^2 - \frac{5}{2N_c^2}a_1^2b_3 + \frac{3}{4N_c^2}a_1^2c_3 - \frac{3(N_c+3)}{N_c^3}a_1b_2b_3 \right] F_{\mathbf{8}}^{(1)} \nonumber \\
& & \mbox{} + \left[ - \frac18 a_1^3 - \frac{N_c+3}{2N_c} a_1^2b_2 - \frac{3}{N_c^2} a_1^2b_3 \right] \frac{\Delta}{N_c} F_{\mathbf{8}}^{(2)} - \frac18 a_1^3 \frac{\Delta^2}{N_c^2} F_{\mathbf{8}}^{(3)},
\end{eqnarray}
\begin{eqnarray}
o_{2} & = & \left[ \frac{5}{36} a_1^3 + \frac{N_c+3}{18N_c} a_1^2b_2 - \frac{3N_c^2+18N_c-8}{12N_c^2} a_1^2b_3
- \frac{N_c+6}{24N_c} a_1^2c_3 \right] F_{\mathbf{8}}^{(1)} \nonumber \\
& & \mbox{} + \left[ - \frac{N_c^2+6N_c-2}{24} a_1^3 - \frac{5N_c+30}{6N_c} a_1^2b_3 - \frac{N_c+6}{4N_c} a_1^2c_3 \right] \frac{\Delta}{N_c} F_{\mathbf{8}}^{(2)} \nonumber \\
& & - \,\frac{11N_c(N_c+6)}{144} a_1^3 \frac{\Delta^2}{N_c^2} F_{\mathbf{8}}^{(3)},
\end{eqnarray}
\begin{eqnarray}
o_{3} & = & \left[ \frac{23}{16N_c} a_1^2b_2 - \frac{3(N_c+3)}{4N_c^2} a_1^2b_3 - \frac{N_c+3}{8N_c^2} a_1^2c_3 - \frac{3}{4N_c^3}b_2^3 - \frac{3}{2N_c^3} a_1b_2b_3 \right.\nonumber \\
& &  \left.+ \frac{9}{N_c^3} a_1b_2c_3 \right] F_{\mathbf{8}}^{(1)} +\left[ - \frac{N_c+3}{8} a_1^3 + \frac{25}{8N_c} a_1^2b_2 - \frac{5(N_c+3)}{2N_c^2} a_1^2b_3\right. \nonumber \\
& &\left. - \frac{3(N_c+3)}{4N_c^2} a_1^2c_3 \right] \frac{\Delta}{N_c} F_{\mathbf{8}}^{(2)}  + \left[ - \frac{11 (N_c+3)}{48} a_1^3 + \frac{3}{N_c} a_1^2b_2 \right] \frac{\Delta^2}{N_c^2} F_{\mathbf{8}}^{(3)},
\end{eqnarray}
\begin{eqnarray}
o_{4} & = & \left[ - \frac{1}{3N_c} a_1^2b_2 - \frac{N_c+3}{12N_c^2} a_1b_2^2 - \frac{N_c+3}{2N_c^2} a_1^2b_3- \frac{N_c+3}{2N_c^2} a_1^2c_3 - \frac{3}{N_c^3} a_1b_2b_3 \right] F_{\mathbf{8}}^{(1)} \nonumber \\
& & \mbox{} + \left[ - \frac{1}{2N_c} a_1^2b_2 - \frac{N_c+3}{N_c^2} a_1^2b_3 \right] \frac{\Delta}{N_c} F_{\mathbf{8}}^{(2)} + \frac{N_c+3}{12} a_1^3 \frac{\Delta^2}{N_c^2} F_{\mathbf{8}}^{(3)},
\end{eqnarray}
\begin{equation}
o_{5} = \left[ \frac{11}{12N_c} a_1^2b_2 + \frac{N_c+3}{6N_c^2} a_1b_2^2 + \frac{4}{N_c^3} a_1b_2b_3 \right] F_{\mathbf{8}}^{(1)} + \frac{3}{4N_c} a_1^2b_2 \frac{\Delta}{N_c} F_{\mathbf{8}}^{(2)},
\end{equation}
\begin{eqnarray}
o_{6} & = & \left[ \frac{9}{16N_c^2} a_1b_2^2 + \frac{65}{48N_c^2} a_1^2b_3 + \frac{1}{6N_c^2} a_1^2c_3
+ \frac{3(N_c+3)}{4N_c^3} a_1b_2b_3 \right.\nonumber \\
& & \mbox{} \left.- \frac{23(N_c+3)}{24N_c^3} a_1b_2c_3\right] F_{\mathbf{8}}^{(1)} + \left[ \frac18 a_1^3 - \frac{N_c+3}{8N_c} a_1^2b_2 + \frac{15}{8N_c^2} a_1^2b_3\right. \nonumber\\
& &\left. + \frac{13}{24N_c^2} a_1^2c_3 \right] \frac{\Delta}{N_c} F_{\mathbf{8}}^{(2)} + \left[ \frac{7}{36} a_1^3 - \frac{53(N_c+3)}{144N_c} a_1^2b_2 \right] \frac{\Delta^2}{N_c^2} F_{\mathbf{8}}^{(3)},
\end{eqnarray}

\begin{eqnarray}
o_{7} & = & \left[ \frac{1}{8N_c^2} a_1b_2^2 - \frac{3}{4N_c^2} a_1^2b_3 + \frac{71}{48N_c^2} a_1^2c_3 - \frac{N_c+3}{6N_c^3} a_1b_2b_3 - \frac{N_c+3}{3N_c^3} a_1b_2c_3 \right] F_{\mathbf{8}}^{(1)} \nonumber \\
& & \mbox{} + \left[ \frac14 a_1^3 - \frac{5}{4N_c^2} a_1b_2^2 - \frac{5}{3N_c^2} a_1^2b_3 + \frac{17}{8N_c^2} a_1^2c_3 \right] \frac{\Delta}{N_c} F_{\mathbf{8}}^{(2)} \nonumber\\
& & + \left[ \frac{23}{72} a_1^3 - \frac{5(N_c+3)}{36N_c} a_1^2b_2 \right] \frac{\Delta^2}{N_c^2} F_{\mathbf{8}}^{(3)},
\end{eqnarray}
\begin{eqnarray}
o_{8} & = & \left[ \frac{1}{4N_c^2} a_1b_2^2 - \frac{1}{6N_c^2} a_1^2b_3 + \frac{5}{2N_c^2} a_1^2c_3
+ \frac{N_c+3}{3N_c^3} a_1b_2b_3 \right] F_{\mathbf{8}}^{(1)} \nonumber \\
& & \mbox{} +\left[ \frac14 a_1^3 - \frac{1}{2N_c^2} a_1b_2^2 + \frac{7}{6N_c^2} a_1^2b_3 + \frac{9}{4N_c^2} a_1^2c_3 \right] \frac{\Delta}{N_c} F_{\mathbf{8}}^{(2)} \nonumber \\
&& + \left[ - \frac{1}{36} a_1^3 - \frac{N_c+3}{36N_c} a_1^2b_2 \right] \frac{\Delta^2}{N_c^2} F_{\mathbf{8}}^{(3)},
\end{eqnarray}
\begin{eqnarray}
o_{9} & = & \left[ \frac{1}{4N_c^2} a_1b_2^2 + \frac{7}{3N_c^2} a_1^2b_3 - \frac{5}{4N_c^2} a_1^2c_3 + \frac{N_c+3}{3N_c^3}a_1b_2b_3 \right] F_{\mathbf{8}}^{(1)} \nonumber \\
& & \mbox{} + \left[ - \frac16 a_1^3 + \frac{N_c+3}{12N_c} a_1^2b_2 + \frac{1}{2N_c^2} a_1b_2^2
+ \frac{17}{6N_c^2} a_1^2b_3 - \frac{9}{4N_c^2} a_1^2c_3 \right] \frac{\Delta}{N_c} F_{\mathbf{8}}^{(2)} \nonumber \\
& & + \left[- \frac{13}{72} a_1^3 + \frac{N_c+3}{36N_c} a_1^2b_2 \right] \frac{\Delta^2}{N_c^2} F_{\mathbf{8}}^{(3)},
\end{eqnarray}
\begin{eqnarray}
o_{10} & = & \left[ \frac{1}{12N_c^2} a_1b_2^2 - \frac{N_c+3}{24N_c^3}b_2^3 - \frac{3}{8N_c^2} a_1^2b_3 - \frac{1}{16N_c^2} a_1^2c_3 + \frac{N_c+3}{4N_c^3} a_1b_2b_3\right. \nonumber \\
& & \mbox{}\left. - \frac{3(N_c+3)}{8N_c^3} a_1b_2c_3 \right] F_{\mathbf{8}}^{(1)} + \left[ - \frac{1}{16} a_1^3 - \frac{N_c+3}{8N_c} a_1^2b_2 - \frac{5}{4N_c^2} a_1^2b_3 \right. \nonumber\\
& & \left. - \frac{3}{8N_c^2} a_1^2c_3 \right] \frac{\Delta}{N_c} F_{\mathbf{8}}^{(2)} + \left[ - \frac{11}{96} a_1^3 - \frac{3(N_c+3)}{16N_c} a_1^2b_2\right] \frac{\Delta^2}{N_c^2} F_{\mathbf{8}}^{(3)},
\end{eqnarray}
\begin{eqnarray}
o_{11} & = & \left[ \frac{2}{N_c^2} a_1^2b_3 - \frac{1}{3N_c^2} a_1^2c_3
- \frac{N_c+3}{3N_c^3} a_1b_2c_3 \right] F_{\mathbf{8}}^{(1)} + \left[ \frac16 a_1^3 - \frac{N_c+3}{12N_c} a_1^2b_2\right.\nonumber \\
& &\left. + \frac{6}{N_c^2} a_1^2b_3  + \frac{7}{6N_c^2} a_1^2c_3 \right]\frac{\Delta}{N_c} F_{\mathbf{8}}^{(2)} + \left[ \frac49 a_1^3 - \frac{N_c+3}{18N_c} a_1^2b_2 \right] \frac{\Delta^2}{N_c^2} F_{\mathbf{8}}^{(3)},
\end{eqnarray}
\begin{eqnarray}
o_{12} & = & \left[ \frac{7}{9N_c^2} a_1^2b_3 + \frac{13}{36N_c^2} a_1^2c_3
+ \frac{N_c+3}{9N_c^3} a_1b_2c_3 \right] F_{\mathbf{8}}^{(1)} \nonumber \\
& & \mbox{} + \left[ \frac{1}{12} a_1^3 - \frac{N_c^2+6N_c-30}{12N_c^2} a_1^2b_3 - \frac{3N_c^2+18N_c-40}{36N_c^2} a_1^2c_3 \right] \frac{\Delta}{N_c} F_{\mathbf{8}}^{(2)}\nonumber\\
& & + \left[ \frac{55}{216} a_1^3 + \frac{N_c+3}{54N_c} a_1^2b_2 \right] \frac{\Delta^2}{N_c^2} F_{\mathbf{8}}^{(3)},
\end{eqnarray}
\begin{eqnarray}
o_{13} & = & \left[ \frac{3}{8N_c^3}b_2^3 + \frac{3}{4N_c^3} a_1b_2b_3 + \frac{3}{N_c^3} a_1b_2c_3 \right] F_{\mathbf{8}}^{(1)} + \left[ \frac{1}{4N_c} a_1^2b_2 - \frac{N_c+3}{4N_c^2} a_1^2b_3 \right. \nonumber \\
& &\left. - \frac{N_c+3}{4N_c^2} a_1^2c_3 \right] \frac{\Delta}{N_c} F_{\mathbf{8}}^{(2)} + \frac{7}{6N_c} a_1^2b_2 \frac{\Delta^2}{N_c^2} F_{\mathbf{8}}^{(3)},
\end{eqnarray}
\begin{eqnarray}
o_{14} & = & \left[ \frac{1}{2N_c^3}b_2^3 - \frac{7}{2N_c^3} a_1b_2b_3 + \frac{10}{3N_c^3} a_1b_2c_3 \right] F_{\mathbf{8}}^{(1)} + \left[ \frac{2}{3N_c} a_1^2b_2 - \frac{N_c+3}{12N_c^2} a_1b_2^2 \right] \frac{\Delta}{N_c} F_{\mathbf{8}}^{(2)} \nonumber\\
& & + \frac{107}{72N_c} a_1^2b_2 \frac{\Delta^2}{N_c^2} F_{\mathbf{8}}^{(3)},
\end{eqnarray}
\begin{eqnarray}
o_{15} & = & \left[ \frac{2}{N_c^3} a_1b_2b_3 - \frac{23}{12N_c^3} a_1b_2c_3 \right] F_{\mathbf{8}}^{(1)} + \left[ - \frac{1}{4N_c} a_1^2b_2 - \frac{N_c+3}{N_c^2} a_1^2b_3\right. \nonumber \\
& & \left.+ \frac{N_c+3}{2N_c^2} a_1^2c_3 \right] \frac{\Delta}{N_c} F_{\mathbf{8}}^{(2)} - \frac{5}{8N_c} a_1^2b_2 \frac{\Delta^2}{N_c^2} F_{\mathbf{8}}^{(3)},
\end{eqnarray}
\begin{equation}
o_{16} = \left[ \frac{1}{6N_c^3} a_1b_2b_3 - \frac{1}{3N_c^3} a_1b_2c_3 \right] F_{\mathbf{8}}^{(1)} - \frac{N_c+3}{2N_c^2} a_1^2c_3 \frac{\Delta}{N_c} F_{\mathbf{8}}^{(2)} - \frac{1}{6N_c} a_1^2b_2 \frac{\Delta^2}{N_c^2} F_{\mathbf{8}}^{(3)},
\end{equation}
\begin{eqnarray}
o_{17} & = & \left[ \frac{5}{3N_c^3} a_1b_2b_3 + \frac{11}{12N_c^3} a_1b_2c_3 \right] F_{\mathbf{8}}^{(1)} + \left[ \frac{1}{12N_c} a_1^2b_2 + \frac{N_c+3}{12N_c^2} a_1b_2^2 \right] \frac{\Delta}{N_c} F_{\mathbf{8}}^{(2)} \nonumber \\
& & + \frac{11}{36N_c} a_1^2b_2 \frac{\Delta^2}{N_c^2} F_{\mathbf{8}}^{(3)},
\end{eqnarray}
\begin{eqnarray}
o_{18} & = & \left[ - \frac{1}{16N_c^3} a_1b_2b_3 - \frac{15}{64N_c^3} a_1b_2c_3 \right] F_{\mathbf{8}}^{(1)} + \left[ - \frac{1}{32N_c} a_1^2b_2 - \frac{3}{32N_c^2} a_1b_2^2 \right. \nonumber \\
& & \left. + \frac{9N_c-91}{384N_c^2} a_1^2b_3 + \frac{42N_c-65}{3072N_c^2} a_1^2c_3 \right] \frac{\Delta}{N_c} F_{\mathbf{8}}^{(2)} \nonumber\\
&& + \left[ \frac{1}{2304} a_1^3 - \frac{1}{24N_c} a_1^2b_2 \right] \frac{\Delta^2}{N_c^2} F_{\mathbf{8}}^{(3)},
\end{eqnarray}
\begin{eqnarray}
o_{19} & = & \left[ \frac{1}{16N_c^3} a_1b_2b_3 + \frac{15}{64N_c^3} a_1b_2c_3 \right] F_{\mathbf{8}}^{(1)} + \left[ \frac{1}{32N_c} a_1^2b_2 + \frac{3}{32N_c^2} a_1b_2^2 - \frac{9N_c-91}{384N_c^2} a_1^2b_3 \right. \nonumber \\
& & \left. - \frac{42N_c-65}{3072N_c^2} a_1^2c_3 \right] \frac{\Delta}{N_c} F_{\mathbf{8}}^{(2)} + \left[ - \frac{1}{2304} a_1^3 + \frac{1}{24N_c} a_1^2b_2 \right] \frac{\Delta^2}{N_c^2} F_{\mathbf{8}}^{(3)},
\end{eqnarray}
\begin{eqnarray}
o_{20} & = & \left[ - \frac{1}{16N_c^3} a_1b_2b_3 - \frac{15}{64N_c^3} a_1b_2c_3 \right] F_{\mathbf{8}}^{(1)}
             + \left[ - \frac{1}{32N_c} a_1^2b_2 - \frac{3}{32N_c^2} a_1b_2^2 + \frac{9N_c-91}{384N_c^2} a_1^2b_3 \right. \nonumber \\
& & \left. + \frac{42N_c-65}{3072N_c^2} a_1^2c_3 \right] \frac{\Delta}{N_c} F_{\mathbf{8}}^{(2)}
 + \left[ \frac{1}{2304} a_1^3 - \frac{1}{24N_c} a_1^2b_2 \right] \frac{\Delta^2}{N_c^2} F_{\mathbf{8}}^{(3)},
\end{eqnarray}
\begin{eqnarray}
o_{21} & = & \left[ \frac{1}{16N_c^3} a_1b_2b_3 + \frac{15}{64N_c^3} a_1b_2c_3 \right] F_{\mathbf{8}}^{(1)} + \left[ \frac{1}{32N_c} a_1^2b_2 + \frac{3}{32N_c^2} a_1b_2^2 - \frac{9N_c-91}{384N_c^2} a_1^2b_3 \right. \nonumber\\
       & & \left. - \frac{42N_c-65}{3072N_c^2} a_1^2c_3 \right] \frac{\Delta}{N_c} F_{\mathbf{8}}^{(2)}
           + \left[ - \frac{1}{2304} a_1^3 + \frac{1}{24N_c} a_1^2b_2 \right] \frac{\Delta^2}{N_c^2} F_{\mathbf{8}}^{(3)},
\end{eqnarray}
\begin{eqnarray}
o_{22} & = & \left[ \frac{1}{16N_c^3} a_1b_2b_3 + \frac{15}{64N_c^3} a_1b_2c_3 \right] F_{\mathbf{8}}^{(1)} +
\left[ \frac{1}{32N_c} a_1^2b_2 + \frac{3}{32N_c^2} a_1b_2^2 - \frac{9N_c-91}{384N_c^2} a_1^2b_3 \right.\nonumber \\
        & & \left. - \frac{42N_c-65}{3072N_c^2} a_1^2c_3 \right]
            \frac{\Delta}{N_c} F_{\mathbf{8}}^{(2)} + \left[ - \frac{1}{2304} a_1^3 + \frac{1}{24N_c} a_1^2b_2 \right] \frac{\Delta^2}{N_c^2} F_{\mathbf{8}}^{(3)},
\end{eqnarray}
\begin{equation}
o_{23} = \left[ \frac{1}{4N_c^2} a_1^2b_3 + \frac{1}{4N_c^2} a_1^2c_3 \right] \frac{\Delta}{N_c} F_{\mathbf{8}}^{(2)},
\end{equation}
\begin{equation}
o_{24} = \frac{3}{4 N_c^2} a_1^2c_3 \frac{\Delta}{N_c} F_{\mathbf{8}}^{(2)},
\end{equation}
\begin{equation}
o_{25} = \frac{7}{4N_c^2} a_1^2c_3 \frac{\Delta}{N_c} F_{\mathbf{8}}^{(2)},
\end{equation}
\begin{equation}
o_{26} = \left[ \frac{1}{3N_c^2} a_1^2b_3 - \frac{2}{3N_c^2} a_1^2c_3 \right] \frac{\Delta}{N_c} F_{\mathbf{8}}^{(2)},
\end{equation}
\begin{equation}
o_{27} = \left[ - \frac{1}{8N_c^2} a_1^2b_3 - \frac{1}{8N_c^2} a_1^2c_3 \right] \frac{\Delta}{N_c} F_{\mathbf{8}}^{(2)},
\end{equation}
\begin{equation}
o_{28} = \left[ \frac{1}{3N_c^2} a_1^2b_3 + \frac{1}{3N_c^2} a_1^2c_3 \right] \frac{\Delta}{N_c} F_{\mathbf{8}}^{(2)},
\end{equation}
\begin{equation}
o_{29} = \left[ \frac{11}{12N_c^2} a_1^2b_3 - \frac{11}{24N_c^2} a_1^2c_3 \right] \frac{\Delta}{N_c} F_{\mathbf{8}}^{(2)},
\end{equation}
\begin{equation}
o_{30} = \left[ \frac{1}{6N_c^2} a_1^2b_3 + \frac{1}{6N_c^2} a_1^2 c_3 \right] \frac{\Delta}{N_c} F_{\mathbf{8}}^{(2)},
\end{equation}

\newpage
Finalmente, para la representaci\'on $\mathbf{27}$ obtenemos
\begin{equation}
\delta A_{\mathbf{27}}^{kc} = \sum_{i=1}^{61} t_i T_i^{kc}, \label{eq:da27}
\end{equation}
donde la base de operadores es
\begin{eqnarray}
\begin{array}{ll}
T_{1}^{kc} = f^{c8e} f^{8eg} G^{kg}, &
T_{2}^{kc} = d^{c8e} d^{8eg} G^{kg}, \\
T_{3}^{kc} = \delta^{c8} G^{k8}, &
T_{4}^{kc} = d^{c88} J^k, \\
T_{5}^{kc} = f^{c8e} f^{8eg} \mathcal{D}_2^{kg}, &
T_{6}^{kc} = \delta^{c8} \mathcal{D}_2^{k8}, \\
T_{7}^{kc} = \delta^{88} \mathcal{D}_2^{kc}, &
T_{8}^{kc} = d^{c8e} \{G^{ke},T^8\}, \\
T_{9}^{kc} = d^{88e} \{G^{ke},T^c\}, &
T_{10}^{kc} = if^{c8e} [G^{ke},\{J^r,G^{r8}\}], \\
T_{11}^{kc} = if^{c8e} [G^{k8},\{J^r,G^{re}\}], &
T_{12}^{kc} = f^{c8e} f^{8eg} \mathcal{D}_3^{kg}, \\
T_{13}^{kc} = d^{c8e} d^{8eg} \mathcal{D}_3^{kg}, &
T_{14}^{kc} = \delta^{c8} \mathcal{D}_3^{k8}, \\
T_{15}^{kc} = \delta^{88} \mathcal{D}_3^{kc}, &
T_{16}^{kc} = f^{c8e} f^{8eg} \mathcal{O}_3^{kg}, \\
T_{17}^{kc} = d^{c8e} d^{8eg} \mathcal{O}_3^{kg}, &
T_{18}^{kc} = \delta^{c8} \mathcal{O}_3^{k8}, \\
T_{19}^{kc} = \delta^{88} \mathcal{O}_3^{kc}, &
T_{20}^{kc} = \{G^{kc},\{T^8,T^8\}\}, \\
T_{21}^{kc} = \{G^{k8},\{T^c,T^8\}\}, &
T_{22}^{kc} = \{G^{kc},\{G^{r8},G^{r8}\}\}, \\
T_{23}^{kc} = \{G^{k8},\{G^{rc},G^{r8}\}\}, &
T_{24}^{kc} = d^{c8e} \{J^k,\{G^{re},G^{r8}\}\}, \\
T_{25}^{kc} = d^{88e} \{J^k,\{G^{rc},G^{re}\}\}, &
T_{26}^{kc} = d^{c8e} \{G^{ke},\{J^r,G^{r8}\}\}, \\
T_{27}^{kc} = d^{c8e} \{G^{k8},\{J^r,G^{re}\}\}, &
T_{28}^{kc} = d^{88e} \{G^{kc},\{J^r,G^{re}\}\}, \\
T_{29}^{kc} = d^{88e} \{G^{ke},\{J^r,G^{rc}\}\}, &
T_{30}^{kc} = d^{c88} \{J^2,J^k\}, \\
T_{31}^{kc} = \epsilon^{kim} f^{c8e} \{T^e,\{J^i,G^{m8}\}\}, &
T_{32}^{kc} = \epsilon^{kim} f^{c8e} \{T^8,\{J^i,G^{me}\}\}, \\
T_{33}^{kc} = f^{c8e} f^{8eg} \mathcal{D}_4^{kg}, &
T_{34}^{kc} = \delta^{c8} \mathcal{D}_4^{k8}, \\
T_{35}^{kc} = \delta^{88} \mathcal{D}_4^{kc}, &
T_{36}^{kc} = \{\mathcal{D}_2^{kc},\{T^8,T^8\}\}, \\
T_{37}^{kc} = \{\mathcal{D}_2^{kc},\{G^{r8},G^{r8}\}\}, &
T_{38}^{kc} = \{\mathcal{D}_2^{k8},\{G^{rc},G^{r8}\}\}, \\
T_{39}^{kc} = d^{c8e} \{J^2,\{G^{ke},T^8\}\}, &
T_{40}^{kc} = d^{88e} \{J^2,\{G^{ke},T^c\}\}, \\
T_{41}^{kc} = d^{c8e} \{\mathcal{D}_2^{k8},\{J^r,G^{re}\}\}, &
T_{42}^{kc} = d^{88e} \{\mathcal{D}_2^{kc},\{J^r,G^{re}\}\}, \\
T_{43}^{kc} = \{\{J^r,G^{rc}\},\{G^{k8},T^8\}\}, &
T_{44}^{kc} = \{\{J^r,G^{r8}\},\{G^{kc},T^8\}\}, \\
T_{45}^{kc} = \{\{J^r,G^{r8}\},\{G^{k8},T^c\}\}, &
T_{46}^{kc} = if^{c8e} \{J^2,[G^{ke},\{J^r,G^{r8}\}]\}, \\
T_{47}^{kc} = if^{c8e} \{J^2,[G^{k8},\{J^r,G^{re}\}]\}, &
T_{48}^{kc} = if^{c8e} \{J^k,[\{J^i,G^{ie}\},\{J^r,G^{r8}\}]\}, \\
T_{49}^{kc} = if^{c8e} \{\{J^r,G^{re}\},[J^2,G^{k8}]\}, &
T_{50}^{kc} = if^{c8e} \{\{J^r,G^{r8}\},[J^2,G^{ke}]\}, \\
T_{51}^{kc} = \delta^{c8} \mathcal{O}_5^{k8}, &
T_{52}^{kc} = \{G^{kc},\{\{J^i,G^{i8}\},\{J^r,G^{r8}\}\}\}, \\
T_{53}^{kc} = \{\mathcal{D}_2^{kc},\{T^8,\{J^r,G^{r8}\}\}\}, &
T_{54}^{kc} = \{\{J^r,G^{rc}\},\{G^{k8},\{J^i,G^{i8}\}\}\}, \\
T_{55}^{kc} = \{J^k,\{\{J^i,G^{ic}\},\{G^{r8},G^{r8}\}\}\}, &
T_{56}^{kc} = \{J^2,\{G^{k8},\{T^c,T^8\}\}\}, \\
T_{57}^{kc} = \{J^2,\{G^{kc},\{G^{r8},G^{r8}\}\}\}, &
T_{58}^{kc} = d^{c8e} \{\mathcal{D}_3^{k8},\{J^r,G^{re}\}\}, \\
T_{59}^{kc} = d^{c8e} \{J^2,\{J^k,\{G^{re},G^{r8}\}\}\}, &
T_{60}^{kc} = d^{c8e} \{J^2,\{G^{k8},\{J^r,G^{re}\}\}\}, \\
T_{61}^{kc} = \epsilon^{kim} f^{c8e} \{J^2,\{T^e,\{J^i,G^{m8}\}\}\}.
\end{array}
\end{eqnarray}\\
y los coeficientes correspondientes son
\begin{eqnarray}
t_{1} &=& \left[ \frac18 a_1^3 - \frac{1}{N_c^2} a_1b_2^2 - \frac{3}{4N_c^2} a_1^2b_3 + \frac{3}{8N_c^2} a_1^2c_3 \right] F_{\mathbf{27}}^{(1)} - \frac{3}{2N_c^2} a_1^2b_3 \frac{\Delta}{N_c} F_{\mathbf{27}}^{(2)}\nonumber \\
&& - \frac{1}{16} a_1^3 \frac{\Delta^2}{N_c^2} F_{\mathbf{27}}^{(3)},
\end{eqnarray}
\begin{equation}
t_{2} = \frac14 a_1^3 F_{\mathbf{27}}^{(1)} + \left[ \frac{1}{4N_c^2} a_1^2b_3 + \frac{1}{4N_c^2} a_1^2c_3 \right] \frac{\Delta}{N_c} F_{\mathbf{27}}^{(2)},
\end{equation}
\begin{equation}
t_{3} = \frac16 a_1^3 F_{\mathbf{27}}^{(1)},
\end{equation}
\begin{equation}
t_{4} = \frac{1}{12} a_1^3 F_{\mathbf{27}}^{(1)} + \left[ \frac{1}{12N_c^2} a_1^2b_3 + \frac{1}{12N_c^2} a_1^2c_3 \right] \frac{\Delta}{N_c} F_{\mathbf{27}}^{(2)},
\end{equation}
\begin{eqnarray}
t_{5} &=& \left[ \frac{7}{8N_c} a_1^2b_2 - \frac{1}{2N_c^3}b_2^3 + \frac{9}{2N_c^3} a_1b_2c_3 \right] F_{\mathbf{27}}^{(1)} + \frac{15}{8N_c} a_1^2b_2 \frac{\Delta}{N_c} F_{\mathbf{27}}^{(2)}\nonumber \\
& & + \frac{7}{4N_c} a_1^2b_2 \frac{\Delta^2}{N_c^2} F_{\mathbf{27}}^{(3)},
\end{eqnarray}
\begin{equation}
t_{6} = \frac{1}{3N_c} a_1^2b_2 F_{\mathbf{27}}^{(1)},
\end{equation}
\begin{equation}
t_{7} = \frac{1}{6N_c} a_1^2b_2 F_{\mathbf{27}}^{(1)},
\end{equation}
\begin{equation}
t_{8} = \frac{1}{2N_c} a_1^2b_2 F_{\mathbf{27}}^{(1)},
\end{equation}
\begin{equation}
t_{9} = \frac{1}{4N_c} a_1^2b_2 F_{\mathbf{27}}^{(1)},
\end{equation}
\begin{equation}
t_{10} = - \frac{2}{N_c^3} a_1b_2b_3 F_{\mathbf{27}}^{(1)} - \frac{1}{4N_c} a_1^2b_2 \frac{\Delta}{N_c} F_{\mathbf{27}}^{(2)},
\end{equation}
\begin{equation}
t_{11} = \left[ \frac{1}{2N_c} a_1^2b_2 + \frac{2}{N_c^3} a_1b_2b_3 \right] F_{\mathbf{27}}^{(1)} + \frac{1}{2N_c} a_1^2b_2 \frac{\Delta}{N_c} F_{\mathbf{27}}^{(2)} + \frac{1}{12N_c} a_1^2b_2 \frac{\Delta^2}{N_c^2} F_{\mathbf{27}}^{(3)},
\end{equation}
\begin{equation}
t_{12} = \left[ \frac{3}{8N_c^2} a_1b_2^2 + \frac{3}{8N_c^2} a_1^2b_3 \right] F_{\mathbf{27}}^{(1)} + \frac{1}{8N_c^2} a_1b_2^2 \frac{\Delta}{N_c} F_{\mathbf{27}}^{(2)},
\end{equation}
\begin{eqnarray}
t_{13} &=& \left[ \frac{1}{4N_c^2} a_1^2b_3 + \frac{1}{4N_c^2} a_1^2c_3 \right] F_{\mathbf{27}}^{(1)} + \left[ \frac{1}{N_c^2} a_1^2b_3 + \frac{1}{2N_c^2} a_1^2c_3 \right] \frac{\Delta}{N_c} F_{\mathbf{27}}^{(2)}\nonumber \\
& & + \frac18 a_1^3 \frac{\Delta^2}{N_c^2} F_{\mathbf{27}}^{(3)},
\end{eqnarray}
\begin{equation}
t_{14} = \left[ \frac{1}{6N_c^2} a_1^2b_3 + \frac{1}{6N_c^2} a_1^2c_3 \right] F_{\mathbf{27}}^{(1)} - \frac{2}{3N_c^2} a_1^2b_3 \frac{\Delta}{N_c} F_{\mathbf{27}}^{(2)} - \frac{1}{36} a_1^3 \frac{\Delta^2}{N_c^2} F_{\mathbf{27}}^{(3)},
\end{equation}
\begin{equation}
t_{15} = \frac{1}{3N_c^2} a_1^2b_3 F_{\mathbf{27}}^{(1)} + \frac{2}{3N_c^2} a_1^2b_3 \frac{\Delta}{N_c} F_{\mathbf{27}}^{(2)},
\end{equation}
\begin{eqnarray}
t_{16} & = & \left[ \frac{1}{4N_c^2} a_1b_2^2 + \frac{3}{8N_c^2} a_1^2c_3 \right] F_{\mathbf{27}}^{(1)} + \left[ - \frac{1}{4N_c^2} a_1b_2^2 - \frac{1}{2N_c^2} a_1^2b_3 \right. \nonumber \\
& & + \, \left. \frac{3}{4N_c^2} a_1^2c_3 \right] \frac{\Delta}{N_c} F_{\mathbf{27}}^{(2)} + \frac18 a_1^3 \frac{\Delta^2}{N_c^2} F_{\mathbf{27}}^{(3)},
\end{eqnarray}
\begin{eqnarray}
t_{17} & = & \left[ \frac{1}{2N_c^2} a_1^2b_3 + \frac{1}{2N_c^2} a_1^2c_3 \right] F_{\mathbf{27}}^{(1)} + \left[ \frac{1}{2N_c^2} a_1^2b_3 + \frac{1}{4N_c^2} a_1^2c_3 \right] \frac{\Delta}{N_c} F_{\mathbf{27}}^{(2)} \nonumber \\
& & + \frac18 a_1^3 \frac{\Delta^2}{N_c^2} F_{\mathbf{27}}^{(3)},
\end{eqnarray}
\begin{eqnarray}
t_{18} & = & \frac{5}{6N_c^2} a_1^2c_3 F_{\mathbf{27}}^{(1)} + \left[ \frac16 a_1^3 - \frac{1}{3N_c^2} a_1^2b_3 + \frac{7}{6N_c^2} a_1^2c_3 \right] \frac{\Delta}{N_c} F_{\mathbf{27}}^{(2)}\nonumber \\
& & + \frac16 a_1^3 \frac{\Delta^2}{N_c^2} F_{\mathbf{27}}^{(3)},
\end{eqnarray}
\begin{equation}
t_{19} = \frac{1}{3N_c^2} a_1^2c_3 F_{\mathbf{27}}^{(1)} + \left[ \frac{1}{3N_c^2} a_1^2b_3 + \frac{1}{6N_c^2} a_1^2c_3 \right] \frac{\Delta}{N_c} F_{\mathbf{27}}^{(2)},
\end{equation}
\begin{equation}
t_{20} = \frac{1}{4N_c^2} a_1b_2^2 F_{\mathbf{27}}^{(1)},
\end{equation}
\begin{equation}
t_{21} = \frac{1}{2N_c^2} a_1b_2^2 F_{\mathbf{27}}^{(1)},
\end{equation}
\begin{equation}
t_{22} = \left[ - \frac{1}{N_c^2} a_1^2b_3 - \frac{1}{2N_c^2} a_1^2c_3 \right] F_{\mathbf{27}}^{(1)} - \frac{2}{N_c^2} a_1^2b_3 \frac{\Delta}{N_c} F_{\mathbf{27}}^{(2)} + \frac{1}{12} a_1^3 \frac{\Delta^2}{N_c^2} F_{\mathbf{27}}^{(3)},
\end{equation}
\begin{equation}
t_{23} = \left[ \frac{1}{N_c^2} a_1^2b_3 - \frac{1}{2N_c^2} a_1^2c_3 \right] F_{\mathbf{27}}^{(1)} + \frac{2}{N_c^2} a_1^2b_3 \frac{\Delta}{N_c} F_{\mathbf{27}}^{(2)} + \frac{1}{12} a_1^3 \frac{\Delta^2}{N_c^2} F_{\mathbf{27}}^{(3)},
\end{equation}
\begin{eqnarray}
t_{24} & = & \left[ - \frac{3}{2N_c^2} a_1^2b_3 - \frac{1}{4N_c^2} a_1^2c_3 \right] F_{\mathbf{27}}^{(1)} + \left[ - \frac14 a_1^3 - \frac{3}{N_c^2} a_1^2b_3 - \frac{1}{N_c^2} a_1^2c_3 \right] \frac{\Delta}{N_c} F_{\mathbf{27}}^{(2)} \nonumber \\
& & - \frac{7}{24} a_1^3 \frac{\Delta^2}{N_c^2} F_{\mathbf{27}}^{(3)},
\end{eqnarray}
\begin{equation}
t_{25} = \left[ \frac{1}{2N_c^2} a_1^2b_3 - \frac{1}{4N_c^2} a_1^2c_3 \right] F_{\mathbf{27}}^{(1)} + \frac{1}{N_c^2} a_1^2b_3 \frac{\Delta}{N_c} F_{\mathbf{27}}^{(2)} + \frac{1}{24} a_1^3 \frac{\Delta^2}{N_c^2} F_{\mathbf{27}}^{(3)},
\end{equation}
\begin{eqnarray}
t_{26} &=& \left[ \frac{2}{N_c^2} a_1^2b_3 - \frac{1}{2N_c^2} a_1^2c_3 \right] F_{\mathbf{27}}^{(1)} + \left[ - \frac{1}{4N_c^2} a_1^2b_3 - \frac{7}{4N_c^2} a_1^2c_3 \right] \frac{\Delta}{N_c} F_{\mathbf{27}}^{(2)} \nonumber \\
& & - \frac16 a_1^3 \frac{\Delta^2}{N_c^2} F_{\mathbf{27}}^{(3)},
\end{eqnarray}
\begin{eqnarray}
t_{27} &=& \left[ - \frac{1}{2N_c^2} a_1^2b_3 + \frac{3}{4N_c^2} a_1^2c_3 \right] F_{\mathbf{27}}^{(1)} + \left[ \frac14 a_1^3 - \frac{1}{N_c^2} a_1^2b_3 +  \frac{3}{2N_c^2} a_1^2c_3 \right] \frac{\Delta}{N_c} F_{\mathbf{27}}^{(2)}\nonumber \\
&&+ \frac18 a_1^3 \frac{\Delta^2}{N_c^2} F_{\mathbf{27}}^{(3)},
\end{eqnarray}
\begin{equation}
t_{28} = \frac{1}{2N_c^2} a_1^2c_3 F_{\mathbf{27}}^{(1)} + \left[ \frac{1}{2N_c^2} a_1^2b_3 + \frac{1}{4N_c^2} a_1^2c_3 \right] \frac{\Delta}{N_c} F_{\mathbf{27}}^{(2)},
\end{equation}
\begin{eqnarray}
t_{29} & = & \left[ \frac{1}{2N_c^2} a_1^2b_3 - \frac{1}{4N_c^2} a_1^2c_3 \right] F_{\mathbf{27}}^{(1)} + \left[ \frac{1}{2N_c^2} a_1^2b_3 - \frac{1}{4N_c^2} a_1^2c_3 \right] \frac{\Delta}{N_c} F_{\mathbf{27}}^{(2)} \nonumber \\
& & - \frac{1}{24} a_1^3 \frac{\Delta^2}{N_c^2} F_{\mathbf{27}}^{(3)},
\end{eqnarray}
\begin{eqnarray}
t_{30} & = & \left[ \frac{1}{6N_c^2} a_1^2b_3 + \frac{1}{6N_c^2} a_1^2c_3 \right] F_{\mathbf{27}}^{(1)} + \left[
\frac{2}{3N_c^2} a_1^2b_3 + \frac{1}{3N_c^2} a_1^2c_3 \right] \frac{\Delta}{N_c} F_{\mathbf{27}}^{(2)} \nonumber \\
& & + \frac{1}{12} a_1^3 \frac{\Delta^2}{N_c^2} F_{\mathbf{27}}^{(3)},
\end{eqnarray}
\begin{eqnarray}
t_{31} & = & \left[ - \frac{1}{4N_c^2} a_1b_2^2 - \frac{1}{4N_c^2} a_1^2b_3 + \frac{3}{8N_c^2} a_1^2c_3 \right] F_{\mathbf{27}}^{(1)} + \left[ \frac18 a_1^3 - \frac{1}{4N_c^2} a_1b_2^2 - \frac{1}{2N_c^2} a_1^2b_3 \right. \nonumber\\
& & + \left. \frac{1}{2N_c^2} a_1^2c_3\right] \frac{\Delta}{N_c} F_{\mathbf{27}}^{(2)} + \frac{1}{48} a_1^3 \frac{\Delta^2}{N_c^2} F_{\mathbf{27}}^{(3)},
\end{eqnarray}
\begin{equation}
t_{32} = \frac{1}{4N_c^2} a_1b_2^2 \frac{\Delta}{N_c} F_{\mathbf{27}}^{(2)},
\end{equation}
\begin{equation}
t_{33} = \left[ \frac{1}{4N_c^3} b_2^3 + \frac{9}{4N_c^3} a_1b_2c_3 \right] F_{\mathbf{27}}^{(1)} + \frac{1}{4N_c} a_1^2b_2 \frac{\Delta}{N_c} F_{\mathbf{27}}^{(2)} + \frac{23}{24N_c} a_1^2b_2 \frac{\Delta^2}{N_c^2} F_{\mathbf{27}}^{(3)},
\end{equation}
\begin{equation}
t_{34} = \frac{2}{3N_c^3} a_1b_2c_3 F_{\mathbf{27}}^{(1)} + \frac{1}{9N_c} a_1^2b_2 \frac{\Delta^2}{N_c^2} F_{\mathbf{27}}^{(3)},
\end{equation}
\begin{equation}
t_{35} = \frac{1}{3N_c^3} a_1b_2c_3 F_{\mathbf{27}}^{(1)} + \frac{1}{6N_c} a_1^2b_2 \frac{\Delta^2}{N_c^2} F_{\mathbf{27}}^{(3)},
\end{equation}
\begin{equation}
t_{36} = \frac{1}{4N_c^3} b_2^3 F_{\mathbf{27}}^{(1)},
\end{equation}
\begin{equation}
t_{37} = - \frac{1}{N_c^3} a_1b_2c_3 F_{\mathbf{27}}^{(1)} - \frac{1}{2N_c} a_1^2b_2 \frac{\Delta}{N_c} F_{\mathbf{27}}^{(2)} - \frac{1}{2N_c} a_1^2b_2 \frac{\Delta^2}{N_c^2} F_{\mathbf{27}}^{(3)},
\end{equation}
\begin{equation}
t_{38} = - \frac{2}{N_c^3} a_1b_2c_3 F_{\mathbf{27}}^{(1)} - \frac{1}{2N_c} a_1^2b_2 \frac{\Delta}{N_c} F_{\mathbf{27}}^{(2)} - \frac{1}{3N_c} a_1^2b_2 \frac{\Delta^2}{N_c^2} F_{\mathbf{27}}^{(3)},
\end{equation}
\begin{equation}
t_{39} = \left[ \frac{1}{N_c^3} a_1b_2b_3 + \frac{1}{2N_c^3} a_1b_2c_3 \right] F_{\mathbf{27}}^{(1)} + \frac{1}{12N_c} a_1^2b_2 \frac{\Delta^2}{N_c^2} F_{\mathbf{27}}^{(3)},
\end{equation}
\begin{equation}
t_{40} = \left[ \frac{1}{2N_c^3} a_1b_2b_3 + \frac{1}{4N_c^3} a_1b_2c_3 \right] F_{\mathbf{27}}^{(1)} + \frac{1}{12N_c} a_1^2b_2 \frac{\Delta^2}{N_c^2} F_{\mathbf{27}}^{(3)},
\end{equation}
\begin{equation}
t_{41} = \left[ - \frac{1}{N_c^3} a_1b_2b_3 + \frac{1}{2N_c^3} a_1b_2c_3 \right] F_{\mathbf{27}}^{(1)} + \frac{1}{12N_c} a_1^2b_2 \frac{\Delta^2}{N_c^2} F_{\mathbf{27}}^{(3)},
\end{equation}
\begin{equation}
t_{42} = \left[ - \frac{1}{2N_c^3} a_1b_2b_3 + \frac{1}{4N_c^3} a_1b_2c_3 \right] F_{\mathbf{27}}^{(1)} + \frac{1}{6N_c} a_1^2b_2 \frac{\Delta^2}{N_c^2} F_{\mathbf{27}}^{(3)},
\end{equation}
\begin{equation}
t_{43} = \frac{1}{N_c^3} a_1b_2b_3 F_{\mathbf{27}}^{(1)} + \frac{1}{4N_c} a_1^2b_2 \frac{\Delta}{N_c} F_{\mathbf{27}}^{(2)} + \frac{1}{12N_c} a_1^2b_2 \frac{\Delta^2}{N_c^2} F_{\mathbf{27}}^{(3)},
\end{equation}
\begin{equation}
t_{44} = \frac{1}{N_c^3} a_1b_2b_3 F_{\mathbf{27}}^{(1)} - \frac{1}{12N_c} a_1^2b_2 \frac{\Delta^2}{N_c^2} F_{\mathbf{27}}^{(3)},
\end{equation}
\begin{equation}
t_{45} = \frac{1}{N_c^3} a_1b_2b_3 F_{\mathbf{27}}^{(1)} + \frac{1}{4N_c} a_1^2b_2 \frac{\Delta}{N_c} F_{\mathbf{27}}^{(2)},
\end{equation}
\begin{equation}
t_{46} = \left[ \frac{1}{N_c^3} a_1b_2b_3- \frac{1}{2N_c^3} a_1b_2c_3 \right] F_{\mathbf{27}}^{(1)} - \frac{1}{12N_c} a_1^2b_2 \frac{\Delta^2}{N_c^2} F_{\mathbf{27}}^{(3)},
\end{equation}
\begin{equation}
t_{47} = \frac{1}{N_c^3} a_1b_2c_3 F_{\mathbf{27}}^{(1)} + \frac{1}{12N_c} a_1^2b_2 \frac{\Delta^2}{N_c^2} F_{\mathbf{27}}^{(3)},
\end{equation}
\begin{equation}
t_{48} = \left[ \frac{1}{N_c^3} a_1b_2b_3 - \frac{1}{N_c^3} a_1b_2c_3 \right] F_{\mathbf{27}}^{(1)} - \frac{1}{4N_c} a_1^2b_2 \frac{\Delta}{N_c} F_{\mathbf{27}}^{(2)} - \frac{5}{8N_c}a_1^2b_2 \frac{\Delta^2}{N_c^2} F_{\mathbf{27}}^{(3)},
\end{equation}
\begin{equation}
t_{49} = \left[ - \frac{1}{N_c^3} a_1b_2b_3 + \frac{1}{2N_c^3} a_1b_2c_3 \right] F_{\mathbf{27}}^{(1)} - \frac{1}{12N_c} a_1^2b_2 \frac{\Delta^2}{N_c^2} F_{\mathbf{27}}^{(3)},
\end{equation}
\begin{equation}
t_{50} = - \frac{1}{2N_c^3} a_1b_2c_3 F_{\mathbf{27}}^{(1)},
\end{equation}
\begin{equation}
t_{51} = \frac{1}{2N_c^2} a_1^2c_3 \frac{\Delta}{N_c} F_{\mathbf{27}}^{(2)},
\end{equation}
\begin{equation}
t_{52} = \frac{1}{2N_c^2} a_1^2c_3 \frac{\Delta}{N_c} F_{\mathbf{27}}^{(2)},
\end{equation}
\begin{equation}
t_{53} = - \frac{1}{4N_c^2} a_1b_2^2 \frac{\Delta}{N_c} F_{\mathbf{27}}^{(2)},
\end{equation}
\begin{equation}
t_{54} = \left[ \frac{1}{N_c^2} a_1^2b_3 - \frac{1}{2N_c^2} a_1^2c_3 \right] \frac{\Delta}{N_c} F_{\mathbf{27}}^{(2)},
\end{equation}
\begin{equation}
t_{55} = \left[ - \frac{1}{N_c^2} a_1^2b_3 + \frac{1}{2N_c^2} a_1^2c_3 \right] \frac{\Delta}{N_c} F_{\mathbf{27}}^{(2)},
\end{equation}
\begin{equation}
t_{56} = \frac{1}{4N_c^2} a_1b_2^2 \frac{\Delta}{N_c} F_{\mathbf{27}}^{(2)},
\end{equation}
\begin{equation}
t_{57} = - \frac{1}{N_c^2} a_1^2c_3 \frac{\Delta}{N_c} F_{\mathbf{27}}^{(2)},
\end{equation}
\begin{equation}
t_{58} = \left[ \frac{1}{4N_c^2} a_1^2b_3 - \frac{1}{8N_c^2} a_1^2c_3 \right] \frac{\Delta}{N_c} F_{\mathbf{27}}^{(2)},
\end{equation}
\begin{equation}
t_{59} = \left[ - \frac{1}{2N_c^2} a_1^2b_3 - \frac{1}{2N_c^2} a_1^2c_3 \right] \frac{\Delta}{N_c} F_{\mathbf{27}}^{(2)},
\end{equation}
\begin{equation}
t_{60} = \frac{3}{8N_c^2} a_1^2c_3 \frac{\Delta}{N_c} F_{\mathbf{27}}^{(2)},
\end{equation}
\begin{equation}
t_{61} = \frac{3}{8N_c^2} a_1^2c_3 \frac{\Delta}{N_c} F_{\mathbf{27}}^{(2)},
\end{equation}
debemos tener en cuenta que en la ecuaci\'on (D.33) las contribuciones de sabor singulete y el octete se debe restar con el fin de obtener verdaderamente la contribuci\'on de sabor {\bf 27}.

\chapter{Elementos de matriz de los operadores de bariones}
A continuaci\'on, presentamos los elementos de matriz que contribuyen a la renor\-ma\-li\-zaci\'on de la
corriente axial, en la Tabla E.1, para la contribuci\'on de sabor octete y para $\bf{10} + \overline{\bf 10}$
y en la Tabla E.2, para la contribuci\'on de sabor $\mathbf{27}$ y ${10} + \overline{\bf 10}$ de las transiciones observadas.
\begin{figure}[h]
\begin{center}
\includegraphics[width=10cm]{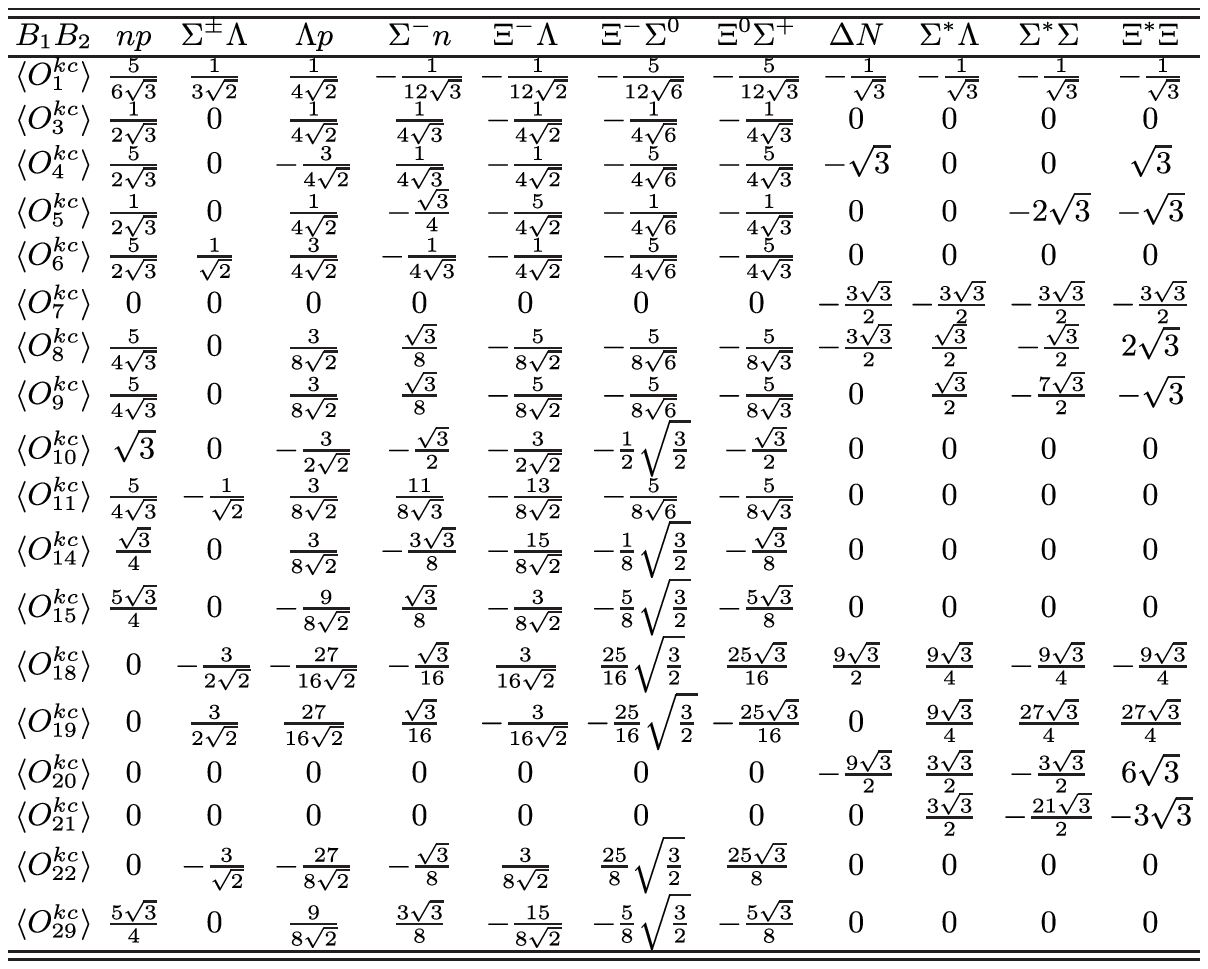}
\vspace{0.5cm}
\end{center}
Tabla E.1: Elementos de matriz de los operadores
contenidos en $g_A$ y $g$: contribuci\'on de sabor $\mathbf{8}$ y ${\bf 10} + \overline{{\bf 10}}$, respectivamente.
\end{figure}

\begin{figure}[ht]
\begin{center}
\includegraphics[width=10.5cm]{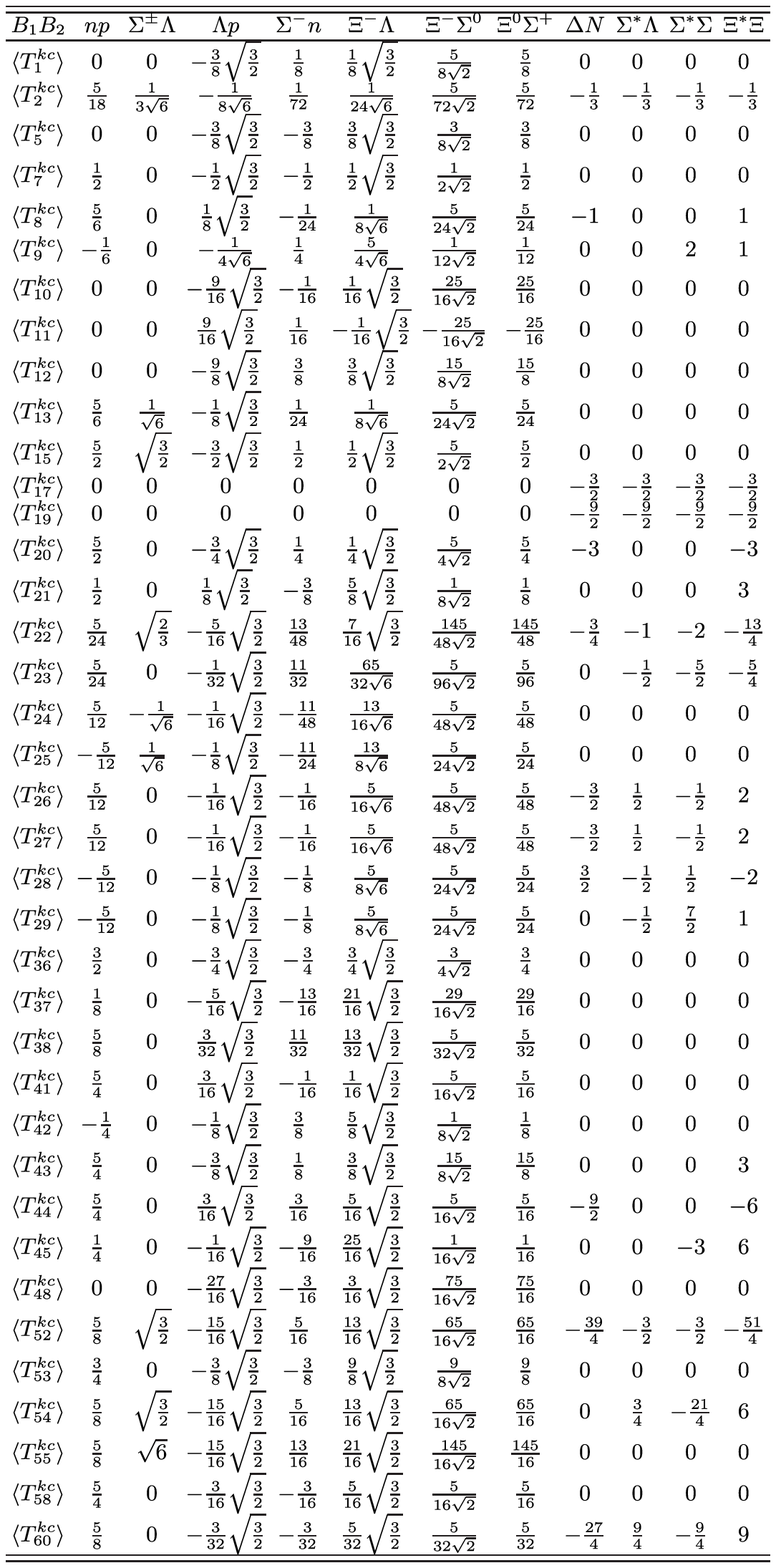}\\
\end{center}
Tabla E.2: Elementos de matriz de los operadores
contenidos en $g_A$ y $g$: contribuci\'on de sabor $\mathbf{27}$ y ${\bf 10} + \overline{\bf 10}$.
\end{figure}

\end{document}